\documentclass{aa}  
\usepackage{graphicx}
\usepackage[figuresright]{rotating}
\usepackage{txfonts}

\newcommand{\reduceme}{\mbox{R\raisebox{-0.35ex}{E}D%
\hspace{-0.05em}\raisebox{0.85ex}{uc}\hspace{-0.90em}%
\raisebox{-.35ex}{{m}}\hspace{0.05em}E}}

\begin{document}
\title{Stellar populations of early-type galaxies in different environments I. 
Line-strength indices}
\subtitle{Relations of line-strengths with $\sigma$}

\author{P. S\'{a}nchez--Bl\'{a}zquez\inst{1,2} \and 
 J. Gorgas\inst{2}
 \and
 N. Cardiel\inst{2,3} 
 \and Gonz\'alez J.J.\inst{4}
\fnmsep\thanks{E-mail:patricia.sanchezblazquez@epfl.ch}
}

\offprints{P. S\'anchez--Bl\'azquez}

\institute{
Laboratoire d'Astrophysique, \'Ecole Polytechnique F\'ed\'erale de Lausanne (EPFL),
Observatoire, 1290 Sauverny, Switzerland\\
\and 
Dpto.\ de Astrof\'{\i}sica, Fac.\ de Ciencias F\'{\i}sicas, Universidad 
Complutense de Madrid, E-28040 Madrid, Spain\\
\and 
Calar Alto Observatory, CAHA, Apartado 511, E-04004 Almer\'{\i}a, Spain\\
\and 
Instituto de Astronom\'{\i}a, Universidad Nacional Aut\'onoma de 
M\'{e}xico, Apdo.\ Postal 70--264, M\'exico D.F., M\'exico}

   \date{Received September 15, 1996; accepted March 16, 1997}

\abstract
{}
{This paper commences a series devoted to the study of the stellar content of
early-type galaxies. The goal of the series is to set constraints on the
evolutionary status of these objects.}
 {In this paper we describe the details of
the galaxy sample, the observations, and the data reduction. Line-strength
indices and velocity dispersions ($\sigma$) are measured in 98 early-type
galaxies drawn from different environments, and the relation of the indices
with the velocity dispersion analysed in detail.}
 {The present sample
indicates that some of the index--$\sigma$ relations depend on galaxy
environment. In particular, the slope of the relation between Balmer lines and
$\sigma$ is steeper for galaxies in the Virgo cluster, small groups, and 
in the field than for galaxies in the Coma cluster. In several
indices there is also a  significant offset in the zero point between 
the relations defined by the different subsamples. 
The slopes of the index--$\sigma$ relation for the Virgo and low-density
environment galaxies are explained by a variation of both age and 
metallicity with velocity dispersion, as previously noted in other studies.
For the galaxies in the Coma cluster, however, the relation of the indices 
with $\sigma$ 
only requires a variation of the abundance along the $\sigma$
sequence. 
In agreement with other studies we find that the 
models that better reproduce the slopes are those in which
the $\alpha$ elements vary more than the Fe-peak elements along the $\sigma$
sequence, while, at a given $\sigma$, older galaxies show 
an higher $\alpha$/Fe ratio.
}
{The results can be explained
assuming that galaxies in the Coma cluster have  
experienced a truncated star formation and chemical
enrichment history compared to a more continuous time-extended history for
their counterparts in lower density environments.}

%%%%%%%%%%%%%%%%%%%%%%%%%%%%%%%%%%%%%%%%%%%%%%%%%%%%%%%%%%%%%%%%%%%%%%%%%%%%%%%%%%%%%%%%%%%%%%%%%%%%%%%%%%%%%%%%%%%%
\keywords
 {galaxies: abundances -- galaxies: formation -- galaxies: elliptical and
 lenticular -- galaxies: kinematics and dynamics.}

\maketitle
%%%%%%%%%%%%%%%%%%%%%%%%%%%%%%%%%%%%%%%%%%%%%%%%%%%%%%%%%%%%%%%%%%%%%%%%%%%%%%%%%%%%%%%%%%%%%%%%%%%%%%%%%%%%%%%%%%%%
\section{Introduction}
There have long been two competing views on the star formation history of the
elliptical galaxies in the present day Universe.
The modern version of the classical ``monolithic collapse'' scenario puts the
stress on elliptical assembly out of gaseous material (that is, with
dissipation), in the form of either a unique cloud or many gaseous clumps, but
not out of preexisting stars. In this scenario, the stars form at high $z$ and
on short timescales relative to spiral galaxies (Chiosi \& Carraro 2002;
Matteucci 2003). The competing ``hierarchical'' scenario (e.g.\ Toomre 1977;
Kauffmann 1996; Somerville et al.\ 1999; de Lucia et al.\ 2006) propounds that galaxies form
hierarchically through successive non-dissipative, random mergers of subunits
over a wide redshift range. 
The first scenario is
favoured by the tight relations followed by the elliptical families, such as
the Fundamental Plane (Djorgovski \& Davis 1987), the colour--magnitude and the
Mg$_{2}$--$\sigma$ relationships (Bender, Burstein \& Faber 1993; J\o rgensen
1999; Kuntschner  2000). The second scenario is favoured by the wide range in
the  apparent age of their stellar populations (Gonz\'alez 1993, hereafter G93;
Faber et al.\ 1995; Trager et al. 2000a; Terlevich \& Forbes 2002; Caldwell; Rose 
\& Concannon 1993; Denicolo et al.
2005), the kinematical and dynamical peculiarities (e.g.\ de
Zeeuw et al.\ 2002) and the presence of  shells and ripples, indicative of
recent interactions, in a large number of elliptical galaxies (Schweizer et
al.\ 1990).

A natural outcome of the hierarchical scenarios is that haloes in regions of the 
Universe that are destined to form a cluster collapse earlier and merge more rapidly
(e.g.  Kauffmann \& Charlot 1998; de Lucia et al.\ 2006).
Therefore, the study of the stellar content
of early-type galaxies in different environments should allow us to test the
hierarchical clustering scenarios of early-type galaxy formation.

Several works have analysed the differences in the evolution of cluster and
field early-type galaxies through the study of the Fundamental Plane (FP), but
the results remain controversial.   
In general, the evolution of the trends in
cluster galaxies suggests an earlier formation for these
systems when compared with their analogs in the field 
(Treu
et al.\ 1999, 2001, 2002; van Dokkum \& Ellis 2003; Gebhardt et al.\ 2003;
Rusin et al.\ 2003; Bernardi et al.\ 2003;  Yi et al.\ 2005), but the
environmental dependencies do not appear to be as large (van Dokkum et al.\
2001) as predicted by some interpretations of hierarchical models (e.g.\
Diaferio et al.\ 2001; de Lucia et al.\ 2004).  However, the evolutionary
trends in the FP can be hidden due to the age-metallicity degeneracy (Worthey
1994), if there is a relation between the age and the metallicity of the
galaxies (Coles et al.\ 1999; Ferreras, Charlot \& Silk 1999).

Another approach, in principle less
affected by this problem, is to compare the absorption spectral features
between galaxies in different environments. 

Kuntschner \& Davies (1998) studied a sample of
galaxies in the Fornax cluster finding that they are mostly coeval, which
contrasts with other studies in which intermediate-age or young populations
have been found in a large fraction of non-cluster luminous elliptical galaxies
(e.g.\ Rose 1985; G93; Forbes, Ponman \& Brown 1998; Trager et al. 2000;
Caldwell et al. 2003; Denicolo et al. 2005).  However,
whether such contrasting results are a product of differences in the
environment is still an unanswered question.  Some authors claim that other
parameters such as the luminosity or the morphological type determine the star
formation history (SFH) of early-type galaxies rather than environment.  In
this context, the differences between cluster and field galaxies found in
several studies could be due to dissimilarities  in the luminosity range of the
different samples (see Poggianti et al.\ 2001b) and/or  different proportions
of S0 with respect to E galaxies. For example, Kuntschner \& Davies (1998)
found that, in the Fornax cluster, luminous elliptical galaxies are old and
coeval, while (less luminous) S0 galaxies display a significant age spread. In
accordance with this result, Smail et al.\ (2001) found, studying a sample of
galaxies drawn from Abell 2218 at $z=0.17$, that elliptical at all magnitudes
and luminous S0's are coeval, while the faintest S0's have younger
luminosity-weighted ages (see also Treu et al.\ 2005). 

Bernardi et al.\ (1998) found a small offset ($0.007\pm0.002$ mag) in the
Mg$_2$--$\sigma$ relation between cluster and field galaxies, but the same
intrinsic scatter about the relation for both subsamples. They also concluded
that the differences are mainly driven by the  faint objects.  These authors
also studied the zero point of the Mg$_2$--$\sigma$ relation among cluster
early-type galaxies and did not find any dependence of this value with 
the cluster richness as measured by cluster X-ray luminosity, temperature of
the ICM or velocity dispersion of member galaxies.

More recently, Denicol\'o et al. (2005a) compared a sample of galaxies
in groups and in the field with the sample of Fornax
galaxies from Kuntschner (2000), finding that
the slope of the index-$\sigma$ relations for ellipticals in  low-density 
regions are not significantly different from those
of cluster E/S0, although the scatter of the relations seems larger for group, 
field, and isolated ellipticals than for cluster galaxies. 
In the second paper of the series (Denicol\'o et al. 2005b), the authors calculate
ages and metallicities, finding that elliptical galaxies in low
density environments are, on average, 3-4 Gyr younger than  their counterparts
in the Fornax cluster. The only caveat in this study is that both samples
 only share one galaxy in common, and the differences can be due to systematic
offsets in the individual indices.

Thomas et al. (2005) also carried out an study of 124 early-type galaxies in high- and low-density environments.
They also found, in agreement with Denicol\'o, that massive early-type galaxies
in low density environment seem, on average, 2 Gyr younger and slightly ($\sim$ 0.05--0.1 dex)
more metal rich than their counterparts in high-density environments, consisting
of galaxies drawed from the Coma and Virgo clusters.
Interestingly, these authors found very massive (M $>$ $10^{11}$M$_{\odot}$) S0
galaxies showing low average ages between 2 and 5 Gyr, and very high metallicities
of [Z/H]$\leq$0.67 dex. These galaxies are  only present in high density environments, 
contrasting the above quoted result by Smail et al.\ (2001).

The lack of conclusive
evidence for or against systematic differences between cluster and field
galaxies has prompted us to carry out a systematic study analysing the stellar
population of galaxies in different environments.

This paper starts a series devoted to the study of the stellar population in
local early-type galaxies. The goal of the series is to shed light on the star
formation history of these systems and the influence of the environment on
their evolution.  Although there have been several studies with larger samples
which have studied the stellar population of the early-type galaxies in the
Coma cluster and in the field, this is the first time that a large number of
spectral features are analysed in a homogenous and high signal-to-noise ratio
sample containing galaxies in both environments.  Furthermore, these features
are compared with the new stellar population synthesis models of Vazdekis et
al.\ (2006, in preparation; hereafter V06), which themselves are an 
updated version of the Vazdekis et al.\ (2003) models with an improved stellar
library, MILES (S\'anchez-Bl\'azquez et al.\ 2006, in preparation). The new
library contains 1003 stars covering the atmospheric parameter space in a
homogenous way, reducing the uncertainties of the predictions at metalicities
departing from solar. Although a more detailed description of these models and
the derivation of age and metallicities procedures will be given in the
Paper II of this series, we discuss here some of the results based on them.
We refer the reader to Vazdekis et al.\ (2003) for details regarding the 
ingredients of the models.

The outline of the paper is as follows: Section 2 presents the sample
description, observations and data reduction. Section 3 presents the
absorption-line measurements. Section 4 shows the relation of the absorption
lines indices with the velocity dispersion of the galaxies. Section 5
summarises our conclusions.

%%%%%%%%%%%%%%%%%%%%%%%%%%%%%%%%%%%%%%%%%%%%%%%%%%%%%%%%%%%%%%%%%%%%%%%%%%%%%%%
\section{Observations and data reduction}

\subsection{Sample selection}

We analyse a sample of 98 early-type galaxies which comprises ellipticals (E)
and S0 galaxies spanning a large range in velocity dispersion (from 
40 km s$^{-1}$ to  400 km s$^{-1}$).

As one of the main goals of this work is to study the influence of the
environment on the star formation history of early-type galaxies, the sample
contains galaxies in the field, poor groups, and in the Virgo, Coma, and some 
Abell  galaxy
clusters. For the purpose of this series we have divided the sample in two main
groups that we call hereafter high density environment galaxies (HDEGs) and low
density environment galaxies (LDEGs).

It is difficult to identify a single, optimal, manner by which to
to separate a sample of galaxies in order
to study the role of environment. In this series of papers, instead of 
using an indicator of the local density, we have chosen to separate 
and delineate galaxies by field, group, or cluster environments, in order
to compare directly with the predictions
of hierarchical models of galaxy formation which separate high- and low-density 
environment galaxies depending upon the mass of the halos in which they reside
(de Lucia et al. 2006).

\begin{figure}
\resizebox{\hsize}{!}{\includegraphics[angle=-90]{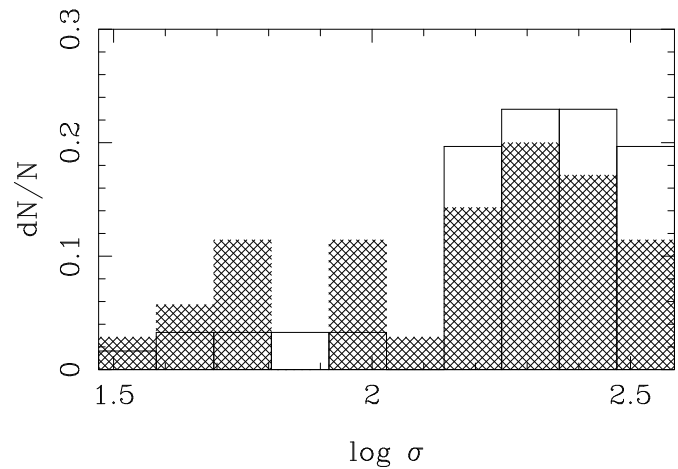}}
\caption{Distribution of central velocity dispersion in HDEGs (grey shaded
histogram) and LDEGs (empty histogram). The mean values of the distributions
are $2.27\pm 0.24$ and $2.16\pm 0.30$ for LDEGs and HDEGs, respectively. The
probability that the differences are due to chance is $\sim$50\%.
\label{distribution}}
\end{figure}

Therefore, in the first group we include galaxies residing in the central regions
of the Coma cluster and one galaxy from the cluster Abell 2199 (NGC 6166). 
We have assigned to the second groups the galaxies in the
field, groups (including here the groups Abell 569 and Abell 779 with very 
low values of velocity dispersion and X-ray luminosity (see, e.g. Xian-Ping \&
Yan-Jie 1999)), and in the Virgo cluster. This assignment of Virgo galaxies
to the LDEG group reflects that our classification should be taken in relative
terms, HDEG meaning in fact Coma-like rich clusters, while the rest of environments
are assigned to the LDEG group.  The Virgo cluster contains significant substructure
(e.g.\ de Vaucouleurs 1961; Binggeli, Sandage \&
Tammann 1985; Binggeli, Popescu \& Tammann 1993; Gavazzi et al.\ 1999), being 
an aggregation of sub-clumps which are very likely in the process of merging
(Binggeli 1997). The clump containing M87 has an estimated mass from  the 
X-ray haloes of $10^{14}$M$_{\odot}$, but the other main sub-clumps, which include
M86 and M49, have masses one order of magnitude less (Bohringer et al. 1994), which is 
the typical mass for a group of galaxies. All this suggests that Virgo can be considered
as a {\it group of groups\/} rather than as a {\it normal\/} cluster. 
 Furthermore, 
and admittedly, this is an ad-hoc argument, as the  
analysis of the stellar populations of Virgo early-type galaxies indicate 
that they 
are not dissimilar to those galaxies in poorer environments (field and small
groups) (e.g., Concannon 2003; S\'anchez-Bl\'azquez 2004)
while they are markedly different from the Coma ellipticals (as it 
will be shown throughout 
this paper). We decided  to group Virgo galaxies together with the rest
of the galaxies in low-density environment in order to maximize the statistical
significance of the sample. Galaxies were consider as group if they were 
included in the  
the Lyon Group of Galaxies catalogue (Garc\'{\i}a 1993). This catalogue
also contain galaxies in the Virgo cluster as belonging to the 
groups  LGG 289 and LGG 292. Galaxies were considered as "Field"
galaxies if they were not listed as a member of a Lyon Galaxy Group 
catalogue or as a member of a known cluster.
Coma clusters were selected from both the GMP catalogue
(Godwin, Metcalfe \& Peach 1983) and the catalogue of galaxies 
and clusters of galaxies (Zwicky et al. 1961). Last column  
Table \ref{tab-sample} list the catalogue reference giving the group 
assigment.
Galaxies in  both subsamples were selected  to span a
wide range in luminosity and observational properties.  Table \ref{tab-sample}
lists the whole sample together with their morphological type and absolute
magnitude.  The distribution in velocity dispersion for LDEGs and HDEGs is
shown in Fig. \ref{distribution}. Although the $\sigma$ range covered by both
subsamples is similar, the LDEG sample is slightly skewed towards more massive
objects. 

\begin{sidewaystable*}
\begin{minipage}[t][180mm]{\textwidth}
\caption{Sample of galaxies in the different observing runs. Type:
morphological type extracted from the NASA/IPAC Extragalactic Database; $M_{\rm B}$: absolute magnitude in the B band, obtained
from the Lyon-Meudon Extragalactic Database; Env.: when
the galaxy is a member of a well known cluster, the cluster name is indicated;
the label Pair indicates that the galaxy is a member of a pair of galaxies;
when the galaxies do not belong to a cluster, group or pair of galaxies, it is
labeled as field galaxy; Run: observation run in which each galaxy was
observed; $t_{\rm exp}$: exposure time (seconds); P.A.: position angle of the
major axis of the galaxy measured from north to east (degrees);
S.A.: orientation of the slit (degrees); Flag: Environment assigned to each
galaxy (L: low-density  environment: H: high-density environment); 
Code: Symbols used in Figure; Ref: Reference used to assigned the environment Flag to each 
galaxy (Gc93: Garc\'{\i}a 1993; VCC85: Binggeli, Sandage \& Tammann 1985; GMP: Godwin, Metcalfe
\& Peach 1983; CGCG: Zwicky et al. 1974). 
\ref{figure.sigma}.   
\label{tab-sample}}
\centering
\begin{tabular}{@{}llllclcrrrrcr@{}}
\hline\hline
  Name      
&\multicolumn{1}{c}{RA (J2000)} 
&\multicolumn{1}{c}{Dec (J2000)}       
&\multicolumn{1}{c}{Type}    
&\multicolumn{1}{c}{$M_{\rm B}$} 
&\multicolumn{1}{c}{Env.}      
&\multicolumn{1}{c}{Run}
&\multicolumn{1}{c}{$t_{\rm exp}$}
&\multicolumn{1}{c}{P.A.} 
&\multicolumn{1}{c}{S.A.} 
&\multicolumn{1}{c}{Flag}
&\multicolumn{1}{c}{Code}
&\multicolumn{1}{c}{Ref}\\
\hline 
 NGC 221    & 00 42 41.87 & +40 51 57.2 & cE2      &$ -17.58 $ & LGG 11        & 1 &    1204  &   170
 &  170.0&L & $\triangle$ & Gc93\\
            &             &             &          &           &               & 3 &    1200  &   170  &   0.0 &  &       \\
 NGC 315    & 00 57 48.88 & +30 21 08.8 &  E+      &$ -22.37 $ & LGG 18        & 1 &    1800  &
 40.0 &   0.0 &L & $\bigcirc$ & Gc93\\
            &             &             &          &           &               & 3 &    2400  &   40.0 &  45.0 &  &        \\
 NGC 507    & 01 23 40.00 & +33 15 21.9 &  SA(r)   &$ -22.15 $ & LGG 26        & 1 &    1585  &
 67.0 &   0.0 &L & $\Box$& Gc93\\
 NGC 584    & 01 31 20.72 &$-$06 52 06.1 &  E4     &$ -20.63 $ & NGC 584 group & 1 &    1800  &  120.0
 &   0.0 &L & $\bigcirc$& Gc93\\
 NGC 636    & 01 39 06.52 &$-$07 30 45.6 &  E3     &$ -19.65 $ & LGG 27        & 1 &    1800  &
 140.0 &   0.0 &L & $\bigcirc$& Gc93\\
 NGC 821    & 02 08 21.04 & +10 59 41.1 &  E6      &$ -20.57 $ & Field         & 1 &    1200  &   25.0
 &   0.0 &L & $\bigcirc$&\\
 NGC 1453   & 03 46 27.22 &$-$03 58 08.9 &  E2--3  &$ -21.52 $ & LGG 103       & 1 &    1800  &
 8.0 &   0.0 &L & $\bigcirc$& Gc93\\
 NGC 1600   & 04 31 39.89 &$-$05 05 10.1 &  E3     &$ -22.31 $ & Field         & 1 &    1800  &  170.0
 &   0.0 &L & $\bigcirc$& \\
            &             &             &          &           &               & 2 &    1800  &  170.0
            & 195.0 &  & &\\
 NGC 1700   & 04 56 56.30 &$-$04 51 52.0 &  E4     &$ -21.80 $ & LGG 123       & 1 &    3000  &
 65.0 &   0.0 &L & $\bigcirc$&Gc93\\
 NGC 2300   & 07 32 21.82 & +85 42 32.2 &  E--S0   &$ -20.85 $ & Abell 569     & 1 &    1500  &   78.0
 &   0.0 &L & $\Box$&Gc93\\
 NGC 2329   & 07 09 08.08 & +48 36 53.3 &  S0$-$   &$-21.73  $ & Abell 569     & 2 &    1800  &
 175.0&  175.0&L & $\square$&Gc93\\
 NGC 2693   & 08 56 59.28 & +51 20 49.5 &  E3      &$ -21.67 $ &  LGG 168      & 1 &    3200  &
 160.0 &   0.0 &L & $\bigcirc$&Gc93\\
 NGC 2694   & 08 56 59.28 & +51 19 55.1 &  E1      &$ -19.15 $ &  LGG 168      & 1 &    3200  &
 &   0.0 &L & $\bigcirc$&Gc93\\
 NGC 2778   & 09 12 24.35 & +35 01 39.4 &  E       &$-19.06  $ &  LGG 171      & 2 &    1800  &
 40.0&  220.0&L & $\bigcirc$&Gc93\\
 NGC 2832   & 09 19 46.89 & +33 45 00.0 &  E+2     &$-22.38  $ & Abell 779     & 2 &    1800  &
 160.0&  160.0&L & $\bigcirc$&CGCG\\
 NGC 3115   & 10 05 13.80 &$-$07 43 08.0&  S0$-$   &$-19.77  $ &  Field        & 2 &    1500  &
 43.0&  310.0&L & $\bigcirc$&\\
 NGC 3377   & 10 47 42.36 & +13 59 08.8 &  E5-6    &$-19.16  $ & Leo group (LGG217)    & 1 &    1800  &   35.0
 &   0.0 &L & $\bigcirc$& Gc93\\
 NGC 3379   & 10 47 49.75 & +12 34 54.6 &  E1      &$-20.57  $ & Leo group (LGG 217)   & 1 &     900  &   71.0
 &   0.0 &L & $\bigcirc$&Gc93\\
            &             &             &          &           &               & 2 &    1200  &
            71.0&  251.0&  & & \\
 NGC 3605   & 11 16 46.69 & +18 01 01.0 &  E4-5    &$-17.07  $ &  LGG 237      & 2 &    1800  &
 17.0&   17.0&L &$\triangle$&Gc93\\
 NGC 3608   & 11 16 59.07 & +18 08 54.6 &  E2      &$ -19.62 $ &  LGG 237      & 1 &    1800  &
 75.0 &   0.0 &L &$\bigcirc$&Gc93\\
 NGC 3641   & 11 21 08.85 & +03 11 40.2 &  E pec   &$-17.91  $ &  LGG 233      & 2 &    1800  &
 3.0&  240.0&L &$\triangle$& Gc93\\
 NGC 3665   & 11 24 43.64 & +38 45 45.0 &  SA(s)   &$ -20.84 $ &  LGG 236      & 1 &    1800  &
 30.0 &   0.0 &L & $\square$& Gc93\\
 NGC 3818   & 11 41 57.50 &$-$06 09 20.0 &  E5     &$-19.11  $ &  Field        & 2 &    1800  &
 103.0&  280.0&L &$\bigcirc$&\\
 NGC 4261   & 12 19 23.21 & +05 49 29.7 &  E2-3    &$-21.32  $ &  LGG 278      & 2 &    1800  &
 160.0&  160.0&L &$\bigcirc$&Gc93\\
 NGC 4278   & 12 20 06.83 & +29 16 50.7 & E1-2     &$-19.26  $ &  LGG 279      & 2 &    1800  &
 22.7&   22.0&L &$\bigcirc$&Gc93\\
 NGC 4365   & 12 24 28.34 & +07 19 04.2 &  E3      &$-20.90  $ & Virgo (LGG 289) & 2 &    1200  &
 40.0&  220.0&L &$\bigcirc$&Gc93\\
 NGC 4374   & 12 25 03.74 & +12 53 13.1 &  E1      &$-20.87  $ & Virgo (LGG 292) & 1 &    1800  &
 135.0 &   0.0 &L &$\bigcirc$&Gc93\\
\hline
\end{tabular}
\vfill
\end{minipage}
\end{sidewaystable*}

\begin{sidewaystable*}
\begin{minipage}[t][180mm]{\textwidth}
\addtocounter{table}{-1}
\caption{\it Continued.}
\centering
\begin{tabular}{@{}llllclcrrrrcr@{}}
\hline\hline
  Name     
&\multicolumn{1}{c}{RA (J2000)}          
&\multicolumn{1}{c}{DEC (J2000)}            
&\multicolumn{1}{c}{Type}    
&\multicolumn{1}{c}{$M_{\rm B}$} 
&\multicolumn{1}{c}{Env.}      
&\multicolumn{1}{c}{Run}
&\multicolumn{1}{c}{$t_{\rm exp}$} 
&\multicolumn{1}{c}{P.A.} 
&\multicolumn{1}{c}{S.A.}
&\multicolumn{1}{c}{Env}
&\multicolumn{1}{c}{Code}
&\multicolumn{1}{c}{Ref}\\
\hline 
 NGC 4415   & 12 26 40.56 & +08 26 07.5 &  S0/a    &$-17.27  $ & Virgo (LGG 292)  & 1 &    4200  &
 0.0  &   0.0 &L &$\bigtriangleup$&Gc93\\
            &             &             &          &           &                  & 2 &    3000  &
            0.0&  180.0&  &       &    \\
 NGC 4431   & 12 27 27.43 & +12 17 23.8 & SA(r)0   &$-16.95  $ & Virgo (LGG 292)  & 1 &    7200  &
 177.0 &   0.0 &L &$\triangle$&Gc93\\
 NGC 4464   & 12 29 21.38 & +08 09 23.1 &  E3      &$-18.12  $ & Virgo (LGG 289)  & 2 &    1800  &
 0.0&  180.0&L &$\square$&Gc93\\
 NGC 4467   & 12 29 30.35 & +07 59 38.3 & E2       &$-16.89  $ & Virgo (LGG 289)  & 2 &    1800  &
 81.2&   99.0&L &$\bigcirc$&Gc93\\
 NGC 4472   & 12 29 46.76 & +07 59 59.9 & E2/S0(2) &$-21.47  $ & Virgo (LGG 292)  & 2 &    1200  &
 155.0&  155.0&L &$\square$&Gc93 \\
 NGC 4478   & 12 30 17.53 & +12 19 40.3 &  E2      &$-19.55  $ & Virgo (LGG 289)  & 2 &    1800  &
 140.0&  140.0&L &$\bigcirc$&Gc93\\
 NGC 4486B  & 12 30 31.82 & +12 29 25.9 & cE0      &$-17.46  $ & Virgo (LGG 289)  & 1 &    1800  &
 117.3 &   0.0 &L &$\bigcirc$&Gc93\\
 NGC 4489   & 12 30 52.34 & +16 45 30.4 & E        &$-18.25  $ & Virgo (LGG 292)  & 1 &    2500  &
 &   0.0 &L &$\triangle$&Gc93\\
 NGC 4552   & 12 35 40.00 & +12 33 22.9 & E        &$-20.91  $ & Virgo            & 1 &    1800  &
 &   0.0 &L &$\bigcirc$&VCC85\\
 NGC 4564   & 12 36 27.01 & +11 26 18.8 & E6       &$-19.43  $ & Virgo (LGG 289)  & 2 &    1500
 &    47.0&   47.0&L &$\bigcirc$&Gc93\\
 NGC 4594   & 12 39 59.43 &$-$11 37 23.0 & SA(s)a  &$-22.36  $ & Field            & 2 &    1013  &
 89.0&  179.0&L &$\bigcirc$&\\
 NGC 4621   & 12 42 02.39 & +11 38 45.1 & E5       &$-20.71  $ & Virgo            & 2 &    1200  &
 165.0&  165.0&L &$\bigcirc$&VCC85\\
 NGC 4636   & 12 42 50.00 & +02 41 16.5 & E--S0    &$-20.78  $ & Field            & 1 &    1200  &  150.0
 &   0.0 &L &$\square$& \\
 NGC 4649   & 12 43 40.19 & +11 33 08.9 & E2       &$-21.53  $ & Virgo (LGG 292) & 1 &     900  &
 105.0 &   0.0 &L &$\bigcirc$&Gc93\\
 NGC 4673   & 12 45 34.77 & +27 03 37.3 & E1--2    &$-20.52  $ &  Coma           & 4 &    1800  & 170.0
 & 170.0 &H &$\bigcirc$&GMP\\
 NGC 4692   & 12 47 55.42 & +27 13 18.3 & E+       &$-21.70  $ &  Coma           & 4 &    1500  &  97.0
 &  97.0 &H &$\bigcirc$&GMP\\
 NGC 4697   & 12 48 35.70 &$-$05 48 03.0 & E6      &$-20.97  $ & LGG 314         & 2 &    1200  &
 70.0&   70.0&L &$\bigcirc$&Gc93\\
 NGC 4742   & 12 51 47.92 &$-$10 27 17.1 & E4      &$-22.40  $ & LGG 307         & 2 &    1800  &
 75.0&   75.0&L &$\bigcirc$&Gc93\\
 NGC 4839   & 12 57 24.31 & +27 29 52.0 & cD;SA0   &$-21.84  $ &  Coma           & 4 &    5400  &  65.0
 &  49.4 &H &$\square$&GMP\\
 NGC 4842A  & 12 57 35.60 & +27 29 36.1 & E/SA0    &$-19.89  $ &  Coma           & 4 &    1800  &
 & 172.4 &H &$\square$&GMP\\
 NGC 4842B  & 12 57 36.14 & +27 29 05.6 & SA0      &$-18.80  $ &  Coma        & 4 &    1800  &
 & 172.4 &H &$\square$&GMP\\
 NGC 4864   & 12 59 13.00 & +27 58 37.2 & E2       &$-20.55  $ &  Coma        & 4 &    1800  &
 & 128.7 &H &$\bigcirc$&GMP\\
 NGC 4865   & 12 59 19.98 & +28 05 02.3 &  E       &$-19.91  $ &  Coma        & 4 &    1800  &
 & 70.0  &H &$\bigcirc$&GMP\\
 NGC 4867   & 12 59 15.00 & +27 58 14.9 & E3       &$-19.01  $ &  Coma        & 4 &    1800  &
 & 128.7 &H &$\bigcirc$&GMP\\
 NGC 4874  & 12 59 35.91 & +27 57 30.8   &cD                 &$-22.53  $ & Coma        & 2 &   1800
 &  39.7 &   79.0&H&$\bigcirc$&GMP\\
           &             &               &                   &           & Coma        & 4 &   3600
           &  39.7 &   39.7& &&\\
 NGC 4875  & 12 59 37.80 & +27 54 26.5   & SAB0              &$-19.72$   & Coma        & 4 &   1800
 &       &   48.5&H&$ \Box$&GMP\\
 NGC 4889  & 13 00 08.03 & +27 58 35.1   & cD                &$-22.46$   & Coma        & 2 &   2178
 &  80.0 &   80.0&H&$\bigcirc$&GMP\\
           &             &               &                   &           &             & 4 &   3800
           &  80.0 &   80.0& & & \\
 NGC 4908  & 13 00 51.55 & +28 02 33.6   & E5                &$-21.01$   & Coma        & 4 &   1800
 &       &  163.0&H&$\square$&GMP\\
 NGC 5638  & 14 29 40.30 & +03 14 00.0   & E1                &$-19.78$   & LGG 386     & 2 &
 1800  & 150.0 &  150.0&L&$\bigcirc$&Gc93\\
 NGC 5796  & 14 59 24.20 &$-$16 37 24.0  & E0--1             &$-20.77$   & LGG 386     & 2 &   1800
 &  95.0 &  194.0&L&$\bigcirc$&Gc93\\
 NGC 5812  & 15 00 55.60 &$-$07 27 26.0  & E0                &$-20.36$   & Field       & 2 &   1800  & 130.0 &  130.0&L&$\bigcirc$\\
 NGC 5813  & 15 01 11.32 & +01 42 06.4   & E1--2             &$-20.99$   & LGG 393     & 2 &   1800
 & 145.0 &  145.0&L&$\bigcirc$&Gc93\\
 NGC 5831  & 15 04 07.10 & +01 13 11.7   & E3                &$-19.72$   & LGG 393     & 2 &   1800
 &  55.0 &  235.0&L&$\bigcirc$&Gc93\\
 NGC 5845  & 15 06 00.90 & +01 38 01.4   & E                 &$-18.58$   & LGG 392     & 2 &   1800
 & 150.0 &  150.0&L&$\bigcirc$&Gc93\\
 NGC 5846  & 15 06 29.37 & +01 36 19.0   & E0--1             &$-21.30$   & LGG 393     & 2 &
 1200  &       &  182.0&L&$\bigcirc$&Gc93\\
           &             &               &                   &           &             & 4 &   1000
           &       &  182.0& & & \\
\hline
\end{tabular}
\vfill
\end{minipage}
\end{sidewaystable*}

\begin{sidewaystable*}
\begin{minipage}[t][180mm]{\textwidth}
\addtocounter{table}{-1}
\caption{\it Continued.}
\centering
\begin{tabular}{@{}llllclcrrrrcr@{}}
\hline
  Name     
&\multicolumn{1}{c}{RA (J2000)}          
&\multicolumn{1}{c}{DEC (J2000)}            
&\multicolumn{1}{c}{Type}    
&\multicolumn{1}{c}{$M_{\rm B}$} 
&\multicolumn{1}{c}{Env.}      
&\multicolumn{1}{c}{Run}
&\multicolumn{1}{c}{$t_{\rm exp}$} 
&\multicolumn{1}{c}{P.A.} 
&\multicolumn{1}{c}{S.A.}
&\multicolumn{1}{c}{Env}
&\multicolumn{1}{c}{Code}\\
\hline 
 NGC 5846A & 15 06 28.90 & +01 35 43.0   & cE2--3            &$-19.83$   & LGG 393        & 2 &
 1200  & 120.0 &  182.0&L&$\bigcirc$&Gc93\\
           &             &               &                   &           &             & 4 &   1000  & 120.0 &  182.0& &\\
 NGC 6127  & 16 19 11.73 & +57 59 02.6   & E                 &$-21.39$   & Field       & 2 &   1800
 &       &   25.0&L&$\bigcirc$&\\
 NGC 6166  & 16 28 38.30 & +39 33 04.7   & S0                &$-23.00$   & Abell 2199  & 2 &   2173
 &  35.0 &   35.0&H&$\square$&CGCG\\
           &             &               &                   &           &             & 3 &   1200  &  35.0 &   35.0& &\\
 NGC 6411  & 17 35 32.82 & +60 48 47.2   & E                 &$-21.07$   & Field       & 3 &   1200
 &  70.0 &   65.0&L&$\bigcirc$&\\
 NGC 6482  & 17 51 48.94 & +23 04 19.1   & E                 &$-22.11$   & Field       & 4 &   1246
 &  70.0 &  160.0&L&$\bigcirc$&\\
 NGC 6577  & 18 12 01.29 & +21 27 49.4   & E                 &$-20.82$   & Field       & 4 &   1800
 & 178.0 &  178.0&L&$\bigcirc$&\\
 NGC 6702  & 18 46 57.64 & +45 42 19.8   & E:                &$-21.42$   & Field       & 2 &   2000
 &  65.0 &   65.0&L&$\bigcirc$&\\
 NGC 6703  & 18 47 18.86 & +45 33 01.0   & SA0--             &$-20.83$   & Field       & 2 &   1800
 &       &   60.0&L&$\square$&\\
           &             &               &                   &           &             & 3 &   1800
           &       &   65.0& &&\\
           &             &               &                   &           &             & 4 &   1200
           &       &   60.0& &&\\
 NGC 7052  & 21 18 33.13 & +26 26 48.7   & E                 &$-21.06$   & Field       & 3 &   1200
 &  64.0 &   65.0&L&$\bigcirc$&\\
 NGC 7332  & 22 37 24.62 & +23 47 55.0   & S0                &$-19.16$   & Field        & 3 &   2400
 & 152.0 &   65.0&L&$\square$&\\
 IC  767   & 12 11 02.80 & +12 06 15.0   & E                &$-18.04$   & Virgo        & 1 &   2000
 &       &    0.0&L&$\triangle$&VCC85\\
 IC  794   & 01 08 49.99 &$-$15 56 54.4  & dE3               &$-18.10$   & LGG 286       & 2 &   2000
 & 110.0 &  110.0&L&$\triangle$Gc93\\
 IC 832    & 12 53 59.10 & +26 26 38.0   & E?                &$-19.96$   & Coma        & 4 &   1800
 & 168.0 &  168.0&H&$\bigcirc$&CGCG\\
 IC 3618   & 12 39 17.12 & +26 40 39.7   & E                 &$-19.23$   & Coma        & 4 &   1800
 &  74.0 &   74.0&H&$\bigcirc$&CGCG\\ % CGCG 159--41
 IC 3957   & 12 59 07.30 & +27 46 02.0   & E                 &$-19.26$   & Coma        & 4 &   1800
 &       &    9.3&H&$\bigcirc$&GMP\\
 IC 3959   & 12 59 08.18 & +27 47 02.7   & E                 &$-20.03$   & Coma        & 4 &   3600
 &       &    9.3&H&$\bigcirc$&GMP\\
 IC 3963   & 12 59 13.48 & +27 46 28.3   & SA0               &$-19.25$   & Coma        & 4 &   1800
 &  84.0 &  296.4&H&$\square$&GMP\\
 IC 3973   & 12 59 30.60 & +27 53 03.2   & S0/a              &$-18.85$   & Coma        & 4 &   1800
 & 142.0 &   48.5&H&$\square$&GMP\\
 IC 4026   & 13 00 22.12 & +28 02 48.8   & SB0               &$-19.84$   & Coma        & 4 &   1750
 &  59.0 &  132.3&H&$\square$&GMP\\
 IC 4042   & 13 00 42.72 & +27 58 16.4   & SA0/a             &$-19.80$   & Coma        & 4 &   1200
 &   2.0 &  177.9&H&$\square$&GMP\\
 IC 4051   & 13 00 54.44 & +28 00 27.0   & E2                &$-20.21$   & Coma        & 4 &   1800
 &  95.0 &  163.0&H&$\bigcirc$&GMP\\

 MCG+05-30-048& 12 39 18.90& +27 46 24.5 & E2                &$-19.63$   & Coma        & 4 &   1800
 &  35.0 &   35.0&H&$\bigcirc$&CGCG\\ % CGCG 159--43
 MCG+05-30-094& 12 49 42.28& +26 53 30.5 & E1:               &$-20.14$   & Coma        & 4 &   1800
 & 149.0 &  149.0&H&$\bigcirc$&CGCG\\%CGCG 159--83
 MCG+05-30-101& 12 50 54.03& +27 50 30.1 & E1                &$-20.03$   & Coma        & 4 &   1800
 &  63.0 &   20.0&H&$\bigcirc$&CGCG\\%CGCG 159--89
 DRCG 27--32  & 12 57 22.75& +27 29 36.3 & S0                &$-19.11$   & Coma        & 4 &   3600
 &       &   49.4&H&$\square$&GMP\\
 DRCG 27--127 & 12 59 40.37& +27 58 05.6 &E--S0              &$-18.49$   & Coma        & 4 &   1800
 &       &   9.0 &H&$\square$&GMP\\
 DRCG 27--128 & 12 59 39.48& +27 57 13.6 & S0A               &$-19.63$   & Coma        & 4 &   1800
 &       &   9.0 &H&$\square$&GMP\\
 GMP 2585     &13 00 35.39 & +27 56 33.9 & dE                &           & Coma        & 4 &   7200
 &       &   67.5&H&$\triangle$&GMP\\
 GMP 3121     &12 59 51.50 &+28 04 25.0  & dE                &           & Coma        & 4 &   3600
 &       &  110.4&H&$\triangle$&GMP\\
 GMP 3131     &12 59 50.25 &+27 54 44.5  & dE                &           & Coma        & 4 &   7200
 &       &   41.1&H&$\triangle$&GMP\\
 GMP 3196     &12 59 44.72 &+27 53 22.8  & dE                &           & Coma        & 4 &   7200
 &       &   41.1&H&$\triangle$&GMP\\
 MCG+05-31-063&12 59 20.50 &+28 03 58.0  & E6                &$-19.68$   & Coma        & 4 &   1800
 &       &   70.0&H&$\square$&GMP\\
 PGC 126756   &13 00 24.78 &+27 55 38.1  &                   &$-17.14$   & Coma        & 4 &   7200
 & 166.0 &   67.5&H&$\triangle$&GMP\\
 PGC 126775   &12 59 57.66 &+28 03 54.7  &                   &$-16.86$   & Coma        & 4 &   3600
 &  52.0 &  110.4&H&$\triangle$&GMP\\
 Rb 091       &13 00 16.90 &+28 03 50.0  &  SB0              &$-18.44$   & Coma        & 4 &   1200
 &  76.0 &  132.3&H&$\square$&GMP\\
 Rb 113       &13 00 42.80 &+27 57 46.5  &  SB0              &$-19.35$   & Coma        & 4 &   1200
 & 101.0 &  177.9&H&$\square$&GMP\\
\hline
\end{tabular}
\vfill
\end{minipage}
\end{sidewaystable*}

\subsection{Observations}

Long-slit spectroscopy was carried out in four observing runs using two
different telescopes.  In Runs 1 and 3 (January 1998 and August 1999) we used
the Cassegrain Twin Spectrograph with a blue coated TEK CCD in the blue channel
on the 3.5 m telescope at the German-Spanish Astronomical Observatory at Calar
Alto (Almer\'{\i}a, Spain).  The observations in Runs 2 and 4 (March 1999 and
April 2001) were carried out with the ISIS double spectrograph mounted at the
f/11 Cassegrain focus on the William Herschel Telescope in  the Roque de los
Muchachos Observatory (La Palma, Spain). Details of the observational
configurations for each run are given in Table \ref{observations}. 
As can be seen,  the spectral resolution of Run 4  
is very different to the rest of the runs. As the Lick indices depend on
the instrumental broadening (as well as velocity dispersion broadening),
special care has to be taken to degrade the spectra to the Lick resolution,
in order to avoid systematic differences between observing Runs.
Typical exposure times varied from 1800 s for the brightest galaxies to 7200 s for the
faintest ones in the Coma cluster. Typical signal-to-noise ratios per \AA,
measured in the range between 3500 and 6500 \AA, are 110 and 50 for the LDEGs
and HDEGs galaxies respectively.  The wavelength coverage varies between
different runs, but all includes the range between 3500 and 5250 \AA, which
allow the measurement of the D4000 break (Bruzual 1983) and 15 Lick/IDS indices
(from H$\delta_A$ to Mgb; Trager et al.\ 1998; hereafter T98). When possible,
the slit was oriented along the major axis of the galaxies. The high quality of
the data allowed us to extract spatially resolved spectra and measured the
indices out to the effective radius  with high signal-to-noise ratio. In this
paper we analyse only the central regions of the galaxies. The analysis of the
gradients will be the objective of Paper III in this series.

\begin{table*}
\caption{Observational configurations employed in the four different observing 
runs.
\label{observations}}
\centering
\begin{tabular}{@{}lllll@{}}
\hline
  & Run 1 & Run 2 & Run 3 & Run 4 \\
\hline
  Dates               & 19--21 Jan 1998  & 15--17 Mar 1999 &18--19 Aug 1999  & 25--27 Apr 2001\\  
  Telescope           & CAHA 3.5 m       & WHT 4.2 m       & CAHA 3.5 m      & WHT 4.2 m\\
  Spectrograph        & CTS              & ISIS blue       & CTS             & ISIS blue\\
  Detector            & CCD TEK 12       & EEV12           &CCD TEK 12       & EEV12\\
  Dispersion          & 1.08 \AA/pixel   & 0.80 \AA/pixel  & 1.08 \AA/pixel  & 1.72 \AA/pixel \\
  Wavelength range    & 3570--5770 \AA   & 3700--5269 \AA  & 3570--5770 \AA  & 3700--6151 \AA \\
  Spectral resolution & 3.6 \AA (FWHM)   & 3.5 \AA         &  3.6 \AA        &  6.56   \AA \\
  Slit width          & 2.1 arcsec       & 2.0 arcsec      & 2.1 arcsec      & 2.0 arcsec \\
  Spatial scale       & 1.1 arcsec/pixel & 0.8 arcsec/pixel& 1.1 arc/pixel   &  0.8 arcsec/pixel \\  
\hline
\end{tabular}
\end{table*}
 
Table \ref{tab-sample} lists the sample of galaxies together with additional
information including total exposure times and position angles of the
spectrograph slit.  Additionally, we observed about $45$ G--K stars to be used
as templates for velocity dispersion measurements as well as to transform our
line-strength indices to the Lick system. Flux standards from Oke (1990) were
observed to correct the continuum shape of the spectra.  In order to check for
the possibility of systematic effects between different runs, several
galaxies were observed in more than one observing period.

\subsection{Data Reduction}

The standard data reduction procedures (flat-fielding, cosmic ray removal,
wavelength calibration, sky subtraction and fluxing) were performed with
\reduceme\ (Cardiel 1999).  This reduction package allows a parallel treatment
of data and error frames and, therefore, produces an associated error spectrum
for each individual data spectrum. We want to stress the importance of
obtaining reliable errors on the measurements of individual features in order
to analyse the effects of the correlated errors between the derived stellar
population parameters. The use of this package allowed us to control errors
more easily than what it would have been possible with other available
software packages.
 
Initial reduction of the CCD frames involved bias and dark current subtraction,
the removal of pixel-to-pixel sensitivity variations (using flat-field
exposures of a tungsten calibration lamp), and correction for two-dimensional
low-frequency scale sensitivity variations  (using twilight sky exposures). 

The dichroics in Runs 2 and 4 produced an intermediate frequency pattern which
varies with the position of the telescope. This pattern was removed during
flat-fielding in the images of Run 4. Unfortunately, in Run 2 this was not
possible as we did not acquire flat field exposures in all the galaxy
positions. Fortunately, several galaxies from Run 2 were also observed in
other runs. By dividing the spectra of these galaxies from Run 2 by the spectra
acquired in other runs (previously resampled to the same instrumental
resolution and dispersion) we  obtained the shape of the
oscillations in Run 2. 
The shape of this pattern turned out to be identical for all the galaxies with 
repeated observations, with  
the exception of small offsets between them. These offsets were quantified with 
respect to a galaxy of reference for all the galaxies of 
Run 2 (given the characteristic
shape of the oscillation pattern, even for galaxies 
without repeated observations, it 
was straightforward to determine this offset using, for this purpose, 
similar spectra from other observing runs).
Finally,  all the galaxies were divided by this reference pattern 
shifted to the previously calculated offset.
(for further details of the procedure see S\'anchez-Bl\'azquez
2004). The uncertainty introduced by this correction was added to the final
error budget, although it is not a dominant error source (see Table
\ref{tipicos}).
 
Prior to the wavelength calibration, arc frames were used to correct from
C-distortion in the images. This rotation correction guaranteed alignment
errors to be below 0.1 pixel. Spectra were converted to a linear wavelength
scale using typically 120 arc lines fitted by 3th--5th order polynomials, with
r.m.s.\ errors of 0.3--0.6 \AA. In addition, all the spectra were corrected for
S-distortion. This correction is performed with a routine that finds the 
maximum corresponding to the center of the galaxy as a function of wavelength
and fits these positions with a low-order polynomial. Finally, the spectra were
displaced with a technique that minimizes the errors due to the discretization
of the signal. This technique does not assume that the signal is constant in a
given pixel but adopts a more realistic distribution by fitting a
second order polynomial using the available information in the adjacent pixels.
The S-distortion can change along the slit, but the correction only affect
the measurements in the central parts, where the signal gradient is important.
Therefore we assume that the error of using the S-distortion pattern derived
from the central galaxy region in the whole extent of the slit can be safely
ignored.

Atmospheric extinction was calculated using the extinction curve of King (1985)
for Runs 2 and 4, and the extinction curve of Calar Alto for Runs 1 and 3. To
correct the effect of interstellar extinction, we used the curve of Savage \&
Mathis (1979). The reddenings were extracted from  the RC3 catalogue of
galaxies (de Vaucouleurs et al.\ 1991).  Relative flux calibration of the
spectra was achieved using exposures of standards stars (3, 4, 5 and 33
exposures were taken in  Runs 1, 2, 3 and 4 respectively). All the calibration
curves of each run were averaged and the flux calibration errors were estimated
by the differences between  the indices measured with different curves.  To
transform our indices into the spectrophotometric system of Lick/IDS we
measured line-strengths in a sample of G--K giant stars in common with Worthey
et al.\ (1994). The comparison is presented in Appendix \ref{appen.broad-lick}.

Since in the outer parts of some galaxies we are measuring indices in spectra
with light levels corresponding to only a few per cent of the sky signal (Paper
III), the sky subtraction is a critical step in our data reduction in order to
obtain reliable gradients. After correcting for both C- and S-distortion, 
a sky image was generated for each galaxy observation. This was done by 
fitting for each channel (pixel in the $\lambda$ direction) 
a low-order polynomial (order zero or one) to regions selected at both sides of the galaxy. A
possible systematic overestimation of the sky level could arise if the galaxy
contribution to the regions from where the sky is extracted was not negligible.
This overestimation of the sky level could increase the measured indices
dramatically in the outer regions. To explore this effect we have fitted de
Vaucouleurs profiles to the surface brightness profiles of our galaxies to
estimate the relative contribution of the galaxy.  When necessary, the effect
has been taken into account, subtracting from the sky spectra a scaled and
averaged galaxy spectrum. 
An underestimation of the sky level  is very unlikely
because  this would imply high systematic errors that should be clearly
detected such as unremoved sky lines (see Cardiel, Gorgas \&
Arag\'on-Salamanca 1995).   

From each fully reduced galaxy frame, the spectra within an equivalent aperture
of 4$^{\prime\prime}$ at a redshift $z=0.016$ were co-added. This corresponds to
a physical aperture size of 0.62 kpc assuming $H_0$=70 km s$^{-1}$ Mpc$^{-1}$ . This aperture was
chosen as a compromise to obtain a fair number of co-added spectra for galaxies
in the Coma and in the Virgo clusters. This method does not fully extract
the same physical area for all galaxies, as the slit width was kept constant. 
To quantify this second-order aperture effect, we also extracted spectra along
the slit, simulating equal-physical-area circular apertures (distance-weighted
co-added spectra), and found no significant differences. Therefore, we chose to
work with the first aperture (without correcting for second-order effects) as
to maximize the S/N of our spectra.
%%%%%%%%%%%%%%%%%%%%%%%%%%%%%%%%%%%%%%%%%%%%%%%%%%%%%%%%%%%%%%%%%%%%%%%%%%%%%%%
\section{Velocity dispersion measurements}
\label{velocitydispersion}

\begin{figure}
\resizebox{\hsize}{!}{\includegraphics[bb=64 50 576 750, angle=-90]{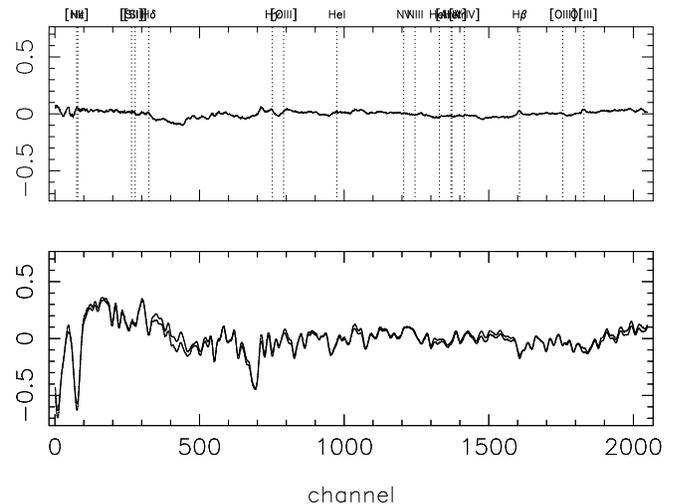}}
\caption{Example of determination of kinematical parameters in the case of the
galaxy NGC 3379. The upper panel shows the residuals of the optimal template
fit, with the vertical dotted lines indicating the position of typical
emission lines. In the lower panel, the thick and thin lines correspond to the
galaxy and optimal template spectra, respectively.\label{movel}}
\end{figure}

Radial velocities and velocity dispersions for the galaxies were measured using
the MOVEL and OPTEMA algorithms described by G93. The MOVEL
algorithm is an iterative procedure based in the Fourier Quotient method
(Sargent et al.\ 1977) in which a galaxy model is processed in parallel to the
galaxy spectrum. The main improvement of the procedure is introduced through
the OPTEMA algorithm, which  is able to overcome the typical template mismatch
problem by constructing for each galaxy an optimal template as a linear
combination of stellar spectra of different spectral types and luminosity
classes (see, for details, G93; Pedraz et al.\ 2002;
S\'anchez-Bl\'azquez 2004). To build the optimal template we made use of 25,
40, 10 and 33 stars in  Runs 1, 2, 3 and 4 respectively. To illustrate the
procedure, Figure \ref{movel} shows a typical fit between the observed central
spectrum of a galaxy and the corresponding optimal template corrected with the
derived kinematic parameters.  The errors in the radial velocity and velocity
dispersion ($\sigma$) were computed through numerical simulations. In each
simulation, a bootstrapped galaxy spectrum, obtained using the error spectrum
provided by the reduction with \reduceme, is fed into the MOVEL and OPTEMA
algorithms   (note that a different optimal template is computed in each
simulation). Errors in the parameters were then calculated as the unbiased
standard deviation of the different solutions. These final errors are expected
to be quite realistic, as they incorporate all the uncertainties of the whole
reduction process, from the first steps (e.g.\ flat-fielding) to the final
measurements of the parameters.

\begin{figure}
\resizebox{\hsize}{!}{\includegraphics[angle=-90]{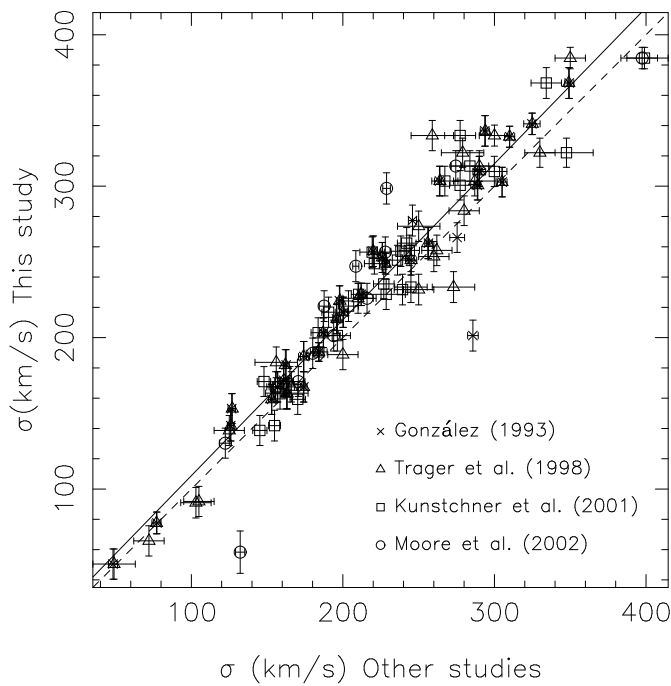}}
\caption{Comparison of the derived velocity dispersions calculated in this work
with the data compiled from the literature. Different symbols represent
different datasets, as indicated in the inset. The 1:1 relation and the
linear fit to the data are drawed with dashed and solid lines respectively.
\label{comparacion.sigmas}}
\end{figure}

In order to check the quality of the measured kinematics, we compared our
derived velocity dispersions with different  data compiled from the literature.
We have chosen four different studies: G93, T98, Kuntschner et al.\ (2001), and 
Moore et al.\ (2002), with 33, 52, 31
and 12 galaxies in common with our sample, respectively.
Fig. \ref{comparacion.sigmas} shows the result of this comparison and, in
Table \ref{tab-comparasigma}, we summarise the mean differences.  As  can be
seen, our velocity dispersions are, on average, $\sim 15$ km s$^{-1}$ larger
than those in other studies. In particular, the differences seem to be  
larger as $\sigma$ increases. Using a linear fit to the data (represented 
with a solid line in 
Fig. \ref{comparacion.sigmas} we calculated that the maximum difference, 
for a galaxy with $\sigma=$400 km$^{-1}$, is 17 km s$^{-1}$. This difference in $\sigma$ 
translates into errors in the indices less than 0.1\%. 
This systematic effect can be the consequence of
template mismatches either in our study or in the others. Nevertheless, our
method makes use of several stars to calculate the optimal template, then
minimising the effect of a poor fit.  Furthermore, the effect of mismatching
tend to underestimate the values of the velocity dispersion (Laird \& Levison
1985; Bender 1990), instead of making them larger.  It is not the purpose of
this section to investigate the real causes of the differences as this
systematic offset does not affect any of our conclusions.

\begin{table}
\caption{Comparison of the velocity dispersions derived in this study and in
previous works. The table columns are: offset: mean offset between this study
and the one in column 1; r.m.s.: dispersion among the mean value; exp.\ r.m.s.:
dispersion expected from the errors; $z$: result of a $z$-test to contrast the
hypothesis of a null offset (a value higher than 1.96 allows to reject
the hypothesis with a significance level lower than 0.05); $N$: number of
galaxies in each comparison. Units are km s$^{-1}$.\label{tab-comparasigma}}
\centering
\begin{tabular}{lccccc} 
\hline
study &   offset     & r.m.s. & exp.\ r.m.s. & $z$ & $N$\\
\hline
G93   &   12.5       &19.5 & $\;\;$9.6   & 3.1& 33\\
T98   &   13.7       &15.1 & 10.7   & 4.8& 52\\      
M02   &   17.4       &21.2 & 11.0   & 1.7& 12\\
K01   &   12.9       &15.8 & 11.9   & 3.5& 31 \\
\hline
\end{tabular}
\end{table}

%%%%%%%%%%%%%%%%%%%%%%%%%%%%%%%%%%%%%%%%%%%%%%%%%%%%%%%%%%%%%%%%%%%%%%%%%%%%%%%%%%%%%%%%%%%%%%%%%%%%%%%%
\section{Line-strength indices}

Line-strength indices in the  Lick/IDS system (e.g.\ T98) and
the D4000 defined by Bruzual (1983) were measured in these spectra.  The errors
were estimated from the uncertainties caused by photon noise, wavelength
calibration and flux calibration. We also added  a residual error based on the
comparison of galaxies observed in several runs and, in Run 2,  the error due
to the dichroic correction.  Table \ref{tipicos} shows the typical values of
the different sources of errors.

Table \ref{tabla.indices} presents the fully corrected line-strength indices
for our sample of galaxies.

\begin{table}
\caption{Typical values of the different considered error sources. Phot.:
photonic noise; Flux: Flux calibration; Wave.: Wavelength calibration; Dichr.:
Typical error in the subtraction of the dichroic pattern (this error is only
added in the measurements of Run 2). Res.: Residual errors, based on repeated
observations of 10 galaxies. \label{tipicos}}
\centering
\begin{tabular}{@{}lccccc@{$\;$}l@{}}
\hline
  Index      & Phot. &  Flux &  Wave.& Dichr. & Res.&\\
             &       &       &       & (Run 2)&    &\\
\hline
D4000        & 0.006 & 0.088 & 0.000 &  0.001&0.000& \AA\\
H$\delta_A$  & 0.156 & 0.038 & 0.005 &  0.053&0.000& \AA\\
H$\delta_F$  & 0.104 & 0.009 & 0.009 &  0.021&0.000& \AA\\
CN$_1$       & 0.004 & 0.002 & 0.000 &  0.003&0.001& mag\\
CN$_2$       & 0.005 & 0.003 & 0.000 &  0.003&0.010& mag\\
Ca4227       & 0.072 & 0.002 & 0.006 &  0.023&0.000& \AA\\
G4300        & 0.121 & 0.014 & 0.013 &  0.138&0.074& \AA\\
H$\gamma_A$  & 0.143 & 0.033 & 0.002 &  0.121&0.302& \AA\\ 
H$\gamma_F$  & 0.087 & 0.012 & 0.003 &  0.062&0.072& \AA\\
Fe4383       & 0.167 & 0.045 & 0.019 &  0.114&0.161& \AA\\ 
Ca4455       & 0.086 & 0.004 & 0.024 &  0.024&0.000& \AA\\
Fe4531       & 0.127 & 0.015 & 0.012 &  0.026&0.000& \AA\\
C4668        & 0.186 & 0.171 & 0.010 &  0.012&0.054& \AA\\ 
H$\beta$     & 0.078 & 0.010 & 0.007 &  0.003&0.029& \AA\\
Fe5015       & 0.167 & 0.081 & 0.022 &  0.025&0.115& \AA\\
Mgb          & 0.087 & 0.035 & 0.006 &  0.108&0.029& \AA\\
Fe5270       &       &       &       &       &0.050& \AA\\
Fe5335       &       &       &       &       &0.000& \AA \\
\hline
\end{tabular} 
\end{table}

%------------------------------------------------------------------------------
\subsection{Conversion to the Lick/IDS system}

The line-strength indices were transformed to the Lick system taking into
account three effects: (a) the differences in the spectral resolution
between the Lick/IDS system and our setups; (b) the internal velocity
broadening of the observed galaxies; and (c) small systematic offsets due to
the continuum shape differences.

(a) In order to account for the differences in spectral resolution, we
broadened the observed spectra with a Gaussian of wavelength-dependent width,
following the prescriptions of Gorgas, Jablonka \& Goudfrooij (2005, in
preparation). These authors estimated the resolution at which each
particular index should be measured by broadening stars in
common with the Lick library to several resolutions in steps of 25 km s$^{-1}$. 
They then calculated an approximate resolution that, changing
gently with wavelength, minimized the residuals. These values are given in 
Table \ref{resolution.lick}. In general, the values calculated by these 
authors agree with the estimates given in 
Worthey \& Ottaviani (1997).

(b) In the second step, we corrected our indices for  velocity dispersion
broadening. We followed the standard procedure of determining correction curves
from artificially broadened stellar spectra.  However, instead of using
individual stars, we broadened the different composite templates obtained in
the determination of the kinematics parameters (Section
\ref{velocitydispersion}). This was done because, in principle, one might
expect a dependence of the correction factor on line-strength. The indices
were then measured for each template and a correction factor of the form
$C(\sigma)=I(0)/I(\sigma)$ (for the atomic indices) and
$C(\sigma)=I(0)-I(\sigma)$ (for the molecular indices, H$\delta_A$ and H$\delta_F$) was determined for
each galaxy, where $I(0)$ represents the index corrected from the broadening,
and $I(\sigma)$ is the index measured in a spectrum broadened by $\sigma$.
Although we derived different polynomial for each galaxy, the final correction
factors were obtained by taking the mean of all the 98 templates in each
$\sigma$-bin.  This allowed us to quantify the errors due to this correction.
Appendix \ref{appen.index-sigma} shows the dependence of the correction factor
on $\sigma$ for all the measured indices and the derived uncertainties.  

(c) The original Lick/IDS spectra were not flux-calibrated by means of a
flux-standard star but normalised to a quartz-iodide tungsten lamp. The
resulting continuum shape cannot be reproduced and causes significant offsets
for indices with a broad wavelength coverage. To calculate these offsets,  we
observed 20, 28, 9 and 30 different Lick stars in the first, second, third and
forth observing runs respectively. By comparing the indices measured in our
stars with those in the Lick/IDS database for the same objects, we derived mean
offsets for all the indices in each observing run. However, the final offsets
were obtained as an average of the offsets in all the runs, weighted by the
number of observed stars. We do not to apply different offsets to  each run, as
they, in principle,  are expected to be the same (as our spectra are flux calibrated). 
Appendix \ref{appen.broad-lick} summarises the comparison with
the Lick/IDS stars. Last column of Table \ref{tab-comp.lick} shows the final
offsets applied and for all the indices analysed.

Differences in the derived offsets
between runs could appear due to systematic errors in the flux calibration, but
these possible errors affect stars and galaxies differently,
and they are taken into account in the comparison between galaxies observed in
different runs. 
Systematic differences  in the offsets between  runs could also 
appear if the Lick resolution was not matched perfectly in all the runs. As 
the resolution in Run 4 is very different from that of the other runs, we have
compared, in Appendix \ref{double.check}, the indices measured from  
stars observed in both Run 4 and other runs. We have not found any 
significative difference.
In the special case of the D4000, we compared the stars in common with
the library of Gorgas et al.\ (1999).  The estimated offset for this index  is
zero.

Finally, we calculated a residual error for each index based on repeated
observations of galaxies in different runs. We have a total of 11 repeated
observations of 10 different galaxies. The residual error was computed by
comparing the r.m.s.\ dispersion with the estimated errors from known sources.
The final residual errors are shown in the last column of Table \ref{tipicos}.
 
\begin{table}
\begin{tabular}{lrlr}
\hline\hline
Index        & Resolution & Index & resolution\\
             & (km/s)     &       &  (km/s)  \\
\hline  
H$\delta_A$  & 325        & Ca4455  & 250 \\
H$\delta_F$  & 325        & Fe4531  & 250\\
CN$_1$       & 325        & C4668   & 250\\
CN$_2$       & 325        & H$\beta$& 225\\
Ca4227       & 300        & Fe5015  & 200\\ 
G4300        & 300        & Mgb     & 200\\
H$\gamma_A$  & 275        & Fe5270  & 200\\
H$\gamma_F$  & 275        & Fe5335  & 200\\
Fe4383       & 250        &         &   \\

\hline
\end{tabular}
\caption{Resolution at which Lick indices are measured.\label{resolution.lick}}
\end{table}

%------------------------------------------------------------------------------
\subsection{Emission correction}

\begin{figure}
\resizebox{\hsize}{!}{\includegraphics[bb=28 9 530 748,angle=-90]{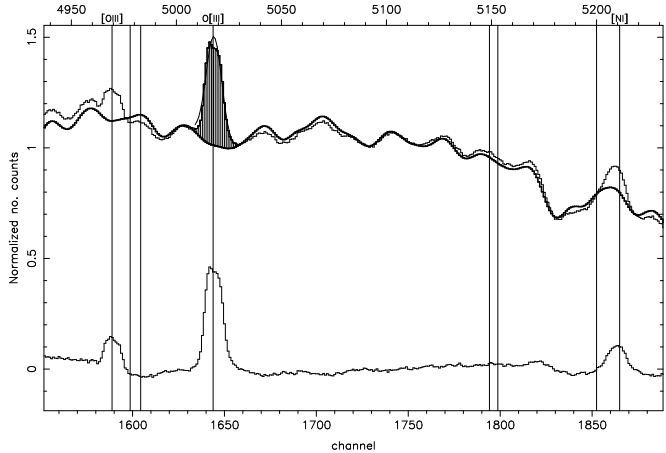}} 
\caption{Illustrative example of the procedure followed to measure
emission-lines. The upper part of the figure shows the galaxy (NGC 4278; thin
line) and the template spectrum corrected for the kinematics parameters (thick
line). The difference between both spectra
\mbox{($\mbox{observed}-\mbox{template}$)} is shown at the bottom of the
figure. The equivalent width of the emission line [O {\sc iii}]$\lambda 5007$
is highlighted in the shaded area. The vertical lines indicate the expected
location of the [O {\sc iii}]$\lambda 5007$, [N {\sc ii}]$\lambda 5199$, and
several sky lines.\label{emision}}
\end{figure}

Elliptical galaxies contain much less dust and ionised gas than spiral
galaxies, and were regarded as dust- and gas- free objects for a long time.
However, surveys of large samples of early-type galaxies (Caldwel 1984;
Phillips et al.  1986; Goudfrooij et al.\ 1994) have revealed that 50--60 per
cent of these galaxies show weak optical emission lines. Some line-strength
indices are  affected by these emission lines, in particular H$\beta$, Fe5015
and Mgb. The effect of the  emission on H$\beta$ is particularly important
because our estimation of ages (Paper II) relies on its strength.  The
emission, when present, tends to fill the line, lowering the value of the index
and, hence, increasing the derived age.
To correct the H$\beta$ index from this effect, we used the calibration of G93,
who showed that H$\beta$ is correlated with [O {\sc iii}] such that
\mbox{EW(H$\beta_{\rm em}$)/EW([O {\sc iii}]$\lambda 5007)\sim 0.7$}. Trager
et al.\ (2000a; hereafter T00a) carefully studied the accuracy of this
correction, finding a better value of 0.6 instead of 0.7. Whilst there is
evidence that this correction factor is uncertain for individual galaxies
(Mehlert et al.\ 2000), it is good enough in a statistical sense. 
 In any case, we have repeated  the analysis eliminating all 
the galaxies with [O {\sc iii}]$\lambda 5007> 0.4$\AA  and none
of the results presented in this work suffer variation.
To determine
the [O {\sc iii}] emission, we subtracted an emission-free template spectrum
from the galaxy spectrum and measured the residual equivalent width.
The zero-emission template was the one calculated during the determination of
the kinematics parameters, as described in Section \ref{velocitydispersion}.
An example of the process is shown in Fig. \ref{emision}. The emission lines
[O {\sc iii}]$\lambda 5007$ and [N {\sc ii}]$\lambda 5199$, which affect the
measurements of Fe5015 and Mgb, were replaced with the optimal template in
these regions. 
A total of 37  galaxies (38\% of our sample) were found to have evidence of
[O {\sc iii}]$\lambda$5007 emission \mbox{(EW([O {\sc iii}])$>0.2$ \AA)}. The
[O {\sc iii}] equivalent widths measured in the central spectra are presented
in the last column of Table \ref{tabla.indices}.
  
%------------------------------------------------------------------------------
\subsection{Comparison with other authors}
\label{sec-comparison}

\begin{figure*}
    \resizebox{0.23\hsize}{!}{\includegraphics[angle=-90]{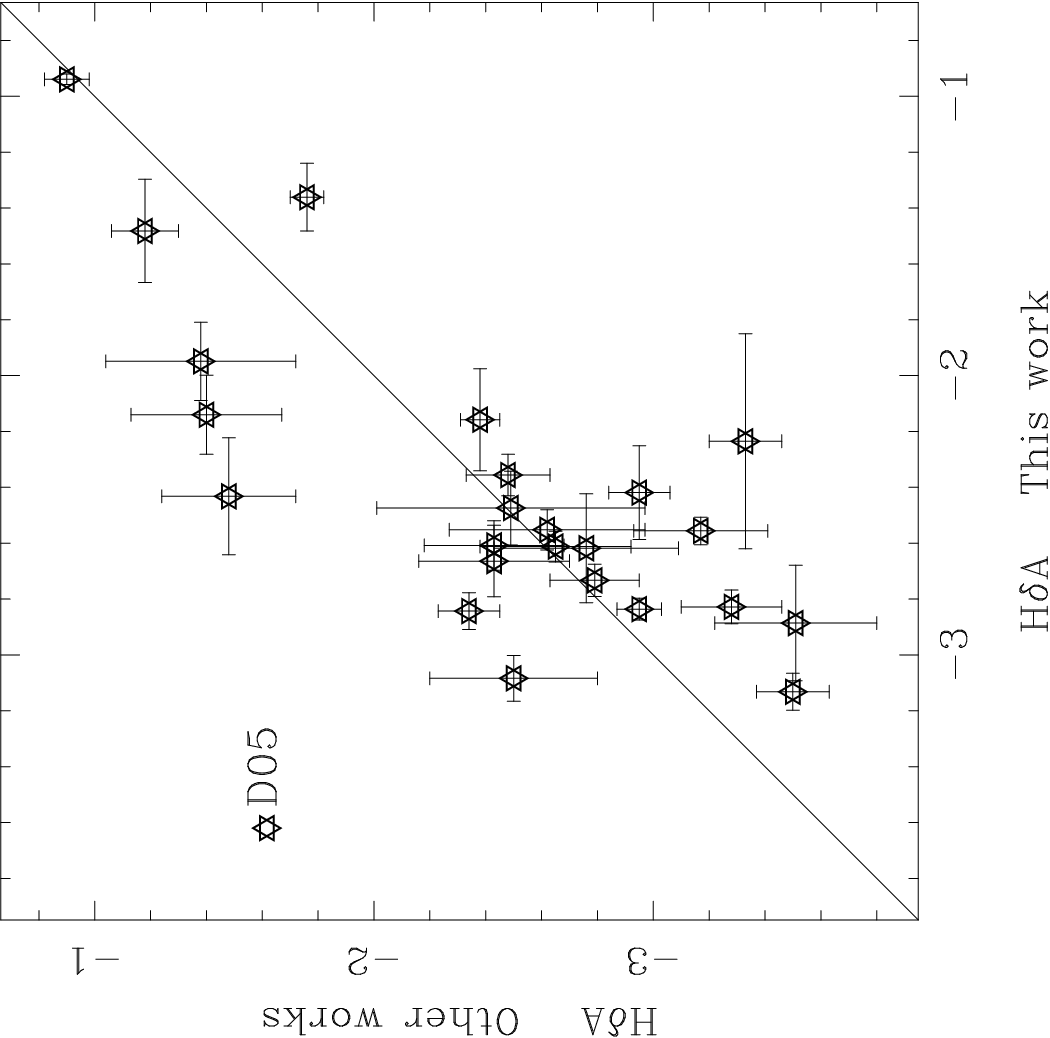}}
    \hfill
    \resizebox{0.23\hsize}{!}{\includegraphics[angle=-90]{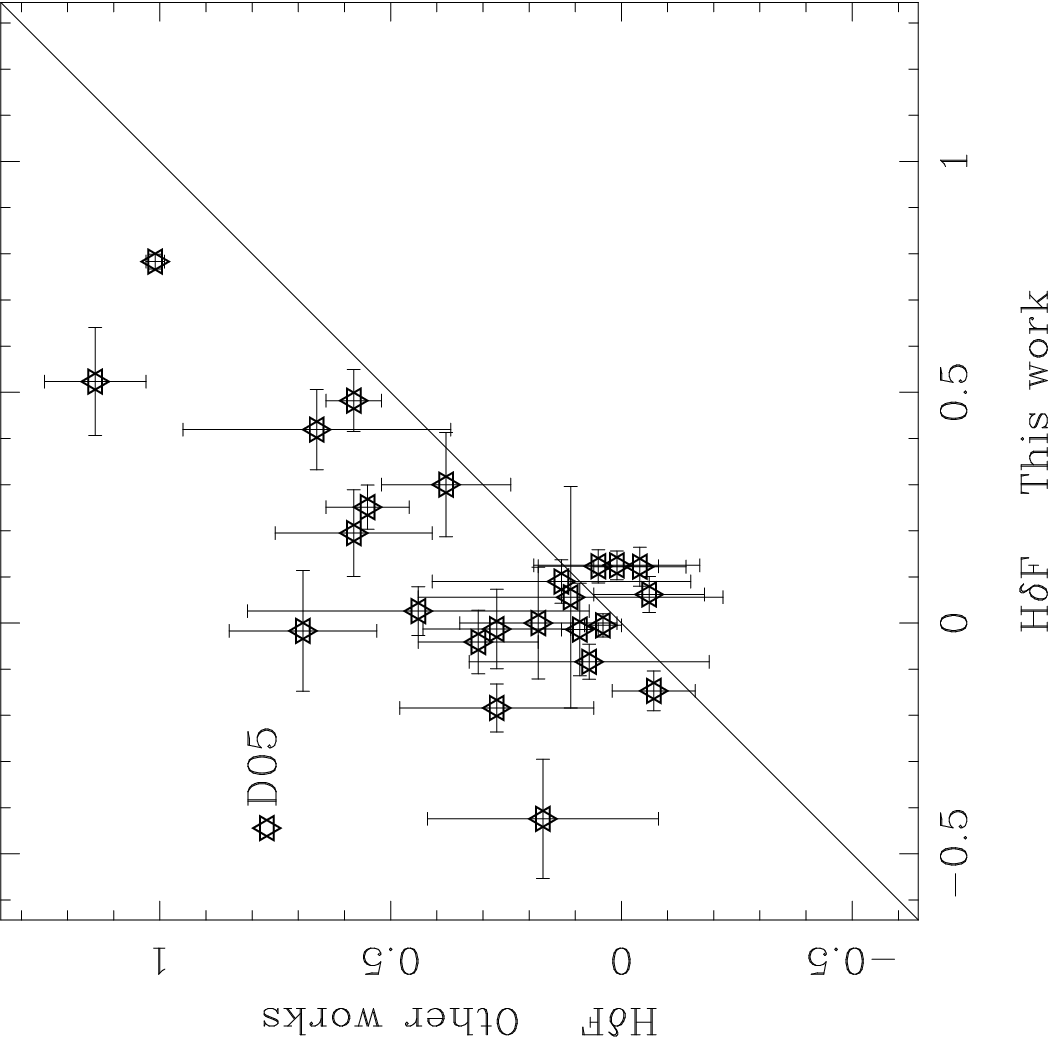}}
    \hfill
    \resizebox{0.23\hsize}{!}{\includegraphics[angle=-90]{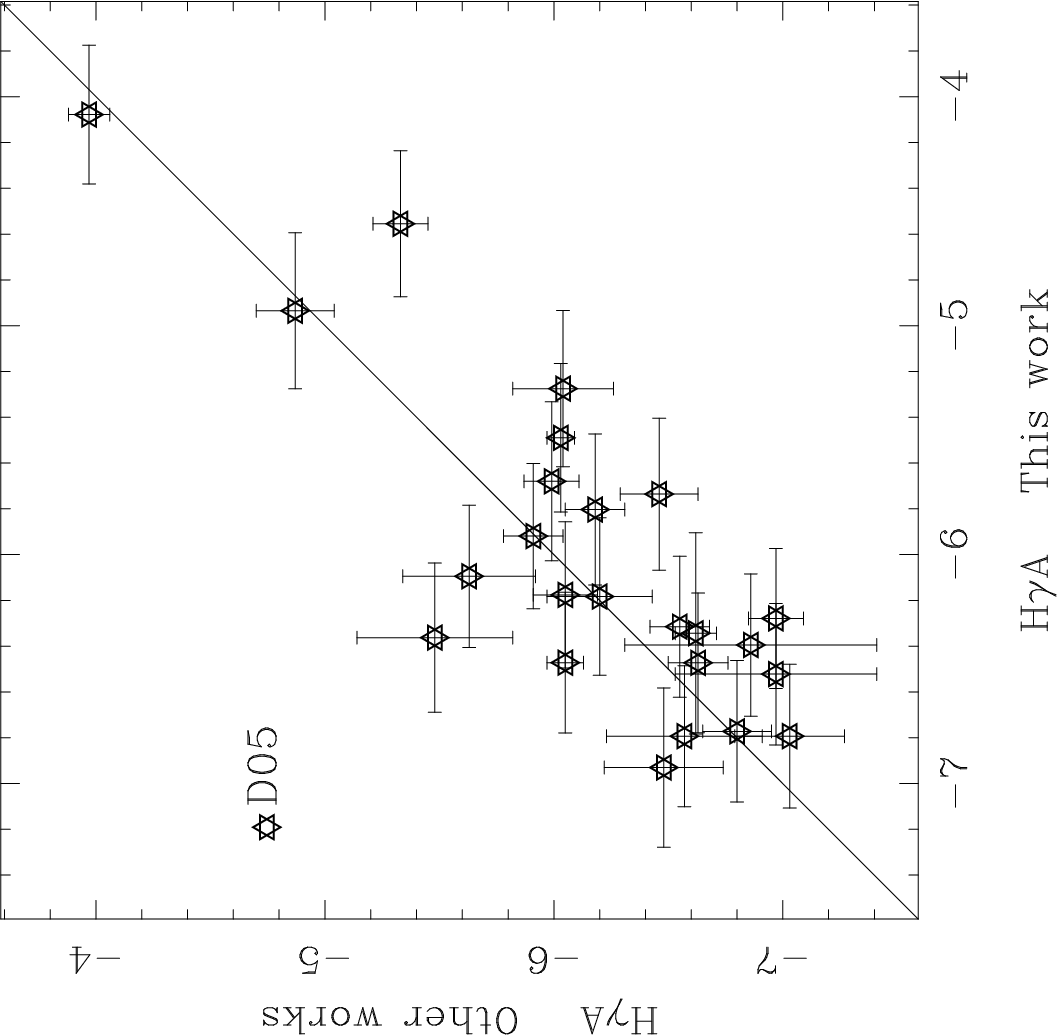}}
    \hfill
    \resizebox{0.23\hsize}{!}{\includegraphics[angle=-90]{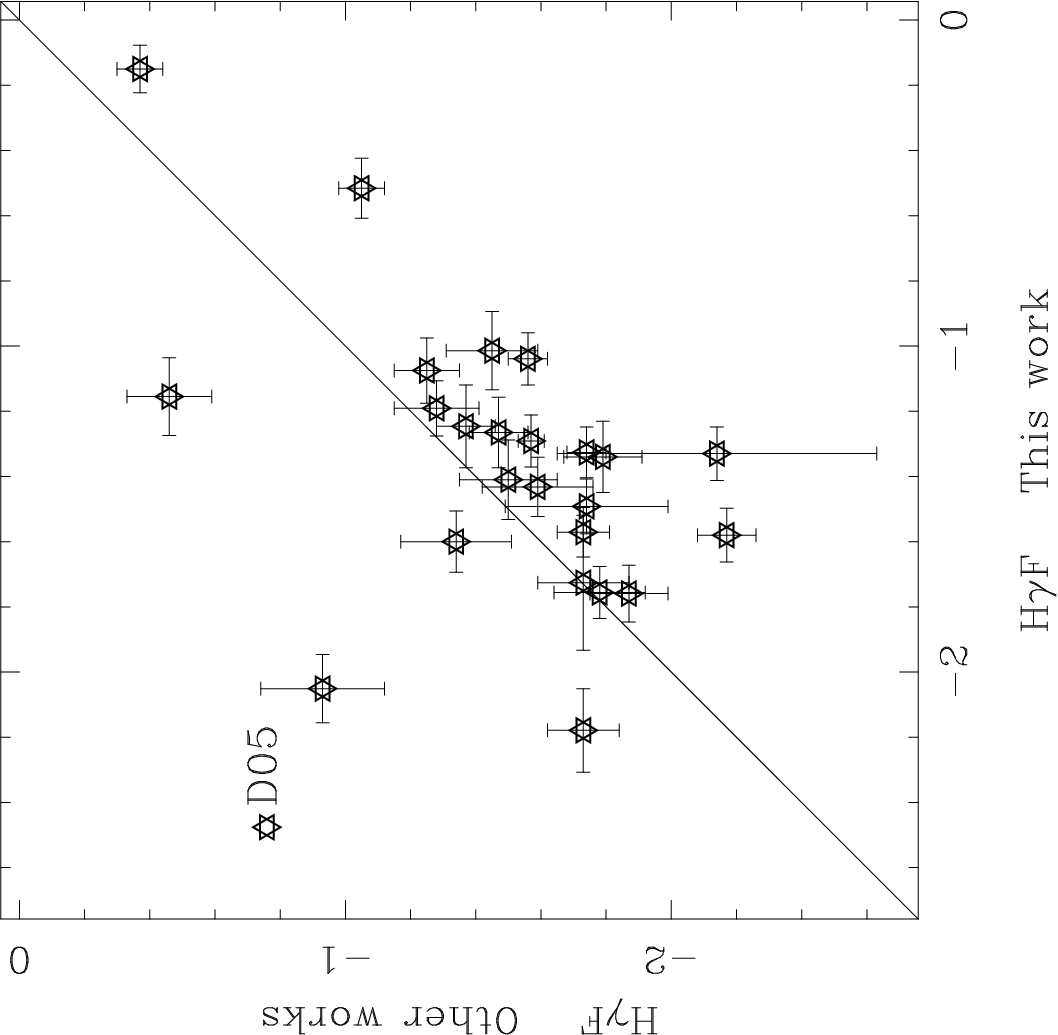}}

 \vspace{4mm}

  \resizebox{0.23\hsize}{!}{\includegraphics[angle=-90]{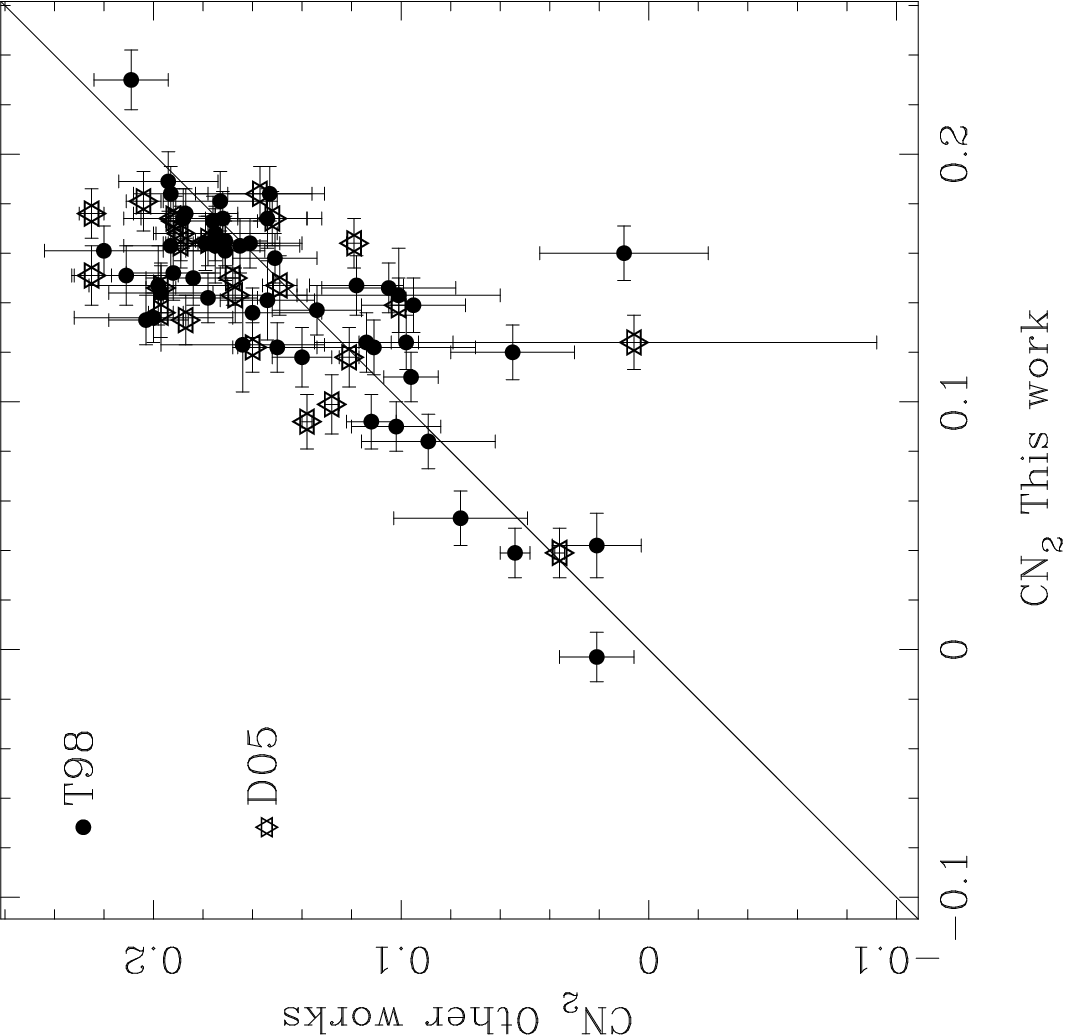}}
  \hfill
  \resizebox{0.23\hsize}{!}{\includegraphics[angle=-90]{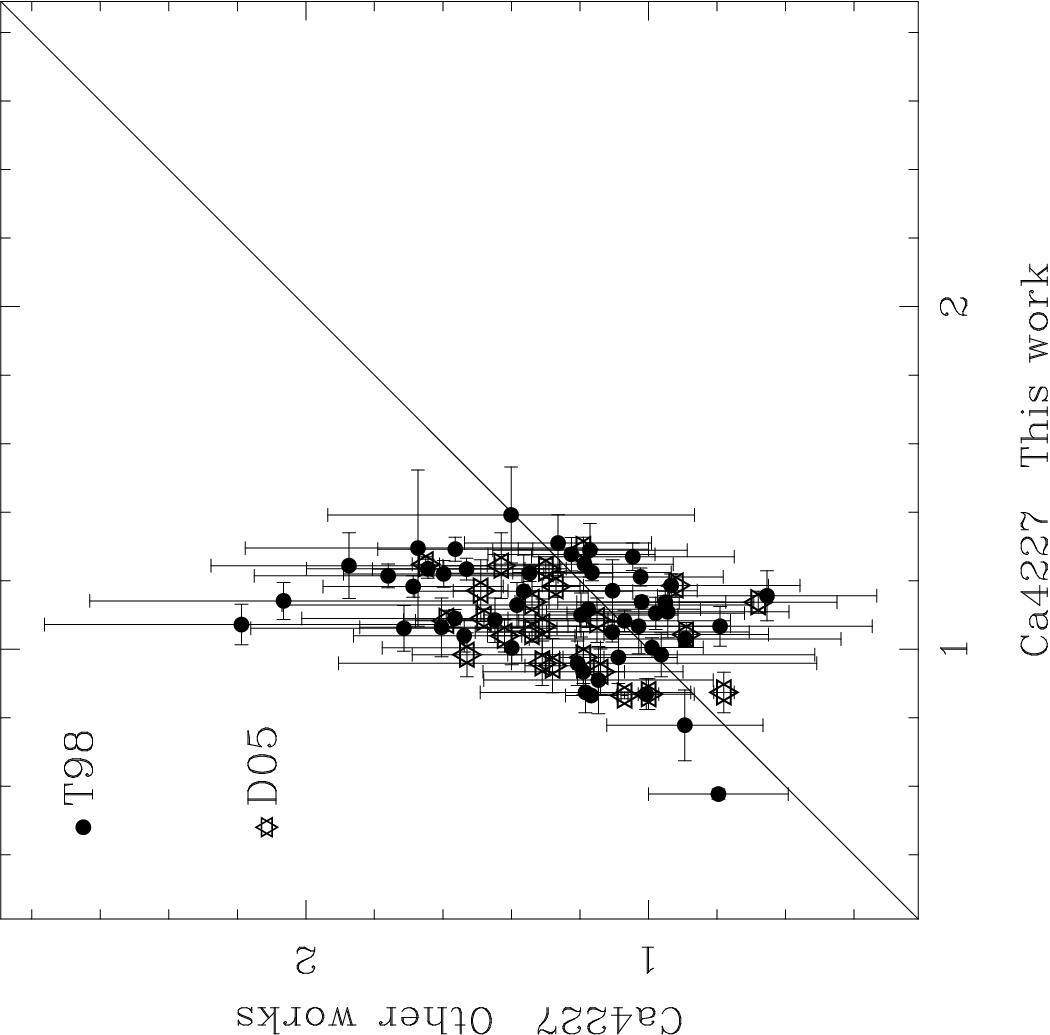}} 
  \hfill
  \resizebox{0.23\hsize}{!}{\includegraphics[angle=-90]{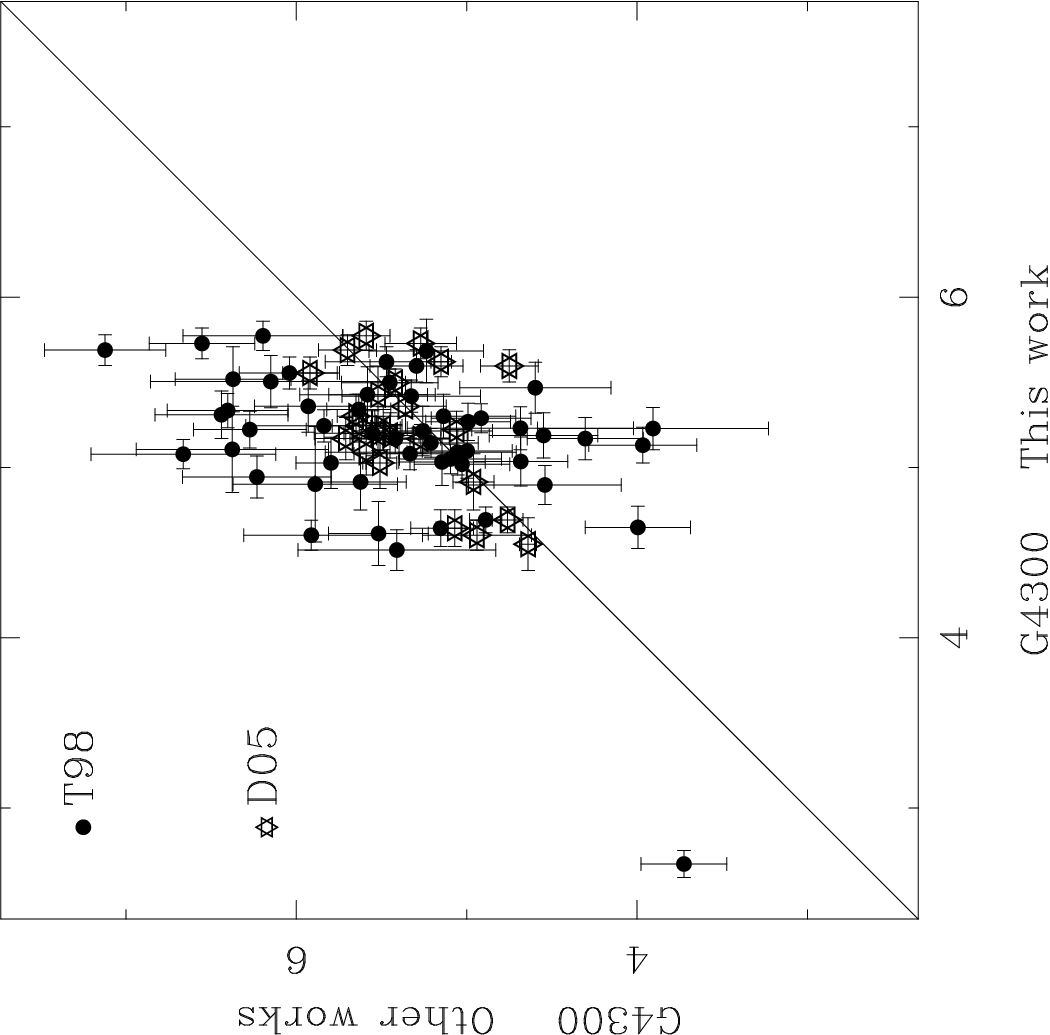}}
  \hfill
  \resizebox{0.23\hsize}{!}{\includegraphics[angle=-90]{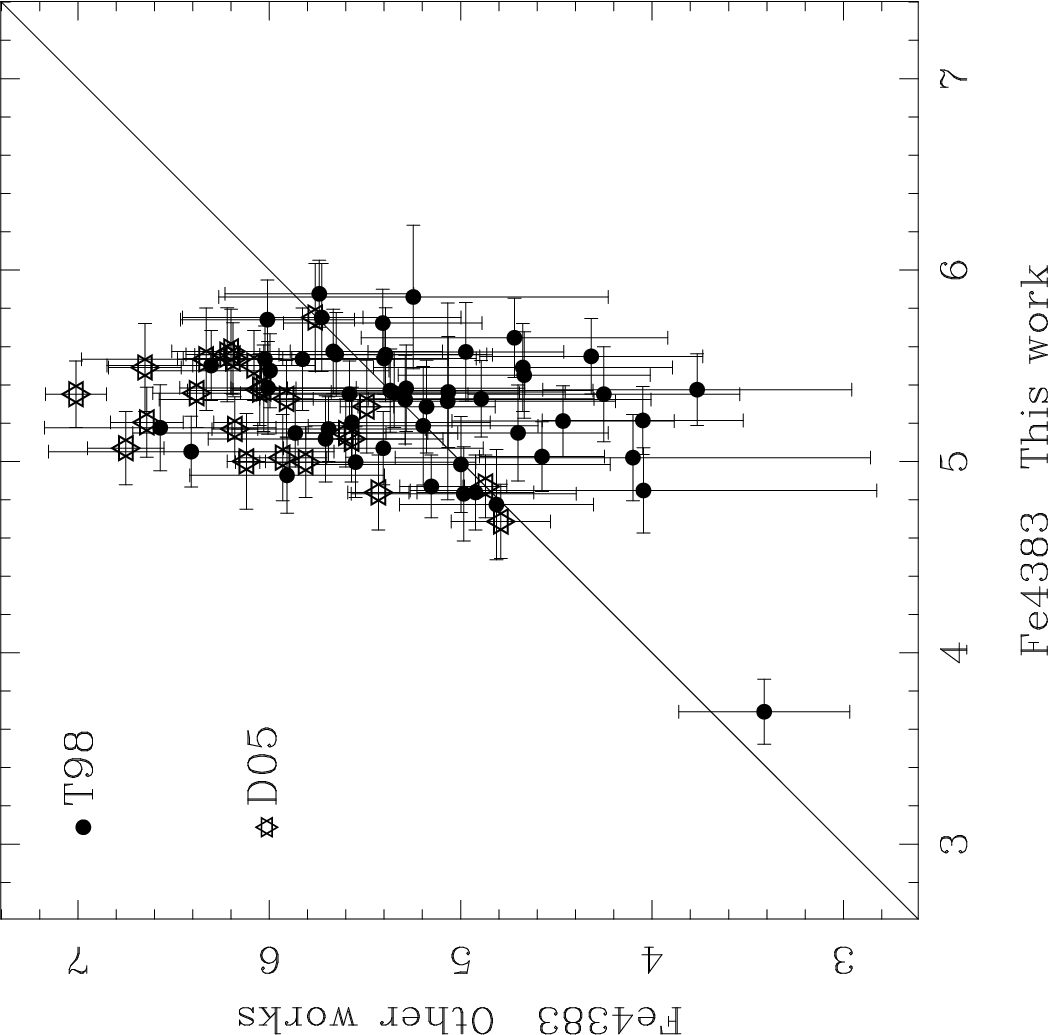}}

  \vspace{4mm}
  
  \resizebox{0.23\hsize}{!}{\includegraphics[angle=-90]{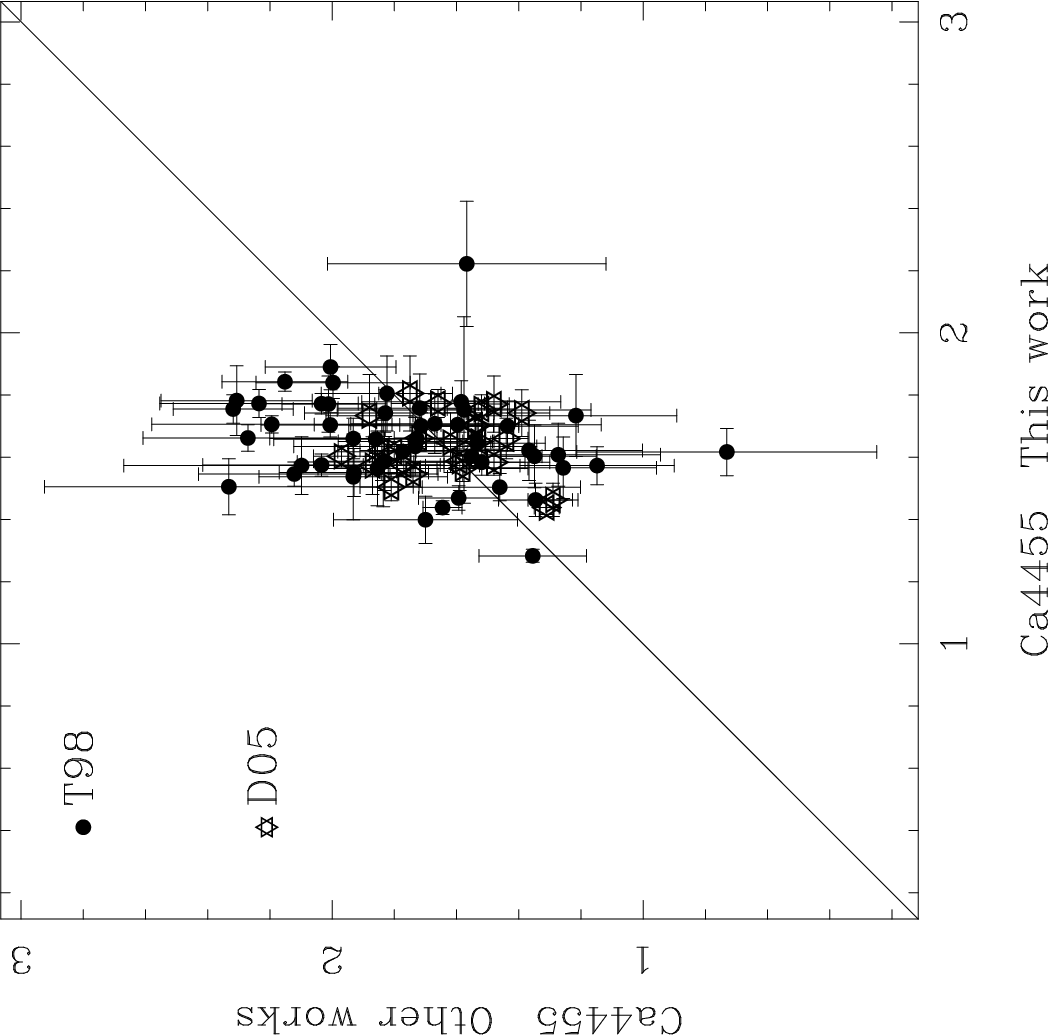}}
  \hfill
  \resizebox{0.23\hsize}{!}{\includegraphics[angle=-90]{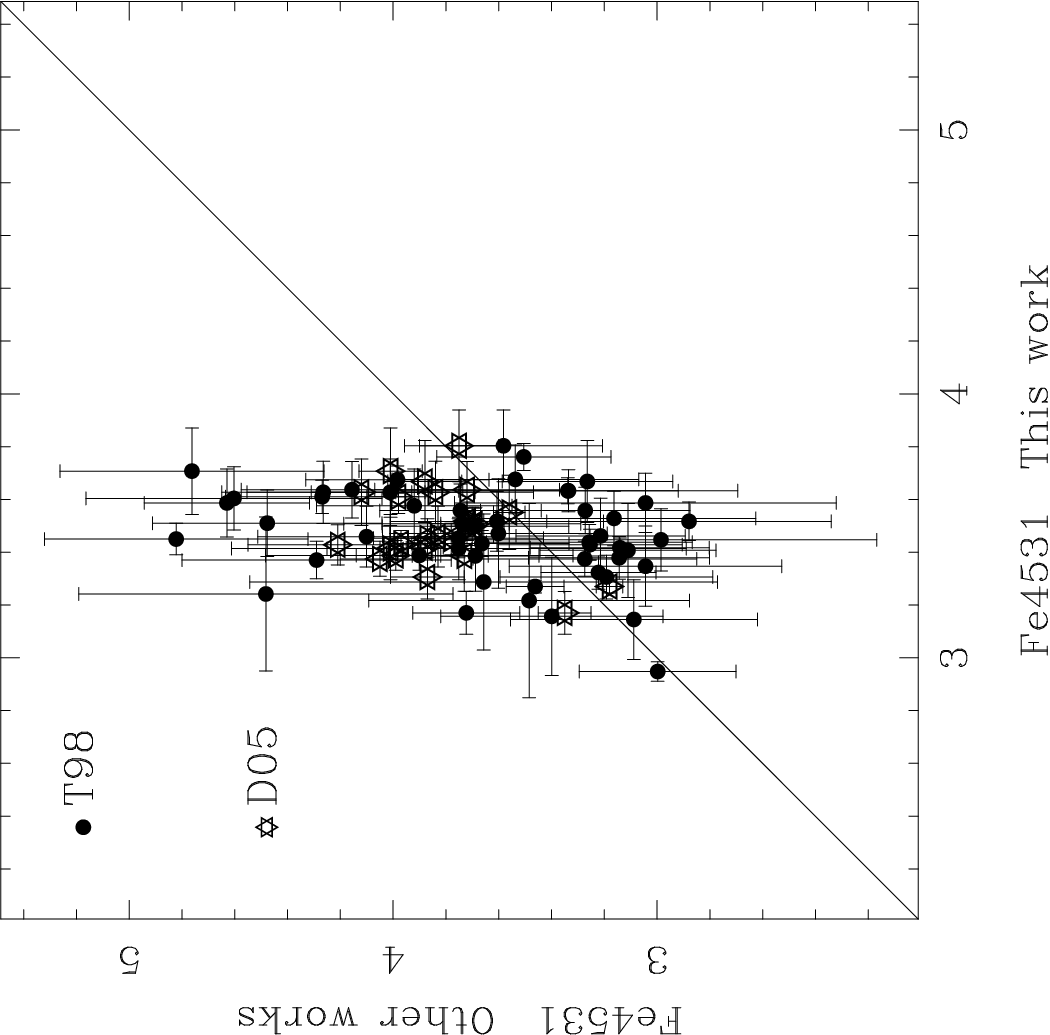}}
  \hfill
  \resizebox{0.23\hsize}{!}{\includegraphics[angle=-90]{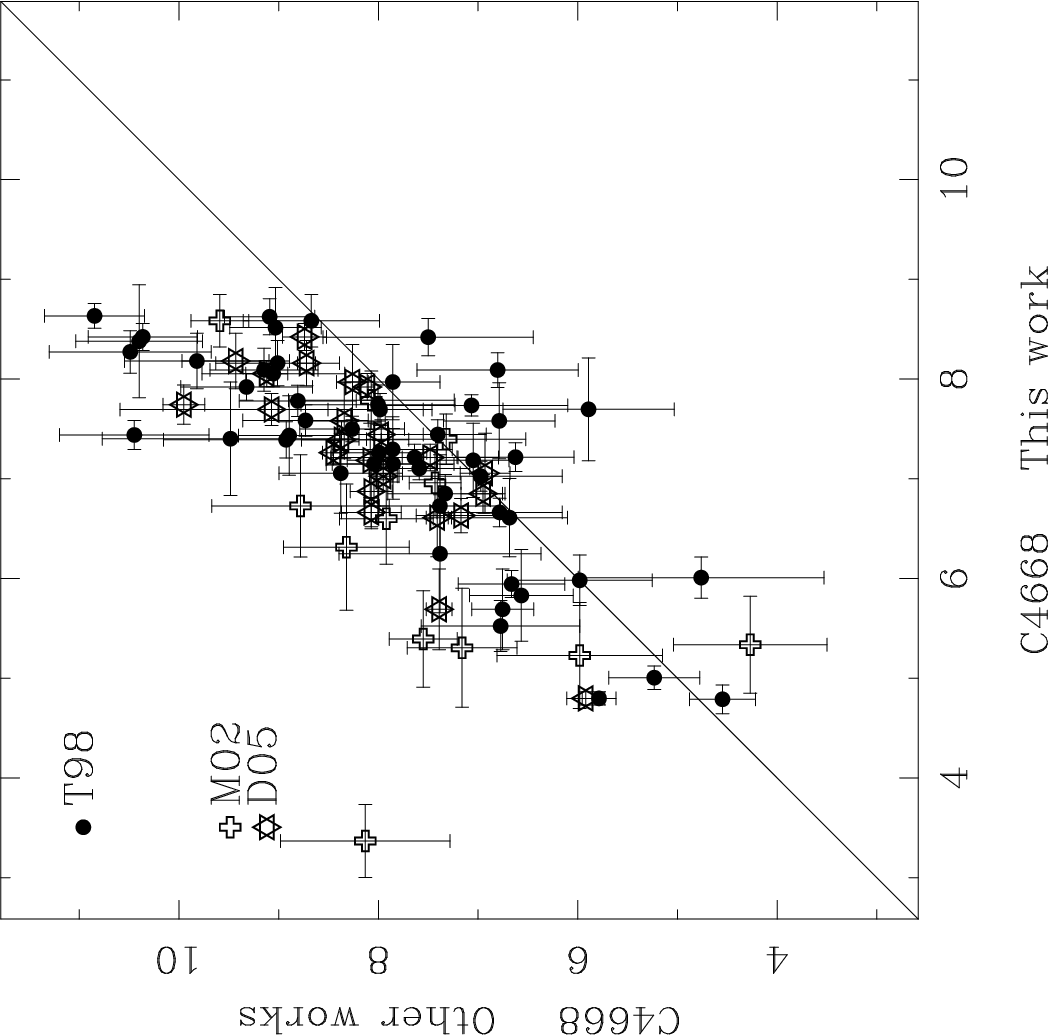}}
  \hfill
  \resizebox{0.23\hsize}{!}{\includegraphics[angle=-90]{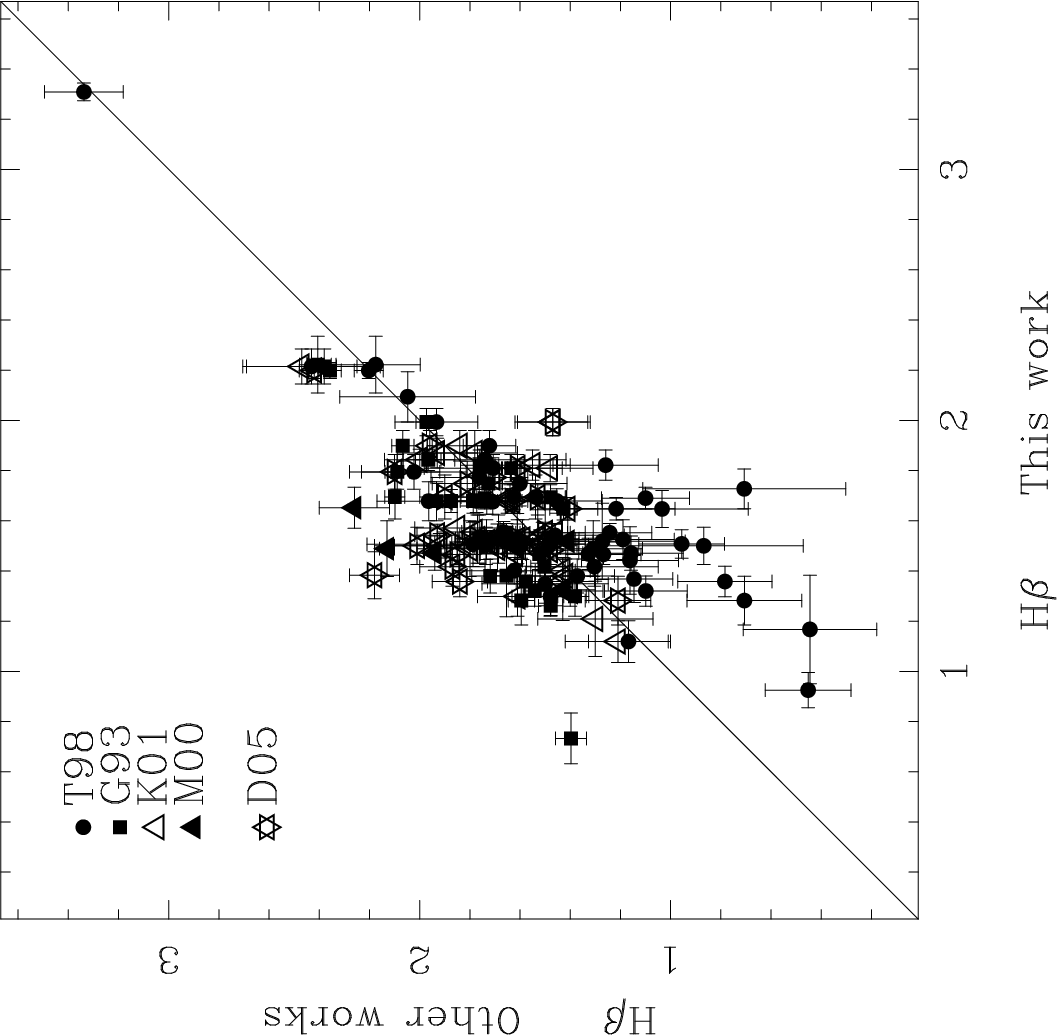}}

  \vspace{4mm}
  
  \resizebox{0.23\hsize}{!}{\includegraphics[angle=-90]{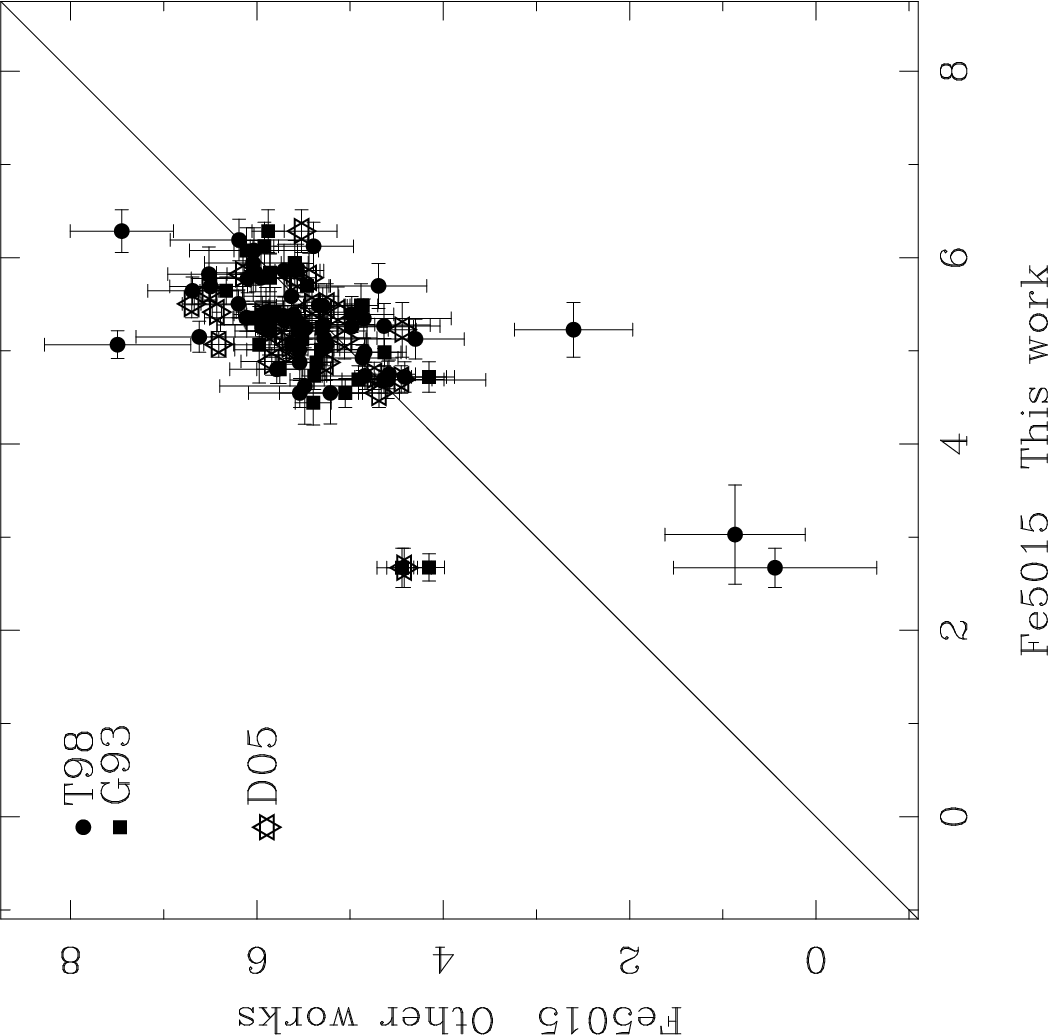}}
  \hfill
  \resizebox{0.23\hsize}{!}{\includegraphics[angle=-90]{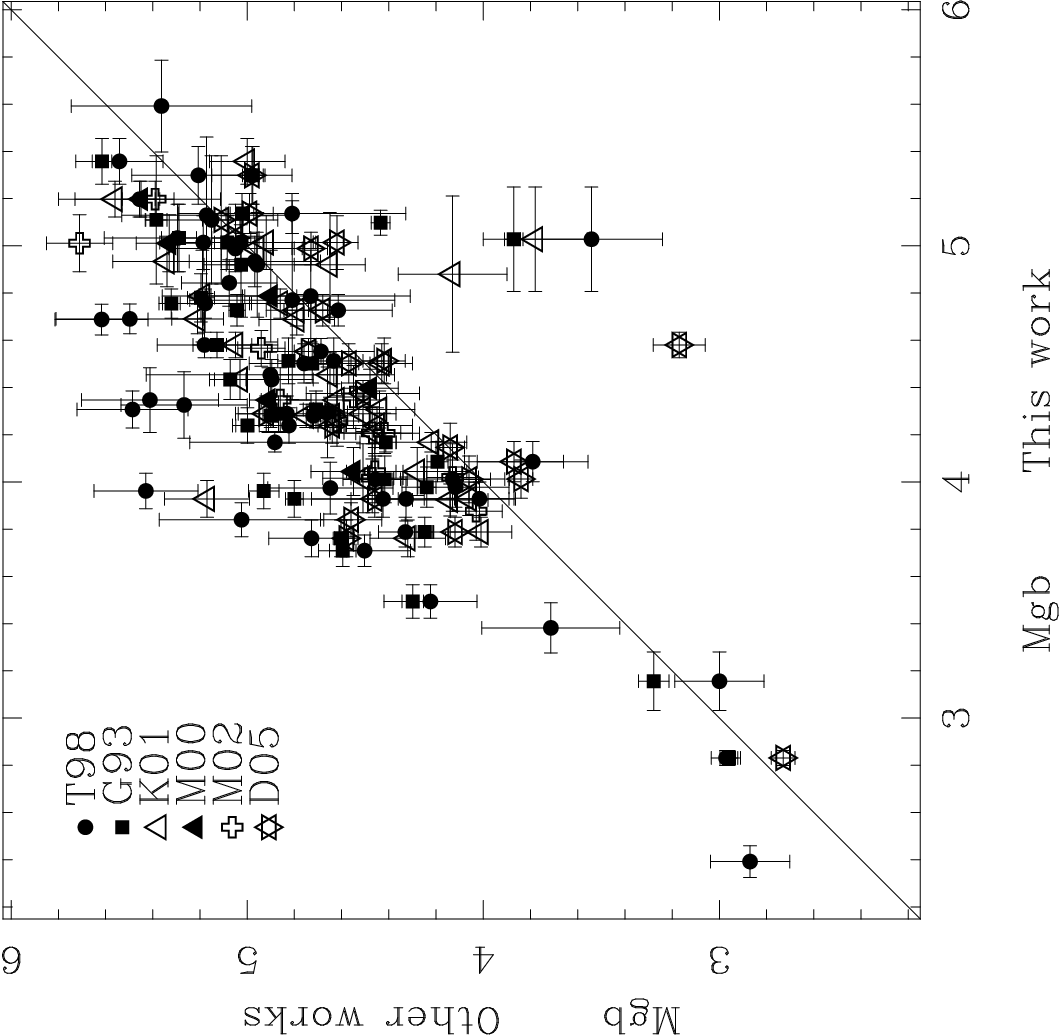}}
  \hfill
  \resizebox{0.23\hsize}{!}{\includegraphics[angle=-90]{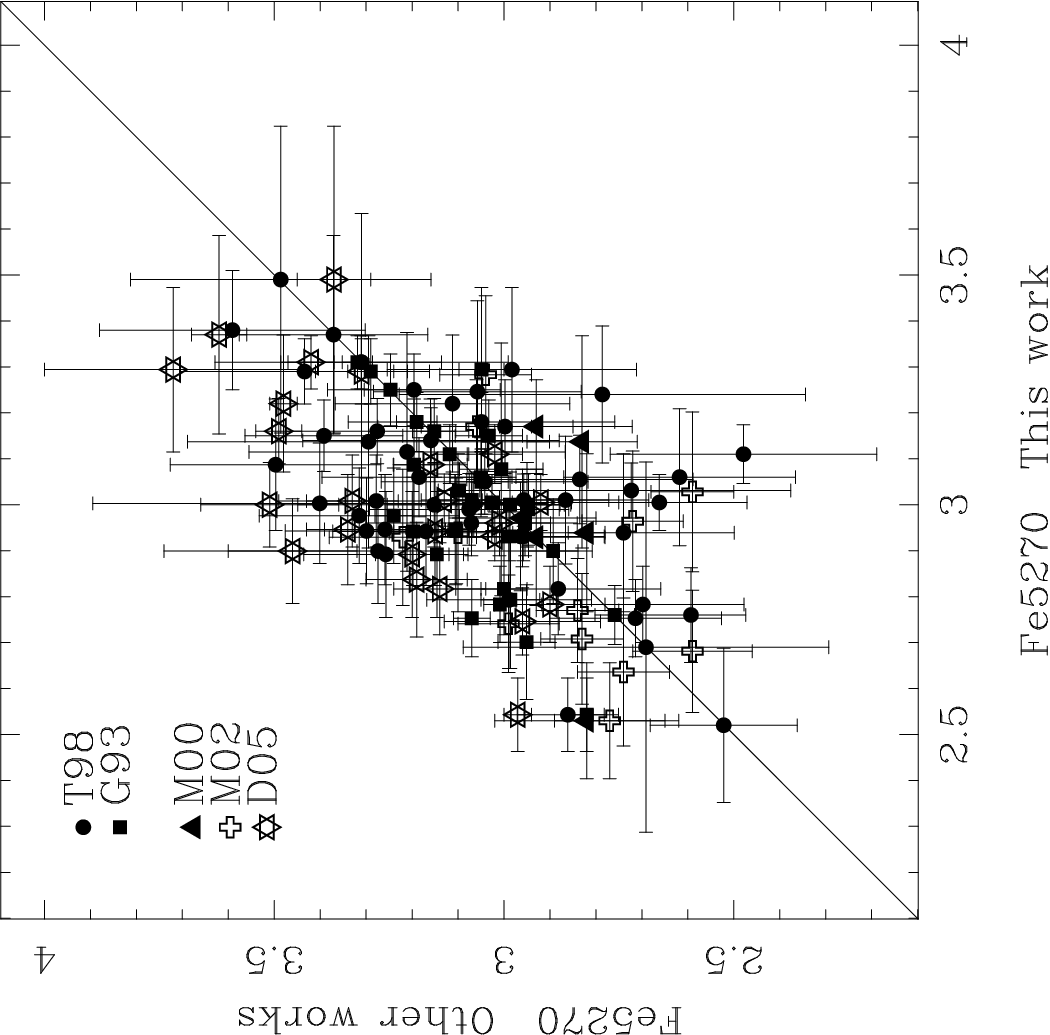}}
  \hfill
  \resizebox{0.23\hsize}{!}{\includegraphics[angle=-90]{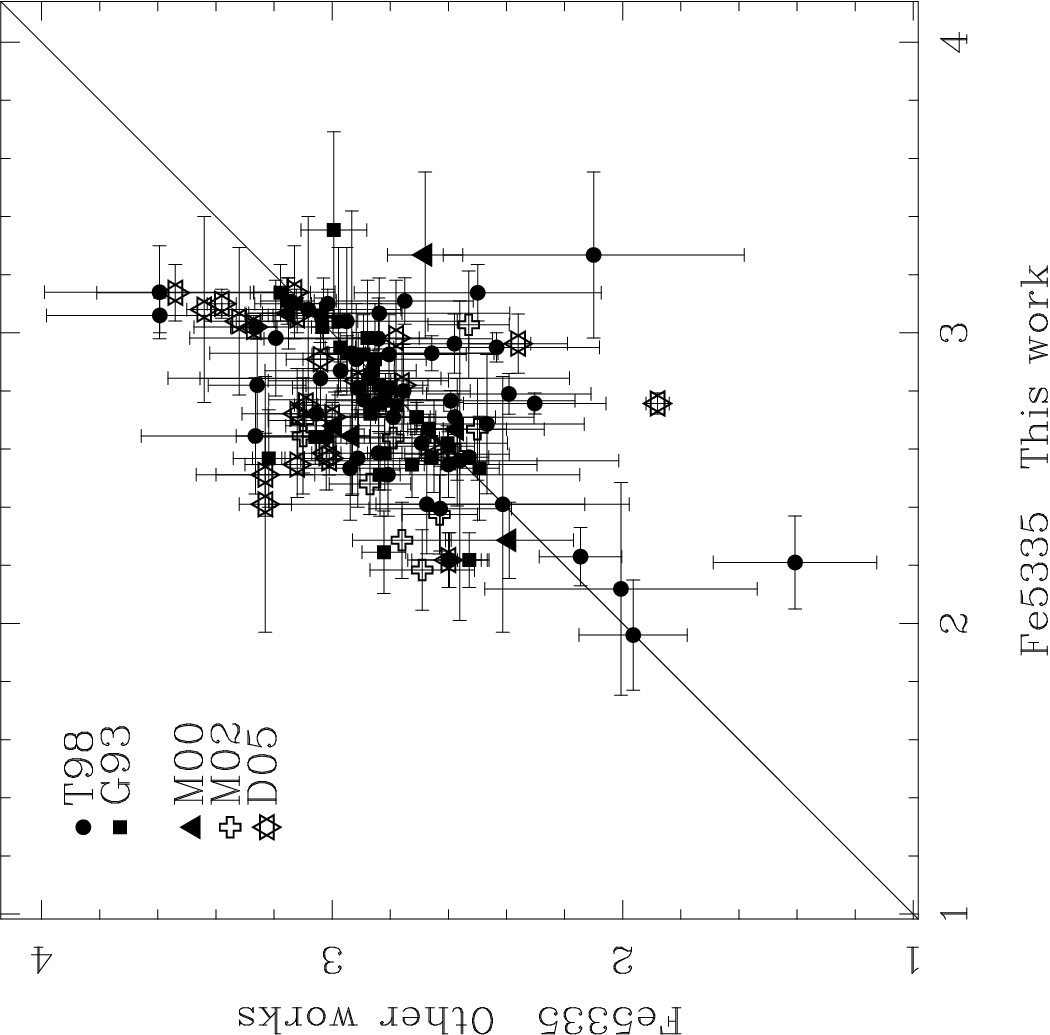}}
\caption{Comparison between  the Lick indices measured in this study with those
in other works for the galaxies in common. Different symbols represent
distinct references as indicated in the insets.\label{comparison.oa}}
\end{figure*}

To verify that our measurements have been  properly transformed into the Lick
system, and that the errors have been properly estimated, we have compared our
line strengths with data from 6 different works with galaxies in common:
(i) G93, (ii) T98, 
(iii) Mehlert et al.\ (2000, hereafter M00),
(iv) Kuntschner et al.\ (2001, hereafter K01),  
(v) Moore et al.\ (2002, hereafter M02) and (vi) Denicol\'o et al. (2005a, 
hereafter D05). Figure \ref{comparison.oa} shows this
comparison and Table \ref{tab.comparison.oa} summarises the mean offsets
between different studies and the root mean square dispersions. 

In general, our measurements agree with other studies fairly well, although we
find significant offsets in the value of some indices.  In particular, Mgb
and C4668 seem to be depressed in our sample compared with the other studies.
We will address throughout the paper the possible variation in our 
conclusions should an offset of $-0.2$ and $-0.848$ in Mgb and C4668 
(corresponding to the mean difference in the comparison with other authors) be 
assumed.

We can obtain a good estimation of the quality of our errors comparing
the dispersion along the 1:1 relation with the one expected from the errors. 
The last column of Table \ref{tab.comparison.oa} shows the probability that
the real scatter is larger than the one expected from the errors by chance. As can be seen, 
in most of the cases the observed scatter agrees with the one predicted from
the errors, which indicates that the errors have been correctly estimated. 
 However, this does not happen in the comparison with D05. As this is  the only 
study in which the real scatter is significatively larger than the 
one expected by the errors, we think that these authors have underestimated
the errors in their measurements.

\begin{table}
\caption{Comparison of line strengths measured in this and other studies. Ref.:
reference of the comparison work (see description in the text); $N$: number of
galaxies in common; $\Delta I$: calculated offset between both studies (this work minus
other study); $\sigma$: r.m.s.\ dispersion; $\sigma_{\rm exp}$:
expected r.m.s.\ from the errors; $t$: $t$-parameter of the comparison of
means; $\alpha$: probability that $\sigma > \sigma_{\rm exp}$ by chance.
\label{tab.comparison.oa}}
\centering
\begin{tabular}{@{}lccccccc@{}}
\hline
Index    &  Ref. &  $N$  & $\Delta I$ & $\sigma$  & $\sigma_{\rm exp}$ & $t$& $\alpha$\\
\hline
H$\delta_A$&D05   &  24         &  +0.305    &  0.442    & 0.569             & 2.76 & 0.006\\
H$\delta_F$&D05   &  24         & $-0.036$   &  0.197    & 0.450              & 0.87 & 0.197\\ 
H$\gamma_A$&D05   &  24         &  +0.550    &  0.428    & 0.555              & 3.82 & 0.001\\
H$\gamma_F$&D05   &  24         &  +0.349    &  0.359    & 0.463              & 3.38 & 0.001\\
CN$_2$   &  T98  &  53          & $-0.010$   &  0.033    & 0.195              & 2.22  & 0.015\\ 
CN$_2$   &  D05  &  24          &  +0.125    &  0.034    & 0.094              & 4.64 & 0.001\\
Ca4427   &  T98  &  53          & $-0.181$   &  0.287    & 0.755              & 3.88 & 0.001\\
Ca4227   &  D05  &  24          & $-0.070$   &  0.170    & 0.341              & 1.85 & 0.039\\
G4300    &  T98  &  53          & $-0.272$   &  0.640    & 0.807              & 2.84 & 0.003\\
G4300    &  D05  &  24          &  +0.061    &  0.310    & 0.506              & 0.94 & 0.179\\
Fe4383   &  T98  &  53          &$-0.051$    &  0.561    & 1.081              & 0.65 & 0.259\\
Fe4383   &  D05  &  24          & $-0.639$   &  0.463    & 0.615              & 3.91 & 0.001\\
Ca4455   &  T98  &  53          & $ -0.150$  &  0.304    & 0.700              & 3.21 & 0.001\\
Ca4455   &  D05  &  24          &  +0.170    &  0.210    & 0.419              & 3.05 & 0.003\\
Fe4531   &  T98  &  53          & $-0.205$   &  0.422    & 0.887              & 3.18 & 0.001\\
Fe4531   &  D05  &  24          & $-0.175$   &  0.217    & 0.411              & 3.05 & 0.003\\
C4668    &  T98  &  53          & $-0.667$   &  0.821    & 1.129              & 4.57 & 0.001\\ 
C4668    &  M02  &  12          & $-1.194$   &  1.332    & 1.143              & 2.19 & 0.027\\
C4668    &  D05  &  24          & $-0.684$   &  0.541    & 0.671              & 3.79 & 0.001\\
H$\beta$ &  T98  &  53          & +0.121     &  0.256    & 0.621              & 3.10 & 0.001\\
H$\beta$ &  G93  &  33          & $-0.057$   &  0.150    & 0.333              & 2.16 & 0.019\\
H$\beta$ &  K01  &  32          &+0.098      &  0.187    & 0.565              & 2.62 & 0.007\\
H$\beta$ &  M00  &  6           & $-0.428$   &  0.316    & 0.418              & 1.86 & 0.068\\
H$\beta$ &  D05  &  24          & +0.004     &  0.491    & 0.357              & 0.04 & 0.484\\
Fe5015   &  T98  &  53          &$-0.187$    &  1.030    & 0.739              & 1.79 & 0.040\\
Fe5015   &  G93  &  33          & $-0.451$   &  0.465    & 0.612              & 2.22 & 0.017\\
Fe5015   &  D05  &  24          & $-0.208$   &  0.653    & 0.574              & 1.48 & 0.076\\
Mgb      &  T98  &  53          &$-0.376$    &  0.478    & 0.717              & 4.49 & 0.001 \\
Mgb      &  G93  &  33          &$-0.535$    &  0.601    & 0.386              & 2.20 & 0.018\\
Mgb      &  K01  &  32          &$-0.047$    &  0.431    & 0.615              & 0.61 & 0.273\\
Mgb      &  M00  &   6          &$-0.158$    &  0.220    & 0.491              & 1.39 & 0.118\\
Mgb      &  M02  &  12          & $-0.160$   &  0.125    & 0.518              & 2.66 & 0.012\\
Mgb      &  D05  &  24          & +0.061     &  0.443    & 0.467              & 0.67 & 0.255\\
\hline
\end{tabular}
\end{table}

\addtocounter{table}{-1}
\begin{table}
\caption{\it Continued}
\centering
\begin{tabular}{@{}lccccccc@{}}
\hline\hline
Index    &  Ref. &  $N$  & $\Delta I$ & $\sigma$  & $\sigma_{\rm exp}$ & $t$& $\alpha$\\
\hline

Fe5270   &  T98  &  53          & +0.009     &  0.340    & 0.761              & 0.19 & 0.425  \\
Fe5270   &  G93  &  33          & +0.004     &  0.027    & 0.322              & 0.61 & 0.273\\
Fe5270   &  M00  &   6          & +0.212     &  0.153    & 0.480              & 1.87 & 0.067\\
Fe5270   &  M02  &  12          & +0.118     &  0.213    & 0.495              & 1.66 & 0.064\\
Fe5270   &  D05  &  24          & $-0.057$   &  0.181    & 0.399              & 1.46 & 0.079\\
Fe5335   &  T98  &  53          &+0.009      &  0.340    & 0.761              & 0.19 & 0.425 \\
Fe5335   &  G93  &  33          &$-0.009$    &  0.018    & 0.340              & 1.79 & 0.042\\
Fe5335   &  M00  &   6          &+0.111      &  0.398    & 0.514              & 0.65 & 0.275\\ 
Fe5335   &  M02  &  12          &+0.035      &  0.373    & 0.598              & 0.32 & 0.378\\ 
Fe5335   &  D05  &  24          &+0.040      &  0.431    & 0.441              & 0.45 & 0.328\\
$<$Fe$>$ &  K01  &  31          &$-0.016$    &  0.183    & 0.159              & 0.65 & 0.110\\
\hline
\end{tabular}
\end{table}

Furthermore, we have used the galaxies in common with other studies to 
double check the possibility of systematic differences
between  different observing runs confirming the absence  of these
differences. This comparison 
can be found in Appendix \ref{appendix.offsets}. 
%%%%%%%%%%%%%%%%%%%%%%%%%%%%%%%%%%%%%%%%%%%%%%%%%%%%%%%%%%%%%%%%%%%%%%%%%%%%%%%
\section{The index--velocity dispersion relations}
\label{sec.indexsigma}

The relations between the central velocity dispersion and the strength of the
integrated stellar Mg and MgH features around 5100 \AA  have been studied in
numerous works (e.g.\ Terlevich et al.\ 1981; Gorgas et al.\ 1990; Guzm\'an et
al.\ 1992; Bender, et al.\ 1993; J\o rgensen et al.\ 1996; Bender et
al.\ 1998; Bernardi et al.\ 1998; Colless et al.\ 1999; J\o rgensen 1999;
Concannon, Rose \& Caldwell 2000; Kuntschner 2000; Poggianti et al.\ 2001a;
Proctor \& Sansom 2002; Worthey \& Collobert 2003; Mehlert et al.\ 2003; Thomas
et al.\ 2004).  Although classically the Mg$_2$--$\sigma$ relation has been
interpreted as the interplay between mass and metallicity (Forbes et al.\ 1998;
Kobayashi \& Arimoto 1999; Terlevich et al.\ 1999), there is still nowadays
much debate as to whether this relation reflects trends in stellar ages,
metallicities or in the relative abundance of different heavy elements (e.g.\
T98; J\o rgensen 1999; Trager et al.\ 2000b; K01; 
Poggianti et al.\ 2001a; Mehlert et al.\ 2003; Caldwell et al.\ 2003; Thomas et al.\ 2005). The
problem is due to the well known age-metallicity degeneracy, which makes very
difficult to separate both effects with the current stellar population models.
However, the use of a larger number of indices can help us to disentangle this
degeneracy thanks to the different sensitivity of each index to variations of
these parameters.

So far, most studies have concentrated on the relation of the Mg indices with
the velocity dispersion. In this section we are showing the relation of 18 Lick
indices with $\sigma$.  Some of the relations have already been presented by
S\'anchez-Bl\'azquez et al. (2003), but they are shown here again for
completeness. Most of them, however, are new.  We aim to answer two questions:
\begin{enumerate}
 \item Which parameter(s) is(are) responsible for the correlation between the
 indices and velocity dispersion?
 \item Which parameter(s) is(are) responsible for the intrinsic scatter in
 these relations?
 \end{enumerate}

Following Colles et al.\ (1999), we will express the atomic indices in 
magnitudes. These indices will be denoted by the name followed by a prime sign, 
and are obtained as
\begin{equation}
 I'=-2.5\log \left( 1-\frac{I}{\Delta\lambda} \right),
\label{magnitudes}
\end{equation} 
where $I$ is the classic index measured in \AA\ (atomic index), and
$\Delta\lambda$ is the width of the index central bandpass.  The D4000 was
transformed  to magnitudes as
\begin{equation}
{\rm D}4000'=-2.5\log ({\rm D4}000).
\end{equation}

\begin{figure*}
 \resizebox{0.31\hsize}{!}{\includegraphics[angle=-90]{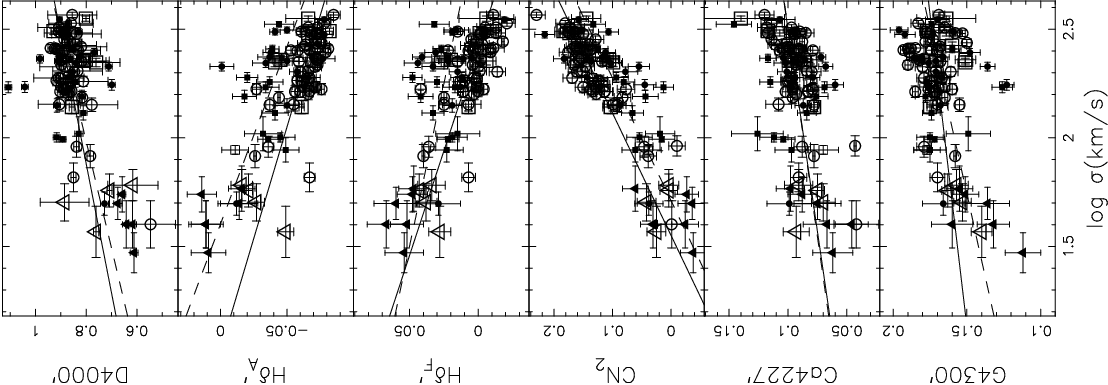}}
 \hfill
 \resizebox{0.31\hsize}{!}{\includegraphics[angle=-90]{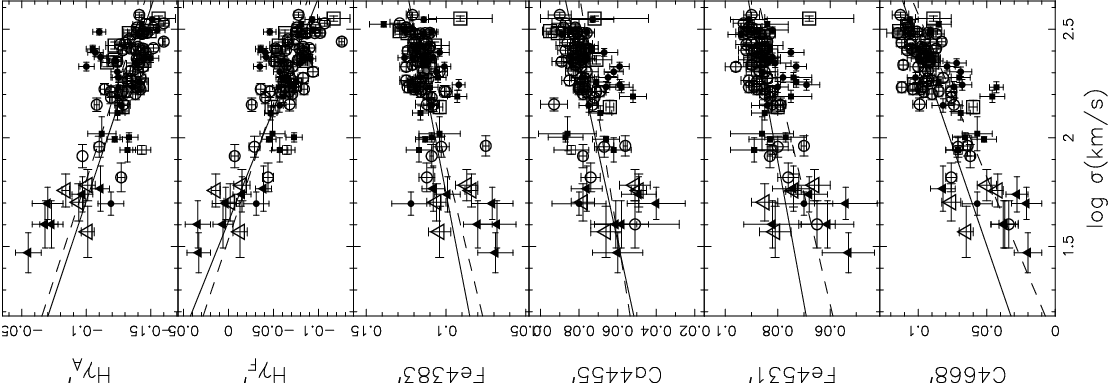}}
 \hfill
 \resizebox{0.31\hsize}{!}{\includegraphics[angle=-90]{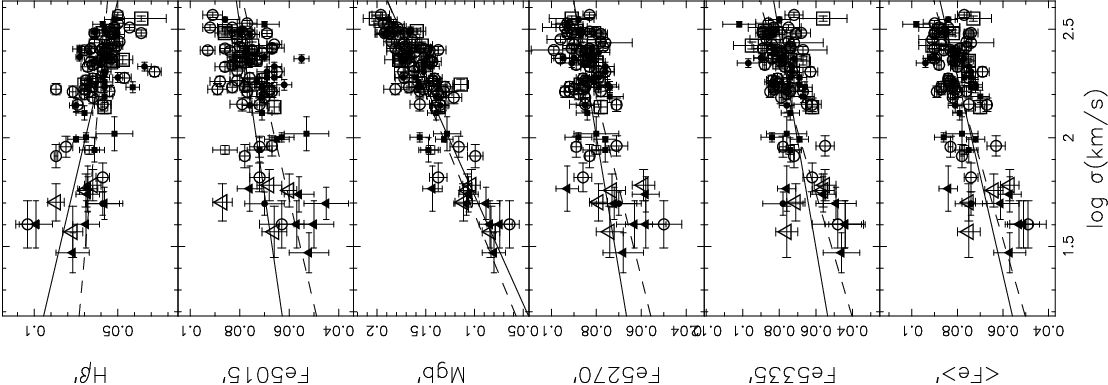}}
\caption{Line-strength indices plotted against $\log\sigma$. Open symbols show
the galaxies in low-density environments and filled symbols galaxies in the
Coma cluster. The different shapes represent distinct morphological types
(triangles: dwarfs; squares: S0; circles: ellipticals). All indices are
measured in magnitudes following the conversion of Eq. (\ref{magnitudes}).
Solid and dashed lines represent error-weighted least-squares linear fits to
LDEGs and HDEGs respectively.\label{figure.sigma}} 
\end{figure*}

Figure. \ref{figure.sigma} shows the index--$\sigma$ relations for all the
indices measured in this work.  The fits were performed using  an ordinary
least squares method, minimising the residuals in both x- and y-direction.
The method initially performs a typical unweighted ordinary least-squares
regression of Y on X  and the 
coefficients from the first fit are then employed to derive (numerically, with a
downhill method) the straight line data fit with errors in both coordinates.
 The
best linear relation \mbox{($I'=a+b\;\log \sigma$)} and the scatter are
summarised in Table \ref{tabla-relaciones} for the two different subsamples
considered in this paper, HDEGs and LDEGs. From the LDEGs sample, two outliers
were eliminated, NGC 4742 and IC 767, and from the HDEGs sample, NGC 6166 was 
also eliminated.  The probability that the parameters ($I'$ and $\sigma$)
are not correlated ($\alpha$) was derived from a (non--parametric) Spearman
rank--order test and is also shown in  Table \ref{tabla-relaciones}. However,
the Spearman test does not take into account the errors of the
individual measurements. For this reason, we also carried out a $t$-test to
check the hypothesis $b=0$, where $b$ represent the slope of the fit. Values of
$t$ larger than 1.96 indicate that the hypothesis can be rejected with a
significance level lower than $\alpha=0.05$. To obtain the errors in the
slope, 1000 Monte Carlo simulations were performed, perturbing the points
randomly in both x- and y-direction with Gaussian distributions of width
given by their errors.
 
\begin{table*}
\caption{Parameters of the linear fits $I'=a+b\;\log\sigma$. First column:
y-intersect of the fit and its error; second column: slope, and corresponding
error, of the linear fit; $\alpha$: probability, given by a non-parametric
Spearman test, that the slope of the fit is significatively different from zero
by chance; $N$: number of galaxies used to make the fit; $t$: $t$ parameter to
test the hypothesis $b=0$ (a $t$ value higher than 1.9 indicates that the slope
is significatively different from zero with a significance level lower than
0.05); $\sigma_{\rm std}$: standard deviation about the fit; $\sigma_{\exp}$:
standard deviation expected from the errors; $\sigma_{\rm res}$: standard
deviation not explained from the errors.\label{tabla-relaciones}}
\centering
\begin{tabular}{@{}lrrrrrrrr@{}}
\hline\hline
                \multicolumn{9}{c}{LDEGs}\\
\hline
  Index        & \multicolumn{1}{c}{$a\pm\sigma(a)$}&\multicolumn{1}{c}{$b\pm\sigma(b)$}&\multicolumn{1}{c}{$\alpha$}&
 \multicolumn{1}{c}{$N$}
  &\multicolumn{1}{c}{$t$}&\multicolumn{1}{c}{$\sigma_{\rm std}$}&
 \multicolumn{1}{c}{$\sigma_{\rm exp}$}&\multicolumn{1}{c}{$\sigma_{\rm res}$}\\
\hline
D4000$'$       &$ 0.484\pm$0.016  &$ 0.1679\pm$0.007    &0.005  & 59& 23.98 & 0.037 &0.012& 0.035\\
H$\delta'_A$   &$ 0.050\pm$0.023  &$-0.049\pm$0.004    &9.22E-07& 59&  4.83 & 0.032 &0.002& 0.032\\
H$\delta'_F$   &$ 0.126\pm$0.016  &$-0.052\pm$0.007    &1.72E-09& 59&  7.66 & 0.038 &0.002& 0.038\\
CN$_2$         &$-0.276\pm$0.035  &$ 0.178\pm$0.007    &3.08E-17& 59& 12.96 & 0.023 &0.006& 0.023\\
Ca4227$'$      &$ 0.033\pm$0.011  &$ 0.027\pm$0.005    &5.93E-06& 59&  2.20 & 0.008 &0.004& 0.007\\ 
G4300$'$       &$ 0.130\pm$0.008  &$ 0.017\pm$0.004    &0.054   & 59&  1.96 & 0.017 &0.003& 0.017\\
H$\gamma'_A$   &$-0.004\pm$0.010  &$-0.056\pm$0.004    &4.16E-09& 59&  6.52 & 0.026 &0.002& 0.026\\
H$\gamma'_F$   &$ 0.157\pm$0.013  &$-0.097\pm$0.006    &9.65E-11& 59&  9.85 & 0.025 &0.004& 0.025\\
Fe4383$'$      &$ 0.054\pm$0.007  &$ 0.026\pm$0.003    &2.97E-08& 59&  3.20 & 0.009 &0.002& 0.009\\  
Ca4455$'$      &$ 0.023\pm$0.008  &$ 0.024\pm$0.003    &3.61E-07& 59&  2.51 & 0.004 &0.003& 0.002\\
Fe4531$'$      &$ 0.051\pm$0.007  &$ 0.015\pm$0.003    &2.53E-07& 59&  3.18 & 0.003 &0.002& 0.003\\   
C4668$'$       &$-0.032\pm$0.001  &$ 0.054\pm$0.004    &2.23E-09& 59&  7.34 & 0.009 &0.003& 0.009\\
H$\beta$'      &$ 0.130\pm$0.014  &$-0.030\pm$0.006    &1.37E-09& 59&  5.25 & 0.012 &0.002& 0.012\\
Fe5015$'$      &$ 0.048\pm$0.005  &$ 0.013\pm$0.002    &1.0E-04 & 59&  2.30 & 0.004 &0.002& 0.004\\
Mgb$'$         &$-0.073\pm$0.012  &$ 0.100\pm$0.005    &1.22E-16& 59& 10.78 & 0.011 &0.002& 0.011\\
Fe5270$'$      &$ 0.047\pm$0.010  &$ 0.016\pm$0.004    &0.001  & 59&  3.60 & 0.004 &0.002& 0.004\\
Fe5335$'$      &$ 0.029\pm$0.013  &$ 0.020\pm$0.006    &0.030  & 59&  3.52 & 0.003 &0.003& 0.000\\
               &                    &                      &        &   &       &        &      & \\
\hline
\end{tabular}
\end{table*}
\addtocounter{table}{-1}
\begin{table*}
\begin{tabular}{@{}lrrrrrrrr@{}}
\hline\hline
               \multicolumn{9}{c}{HDEGs}\\
\hline
  Index        & \multicolumn{1}{c}{$a\pm\sigma(a)$}&\multicolumn{1}{c}{$b\pm\sigma(b)$}&\multicolumn{1}{c}{$\alpha$}&
 \multicolumn{1}{c}{$N$}
  &\multicolumn{1}{c}{$t$}&\multicolumn{1}{c}{$\sigma_{\rm std}$}&
 \multicolumn{1}{c}{$\sigma_{\rm exp}$}&\multicolumn{1}{c}{$\sigma_{\rm res}$}\\
\hline
 D4000$'$      &$ 0.397\pm$0.019&$ 0.203\pm$0.008 & 0.071     & 36 &  23.87  &  0.098 & 0.014& 0.097\\
H$\delta'_{A}$ &$ 0.098\pm$0.018&$-0.061\pm$0.009 & 1.00E-04  & 36 &   7.03  &  0.018 & 0.008& 0.017\\
H$\delta'_{F}$ &$ 0.100\pm$0.0205&$-0.033\pm$0.009& 0.010     & 36 &   3.59  &  0.015 & 0.008& 0.013\\
CN$_{2}$       &$-0.303\pm$0.043&$ 0.177\pm$0.020 & 1.44E-10  & 36 &   8.81  &  0.034 & 0.013& 0.031\\
Ca4227$'$      &$ 0.035\pm$0.023&$ 0.026\pm$0.010 & 0.125     & 36 &   2.51  &  0.011 & 0.008& 0.007\\
G4300$'$       &$ 0.093\pm$0.021&$ 0.033\pm$0.010 & 0.002     & 36 &   3.41  &  0.016 & 0.006& 0.014\\
H$\gamma'_{A}$ &$-0.001\pm$0.026&$-0.055\pm$0.012 & 0.020     & 36 &   4.70  &  0.016 & 0.006& 0.015\\
H$\gamma'_{F}$ &$ 0.122\pm$0.030&$-0.080\pm$0.013 & 3.00E-04  & 36 &   5.95  &  0.017 & 0.070& 0.016\\
Fe4383$'$      &$ 0.038\pm$0.023&$ 0.033\pm$0.010 & 0.006     & 36 &   3.29  &  0.007 & 0.005& 0.005\\
Ca4455$'$      &$ 0.032\pm$0.014&$ 0.018\pm$0.006 & 0.011     & 36 &   2.87  &  0.006 & 0.005& 0.002\\
Fe4531$'$      &$ 0.036\pm$0.0120&$ 0.020\pm$0.005 & 5.0E-04  & 36 &   3.70  &  0.006 & 0.005& 0.0027\\
Fe4668$'$      &$-0.071\pm$0.018&$ 0.067\pm$0.008 & 2.19E-09  & 36 &   8.25  &  0.012 & 0.006& 0.010\\
H$\beta'$      &$ 0.087\pm$0.009&$-0.012\pm$0.004 & 0.189     & 36 &   2.85  &  0.009 & 0.004& 0.008\\
Fe5015$'$      &$ 0.025\pm$0.011&$ 0.021\pm$0.005 & 1.46E-05  & 36 &   4.38  &  0.004 & 0.003& 0.002\\
Mgb$'$         &$-0.050\pm$0.019&$ 0.091\pm$0.008 & 3.15E-10  & 36 &  10.96  &  0.009 & 0.004& 0.008\\
Fe5270$'$      &$ 0.027\pm$0.007&$ 0.024\pm$0.003 & 0.001     & 36 &   7.27  &  0.004 & 0.004& 0.001\\
Fe5335$'$      &$-0.001\pm$0.013&$ 0.035\pm$0.006 & 1.79E-05  & 36 &   6.02  &  0.004 & 0.004& 0.000\\
 \hline
\end{tabular}
\end{table*}

%------------------------------------------------------------------------------
\subsection{The slope of the relations}
\label{sec.slopes}

\begin{figure}
\resizebox{\hsize}{!}{\includegraphics[angle=-90]{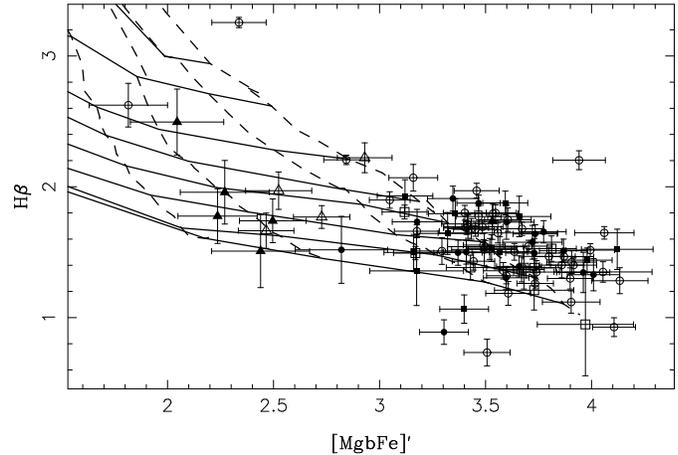}}
\caption{[MgFe]$'$ {\it versus} H$\beta$ for the LDEG (open symbols) and HDEG (filled symbols)
Model grids from V06 are superposed: solid lines are contours of constant
age ({\it top to bottom}, 1.41, 2.00, 2.82, 3.98, 5.62, 7.98, 11.22, 15.85 Gyr), 
and dotted lines are contours of constant [M/H] ({\it left to right}, [M/H]= $-0.68$,
$-0.38$,$0.0$,$+0.2$). 
\label{index-index}}
\end{figure}

\begin{figure}
\resizebox{\hsize}{!}{\includegraphics[angle=-90]{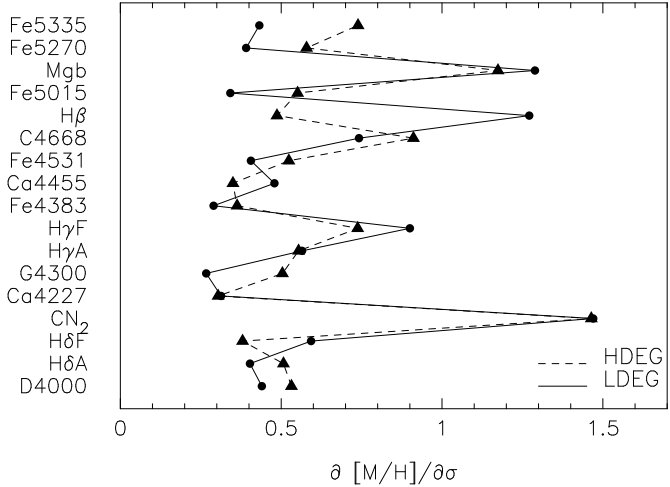}}
\caption{Variation in metallicity required to explain the slope in the
index--$\sigma$ relations for 17 indices, assuming that the metallicity is the
only parameter changing with $\sigma$. The solid line represents the change in
metallicity for the LDEGs, while the dashed line refers to the galaxies in the
Coma cluster.\label{derivatives.label}}
\end{figure}

To study which parameters drive the relation of the indices with $\sigma$, we
have parameterised the models of V06 as a function of age and metallicity. The
relations are shown in Table \ref{parametrization}. In estimating these
expressions we have restricted the age to the interval 4.7--17.78 Gyr, and the
metallicity range to $-0.68<[$M/H$]<+0.2$, which is the range covered
by the galaxies in our sample. To show that,  we 
have plotted, in Figure \ref{index-index} an index-index diagram
comparing the [MgFe]$'$ index (Thomas, Maraston \& Bender 2003, TMB03 hereafter) versus H$\beta$ diagram, including
the LDEGs (open symbols) and HDEG (filled symbols). We have also over-plotted
the grid of models by V06. The caption of the figure indicate
the corresponding values of age and metallicity for each model. 
The two last columns of Table \ref{parametrization} give, 
for the LDEGs and HDEGs respectively, the required
dependence of metallicity on ($\log\sigma$) if the observed slope of the
particular index in the index-$\log\sigma$ diagram were to be driven 
entirely by a metallicity dependence on $\log\sigma$.
 This is also illustrated in 
Figure \ref{derivatives.label} for both subsamples, LDEGs and HDEGs. 
As can be seen
in Table \ref{parametrization}, the metallicity 
dependence required to explain the slope of
the  CN$_2$, C4668 and Mgb indices with $\sigma$ is much larger than 
that required to explain
the slope of the other indices.
Therefore, the metallicity can not be the only parameter varying with the 
velocity dispersion. 
It does not seem probably, neither, that a unique combination of age and
metallicity is responsible for the differences in the slope between indices,
since CN$_2$, C4668 and Mgb are not especially sensitive to age variations
(e.g.\ Worthey 1994). A dependence of age on velocity dispersion in the
galaxies would produce steeper slopes in other indices, such as G4300
or Fe4531.
 
The most plausible explanation is the existence of systematic changes 
in the chemical
abundance ratios along the $\sigma$ sequence. That is, not all the elements are
changing in lockstep with velocity dispersion. As different indices have different
sensitivities to changes in the chemical composition, a variation of the later 
with $\sigma$ would 
produce differences in the slope of the relation. Although we do not have the tools to
derive detailed chemical abundances in early-type galaxies, we know that
the index CN$_2$ is especially sensitive to variations of C and N, C4668 to variations 
in the C abundance, while the Mgb index increases with the Mg abundance, and have an 
inverse dependence with the C abundance 
(Tripicco \& Bell 1995, TB95 hereafter; Korn, Maraston \& Thomas 2005). 
Therefore, the explanation of the differences in the slopes of these indices
with respect to the others would require, {\it in principle},  an increase of the
[Mg/Fe]  (already noted by a number of authors, e.g. Worthey et al. 1992; 
Greggio 1997, J\o rgensen 1999,
Kuntschner 2000; Trager et al. 2000b; Thomas et al. 2002; M03;  Thomas et al. 2005), 
[C/Fe] and [N/Fe]  (see  Worthey 1998) ratios with $\sigma$.

However, there are other indices which also depend strongly on these elements
and, thus, before reaching the above conclusion we should study if the
variations in the relative abundance of these elements with $\sigma$ are
compatible with the slopes obtained for the rest of the indices.
%%%%%%%%%%%%%%%%%%%%%%%%%%%%%%%%%%%%%%%%%%%%%%%%%%%%%%%%%%%%%%%%%%%%%%%%%

Using a similar procedure than T00a, we have used the response functions of
TB95 to calculate the variation of the Lick indices to changes in the
abundances of different elements.
 
Instead of changing all the elements separately we assume that some elements
are linked nucleosynthetically, so we  vary them in lockstep. Following T00a we
separate the elements into two different groups: (G1) Na, Si, Ca, O, and
Ti\footnote{Although the nucleosynthesis theory of Ti is not well understood,
this element appears to be enhanced in most bulge stars by $\sim 0.3$ dex
(e.g.\ see McWilliam \& Rich 2004).}; (G2) Fe and Cr. In addition, we allow
the abundance of C, N and Mg to vary independently. TB95 response functions are
for enhancement values corresponding to $[$X/H$]=+0.3$ dex. The fractional
response of an index $I$ to other arbitrary values is (see T00a)
\begin{equation} 
\frac{\Delta I}{I_{0}}= 
  \left\{ 
    \prod_{i} \left[ 1+R_{0.3}(X_{i})\right]^{([X_{i}/H]/0.3)}
  \right\} -1,
\label{response}
\end{equation}   
where $R_{0.3}(X_{i})$ is the TB95 response function for the $i^{\rm
th}$ element at $[$X$_{i}$/H$]=+0.3$ dex.
  
To compute the final index variations we have assumed the following
composition: 53\% of cool giants, 44\% of turn off stars, and a 3\% of cool
dwarfs stars, as in T00a.

We calculated the changes in the indices that are induced by changing the
abundance ratio patterns according to the following four different
prescriptions:\\

\begin{tabular}{@{}l@{}c@{$\;$}l@{}}
 \raisebox{0ex}[3ex]{\mbox{}}
 Model (i)  &:& \parbox[t]{0.75\columnwidth}{G1 [X/H]=+0.5 (including C, N and 
 Mg) and G2 [X/H]=+0.3}\\
 \raisebox{0ex}[3ex]{\mbox{}}
 Model (ii) &:& \parbox[t]{0.75\columnwidth}{G1 [X/H]=+0.5 (including C, N and 
 Mg) and G2 [X/H]=+0.0}\\
 \raisebox{0ex}[3ex]{\mbox{}}
 Model (iii)&:& \parbox[t]{0.75\columnwidth}{G1 [X/H]=+0.5 (including N and Mg)
 and G2 [X/H]=+0.3 (including C)}\\
 \raisebox{0ex}[3ex]{\mbox{}}
 Model (iv) &:& \parbox[t]{0.75\columnwidth}{G1 [X/H]=+0.5, G2 [X/H]=+0.3, 
 [C/H]=+0.43, [N/H]=+0.63, and [Mg/H]=+1.2}\\
\end{tabular}

\begin{figure}
\resizebox{0.35\hsize}{!}{%
\includegraphics[bb=94 149 529 641,angle=-90]{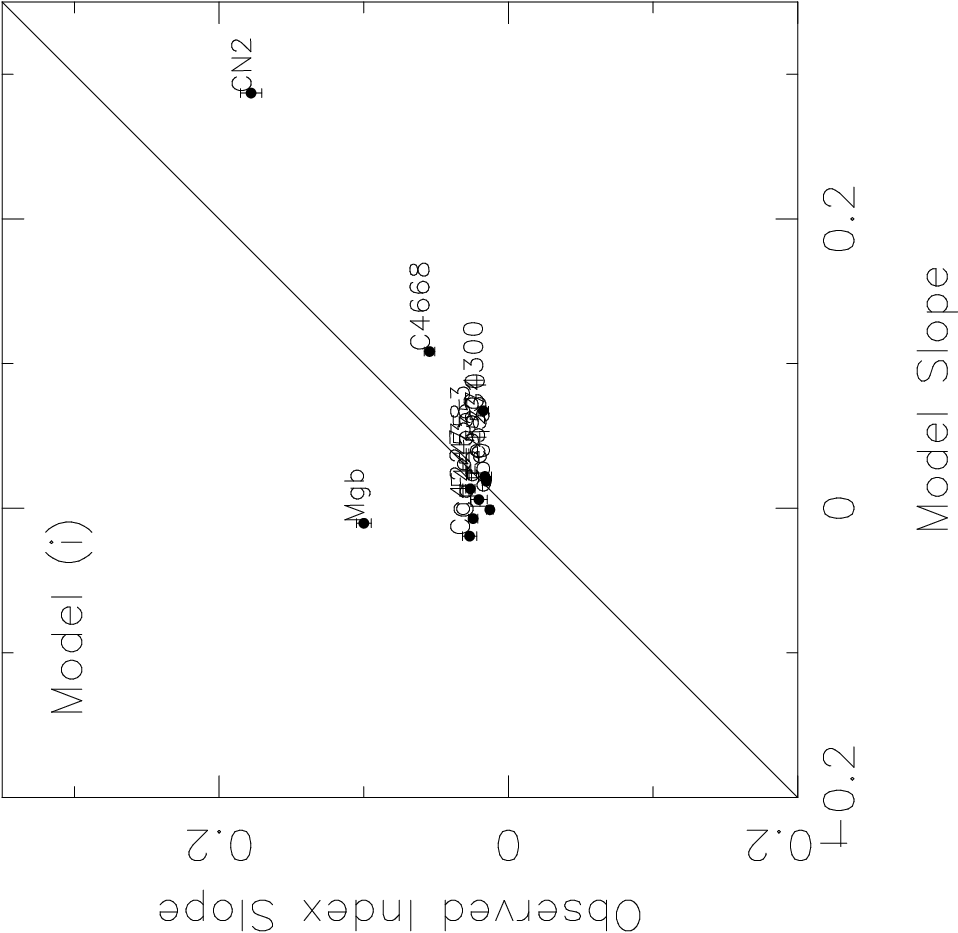}}
\hfill
\resizebox{0.35\hsize}{!}{%
\includegraphics[bb=94 149 529 641,angle=-90]{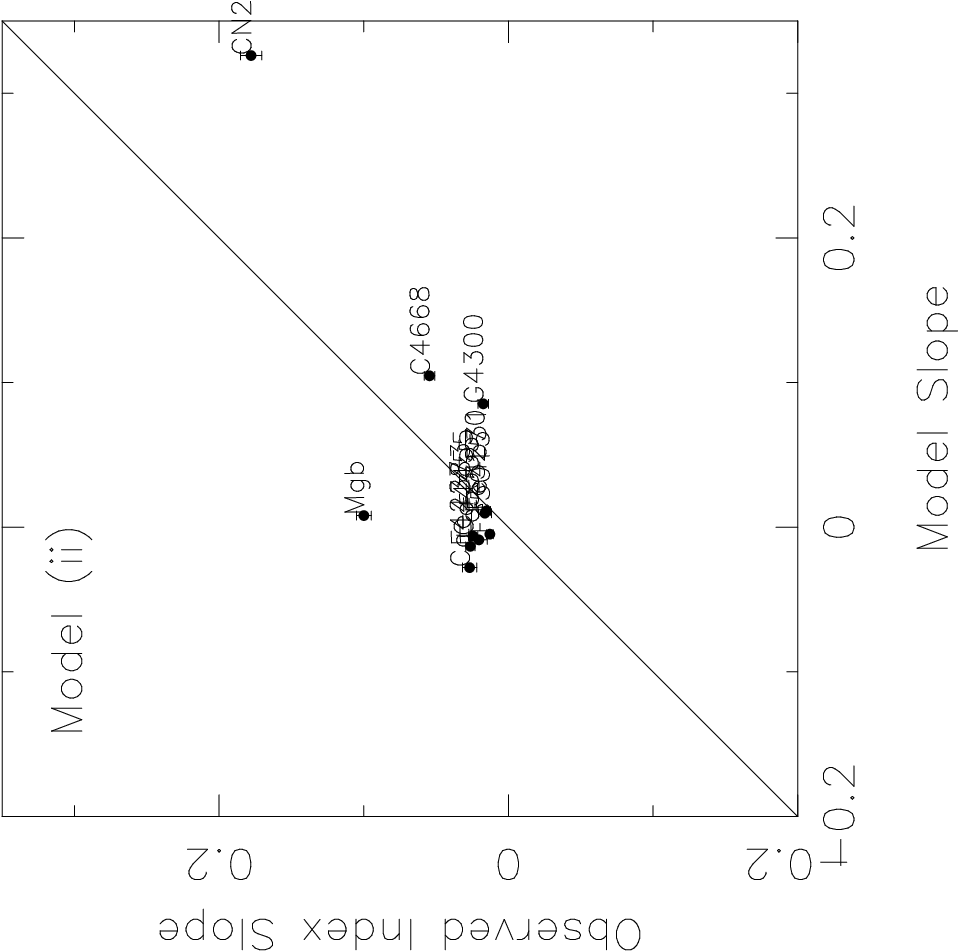}}

\vspace{3mm}

\resizebox{0.35\hsize}{!}{%
\includegraphics[bb=94 149 529 641,angle=-90]{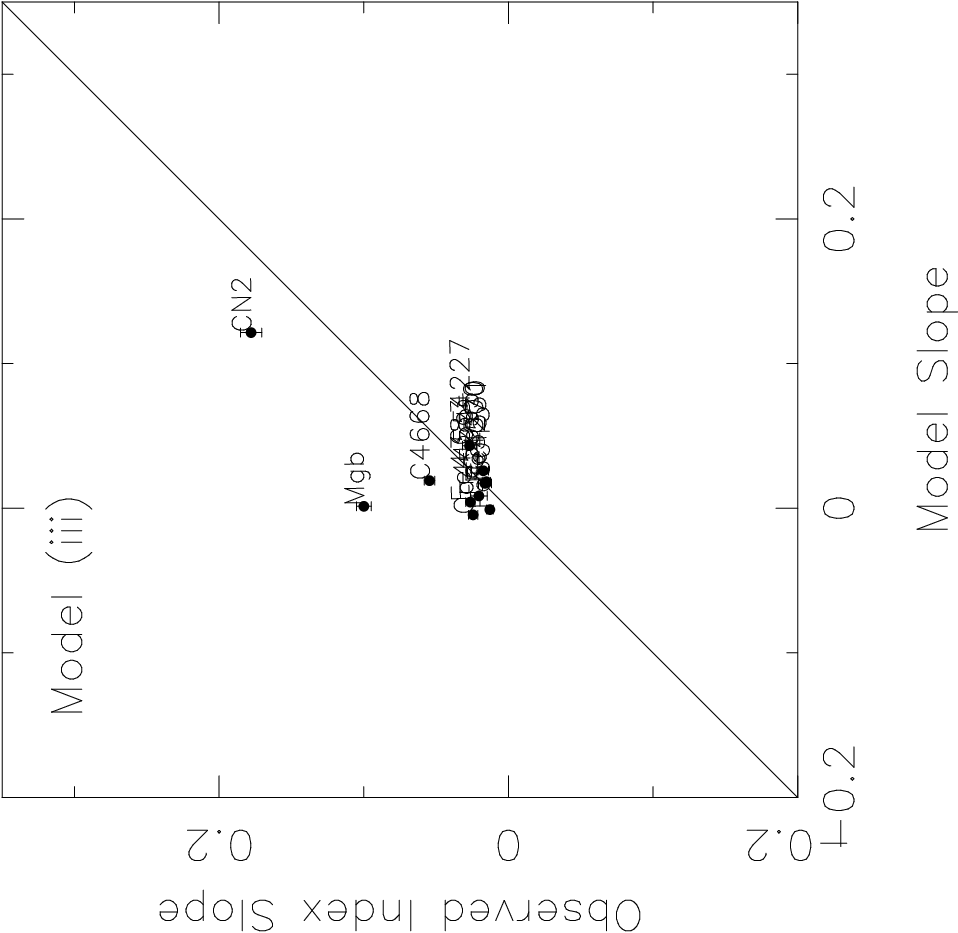}}
\hfill
\resizebox{0.35\hsize}{!}{%
\includegraphics[bb=94 149 529 641,angle=-90]{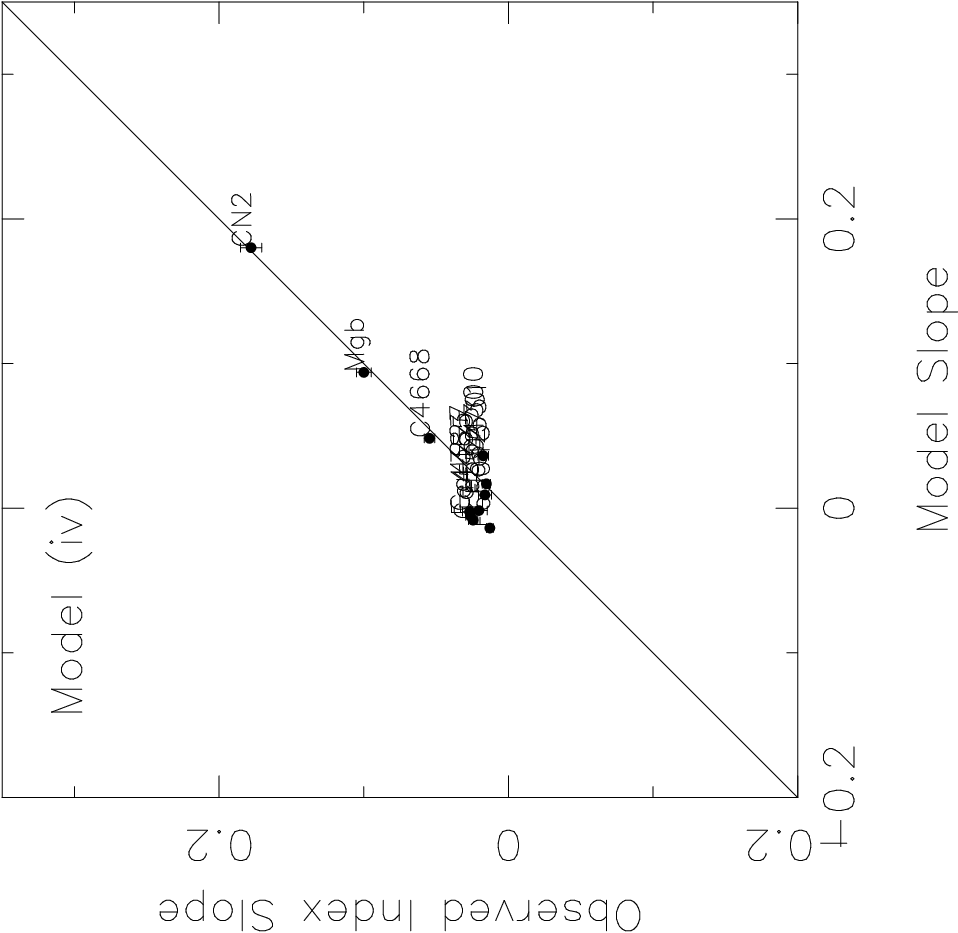}}
\caption{Comparison of the observed slopes in the index-$\sigma$ relations for the LDEGs
(y-axis) with the expected slopes caused by changes in abundance ratio
patterns given in the four models discussed in the text
 (x-axis). The errors bars show the 
error in the slope of the relations.  (see text for details).\label{fig-tripicco}}
\end{figure}

\begin{table*}
\caption{Line-strength indices used in this work as a function of the logarithm
of age and metallicity using the models by V06. The two last columns express
the required variation in metallicity with velocity dispersion to explain the
slopes of the index--$\sigma$ relations, assuming that the slope is due to an
exclusive variation of this parameter with the velocity
dispersion.\label{parametrization}}
\centering
\begin{tabular} {@{}l r c c@{}}
\hline\hline
                & &\multicolumn{1}{c}{(LDEG)}& \multicolumn{1}{c@{}}{(HDEG)}\\
  Index         & &$\displaystyle\frac{\partial [M/H]}{\partial\log\sigma}$&$\displaystyle\frac{\partial [M/H]}{\partial\log\sigma}$\\
\hline
  D4000$'$       & $-2.7672 + 0.3817\;$[M/H]$+0.3724\log$(age)&   0.440  &  0.532 \\
  H$\delta_A'$   & $+1.0737 - 0.1206\;$[M/H]$-0.1155\log$(age)&   0.403  &  0.507\\
  H$\delta_F'$   & $+1.0202 - 0.0878\;$[M/H]$-0.1029\log$(age)&   0.593  &  0.380\\
  CN$_2$         & $-0.5655 + 0.1210\;$[M/H]$+0.0640\log$(age)&   1.470  &  1.464\\
  Ca4227$'$      & $-0.6957 + 0.0857\;$[M/H]$+0.0845\log$(age)&   0.313  &  0.304\\
  G4300$'$       & $-0.4797 + 0.0651\;$[M/H]$+0.0669\log$(age)&   0.267  &  0.504\\
  H$\gamma_A'$   & $+0.9424 - 0.0993\;$[M/H]$-0.1092\log$(age)&   0.565  &  0.554\\
  H$\gamma_F'$   & $+1.3029 - 0.1080\;$[M/H]$-0.1388\log$(age)&   0.900  &  0.738 \\
  Fe4383$'$      & $-0.4544 + 0.0907\;$[M/H]$+0.0583\log$(age)&   0.290  &  0.363\\
  Ca4455$'$      & $-0.1952 + 0.0509\;$[M/H]$+0.0282\log$(age)&   0.479  &  0.350\\
  Fe4531$'$      & $-0.1611 + 0.0374\;$[M/H]$+0.0252\log$(age)&   0.406  &  0.524\\
  C4668$'$       & $-0.1022 + 0.0734\;$[M/H]$+0.0171\log$(age)&   0.742  &  0.911\\
  H$\beta'$      & $+0.5974 - 0.0240\;$[M/H]$-0.0538\log$(age)&   1.271  &  0.487\\
  Fe5015$'$      & $-0.0133 + 0.0374\;$[M/H]$+0.0097\log$(age)&   0.342  &  0.551\\
  Mgb$'$         & $-0.4036 + 0.0775\;$[M/H]$+0.0545\log$(age)&   1.289  &  1.174\\
  Fe5270$'$      & $-0.1415 + 0.0414\;$[M/H]$+0.0229\log$(age)&   0.391  &  0.579\\
  Fe5335$'$      & $-0.1561 + 0.0472\;$[M/H]$+0.0236\log$(age)&   0.432  &  0.739\\
  \hline
\end{tabular}
\end{table*}

\begin{table*}
\caption{Difference between the slope of the index--$\sigma$ relation and the
expected index variation due to changes in the chemical composition.
\label{table-tripicco}}
\begin{tabular}{@{}l rrrrrrrrrr@{}}
\hline\hline
Model & \multicolumn{1}{c}{$\Delta$CN$_2$}&\multicolumn{1}{c}{$\Delta$ Ca4227}&
\multicolumn{1}{c}{$\Delta$G4300}&\multicolumn{1}{c}{$\Delta$Fe4383}&
\multicolumn{1}{c}{$\Delta$Ca4455}&
\multicolumn{1}{c}{$\Delta$Fe4531}&
\multicolumn{1}{c}{$\Delta$C4668}&
\multicolumn{1}{c}{$\Delta$Fe5015}&
\multicolumn{1}{c@{}}{$\Delta$Mgb}&r.m.s. \\ 
\hline
 i    & $-0.109$ & 0.046  & $-0.050$ & 0.013 &0.032 & $-0.003$&$-0.054$ &  0.014 &0.110&0.040 \\
 ii   & $-0.148$ & 0.055  & $-0.068$ & 0.039 &0.030 & 0.004   &$-0.050$ &$-0.050$&0.092&0.045 \\
 iii  & $ 0.056$ &$-0.016$& $-0.008$ & 0.022 &0.029 &$-0.003$ & 0.035   & 0.014 & 0.098&0.029 \\ 
 iv   & $-0.002$ & 0.029  & $-0.019$ & 0.031 &0.033 &$-0.002$ & 0.006   & 0.026 & 0.006&0.023 \\
\hline
\end{tabular}
\end{table*}
  The first three models are simple variations of the different proposed groups, and are
 shown here to show the influence  that  the variation of  different  groups of elements  
 have  over the slopes of the indices with $\sigma$. In the forth 
 model, on the other hand,  we have fixed the values for G1 and G2 as in  model (iii),
 which is the one with lower rms of the first three models, and we have 
 fitted the values of C, N and Mg to  reproduce the observed slopes in 
 CN$_2$, C4668 and Mgb.

Figure \ref{fig-tripicco} shows the comparison of the slopes with the variation
of the indices in the different proposed models. The r.m.s.\ of the dispersion
and the mean differences in the indices modeled by TB95 are summarised in Table
\ref{table-tripicco}. As it is apparent, the model that best reproduces the
observed slopes is model (iv), in which the Mg and the N change more than the
other $\alpha$-elements, while the C changes more than the Fe-peak elements but
less than the $\alpha$-elements. However, this variation of C gives slopes in
the G4300 index (also very C sensitive) steeper than observed. The observations
of this index do not fit well with the stellar atmosphere model of the coolest
giant star of TB95.
Therefore the differences may be also due to a problem in the modeling.  On the
other hand, there is a possibility that the C4668 index is not as sensitive to
variations in the C abundance as predicted by TB95 (Worthey 2004).  Therefore,
we cannot make firm conclusions about the variation of the [C/Fe] abundance
with the velocity dispersion of the galaxies. 
 The reported values for the different models are only illustrative. We do not
pretend to quantify the real variations in the chemical abundances ratios with 
the velocity dispersion, but only to obtain relative variation of some elements 
with respect to the others.
To explain the differences in the slope of the relations of Mgb and
$\langle$Fe$\rangle$ with $\sigma$, some authors have already proposed the
existence of an increase in the Mg/Fe ratio with the velocity dispersion
(K01; Proctor \& Sansom 2002; Mehlert et al.\ 2003; but,
for an opposing alternative, see Proctor et al.\ 2004). In this work we propose
that, apart from the variation in this ratio, there is also a variation in the
[N/Fe] ratio with the velocity dispersion. On the other hand, although all the
$\alpha$ elements change along the $\sigma$ sequence more than the Fe-peak elements,
the variation of the ratio [Mg/Fe] is larger than [X/Fe] for the rest of
$\alpha$ elements. 

Although this experiment has been done using the calculated  slopes for the
LDEG sample, the conclusions are the same for the HDEGs, as the slopes of the
metal-sensitive indices in both subsamples are very similar (see Figure
\ref{derivatives.label}).
This is not true, however, for the slopes of the age-sensitive indices.
%%%%%%%%%%%%%%%%%%%%%%%%%%%%%%%%%%%%%%%%%%%%%%%%%%%%%%%%%%%%%%%%%%%%%%%%%
In particular, the relation of H$\beta$ with $\sigma$ is much steeper for the first
subsample of galaxies (see Table \ref{tabla-relaciones}, and note the striking
difference for H$\beta$ in Figure \ref{derivatives.label}).  For  LDEGs these
indices are strongly correlated with $\sigma$ while for HDEGs the statistical
significance of the correlation is lower.  There exists in the literature
different points of views about the relation of the H$\beta$ index with
velocity dispersion.  While some authors report a strong correlation between
these two parameters (e.g.\ J\o rgensen 1997; T98; Caldwell et al. 2003), others find
a weak or null correlation (e.g.\ Mehlert et al.\ 2003). These discrepancies
could be due to differences in the relation as a function of the environment.
In fact, the sample of T98 is composed mainly of field and
low density environment galaxies, while the sample of  Mehlert et al.\ (2003)
consists of  galaxies in the Coma cluster. J\o rgensen's (1997) sample is more
heterogeneous, containing galaxies belonging to 11 different clusters of
variable density and field galaxies, while the Caldwell et al. (2003) 
sample consists of galaxies in the field and in the Virgo cluster, therefore, very 
similar to our LDEG sample.  In general, it seems that all the studies
analysing galaxies in low density environments find a correlation between
H$\beta$ and $\sigma$, while those analysing exclusively galaxies in the Coma
cluster do not. However, Kuntschner (1998) finds a significant correlation
between H$\beta$ and $\sigma$ in his sample of nearly coeval Fornax galaxies,
although he claims that the slope in this relation is driven mainly by
variations in metallicity, as the galaxies in the Fornax cluster span a very
broad metallicity range.  To study the possibility of an age variation along
the $\sigma$ sequence, we have derived the change in age that would explain the
slope of the H$\beta$--$\sigma$ relation, assuming that the age is the only
parameter varying with $\sigma$.  For the LDEGs this value is $\partial
\log(\rm age)/\partial \log\sigma =0.8$, while for HDEGs is only $\partial
\log(\rm age)/\partial \log\sigma =0.2$. In fact, the metallicity
variation that explains the slope in the Fe4383--$\sigma$ relation ($\partial
[M/H]/\partial\log\sigma \sim 0.36$) can account for the slope of the
H$\beta$--$\sigma$ relation obtained for HDEGs, without any further variation in
age. For the LDEGs it is however  required an additional age variation of
$\partial \log(\rm age)/\partial \log\sigma =0.4$. 

Summarising, we conclude that there exists an increase of the overall
metallicity with the velocity dispersion for both, HDEGs and LDEGs. However, not
all the elements change in lockstep along the $\sigma$ sequence. The models
that best reproduce the slopes in all the indices are those in which the
$\alpha$ elements change more than the Fe-peak elements and, furthermore, the
$[$Mg/Fe$]$ and $[$N/Fe$]$ ratios change more than the rest of the $\alpha$
elements with the velocity dispersion of the galaxies.  In the case of LDEGs,
the slopes of the index--$\sigma$ relations also reflect a trend in the mean age
of the galaxies in the sense that less massive galaxies are also younger.  We
will analyse this relation  in more detail in Paper II, where we derive ages
and metallicities for our sample of galaxies.  None of these conclusions
would change if we increased the Mgb and C4668 indices by a constant value, as
all the results are based in the slope of the index-$\sigma$ relations.

%------------------------------------------------------------------------------
\subsection{The dispersion of the relations}
\label{sec.dispersion}

The  low dispersion  in the Mg$_2$--$\sigma$ relation and the fact that it is
distance independent make it a powerful tool to constrain the models of
formation and evolution of galaxies.  In fact, the low dispersion  has been
used as an argument to evidence that all ellipticals have nearly coeval stellar
populations (Bender, Burstein \& Faber 1993; Bernardi et al.\ 1998) in clear
contradiction with other studies (G93; T98; Trager et al.\ 2001; Caldwell et al.\ 2003).  Some authors
have studied this problem, concluding that the low scatter is due  to a
conspiracy between the age and the metallicity, in the sense that younger
galaxies are also more metal rich, canceling the deviations with respect to the
mean relation (e.g.\ Worthey, Trager \& Faber 1995; Trager 1997; Pedraz et al.\
1998; J\o rgensen 1999; K01). 

However, though small, an intrinsic dispersion does exists in the
Mg$_2$--$\sigma$ relation. To understand  the cause of this scatter, many
authors have tried to find correlations between the deviation from the relation
and other parameters.  Bender, Burstein \& Faber (1993) did not find any
correlation between the residuals of this relation and other structural
parameters of the galaxies such as effective radius, surface brightness or
mass. Neither did they find any correlation between the residuals and the
position of the objects in the Fundamental Plane.  On the other hand, Schweizer
et al.\ (1990) found a correlation between the deviation of the
Mg$_2$--$\sigma$ relation and the fine structure parameter $\sum$, which is an
indicator of recent interactions. Gonz\'alez \& Gorgas (1996, hereafter GG96)
found an anti-correlation between the residuals  and the H$\beta$ index, and
concluded that recent episodes of star formation could explain, at least
partially, the present scatter in the relation (see also Worthey \& Collobert
2003).

\begin{figure}
\resizebox{1.0\hsize}{!}{%
\includegraphics[angle=-90]{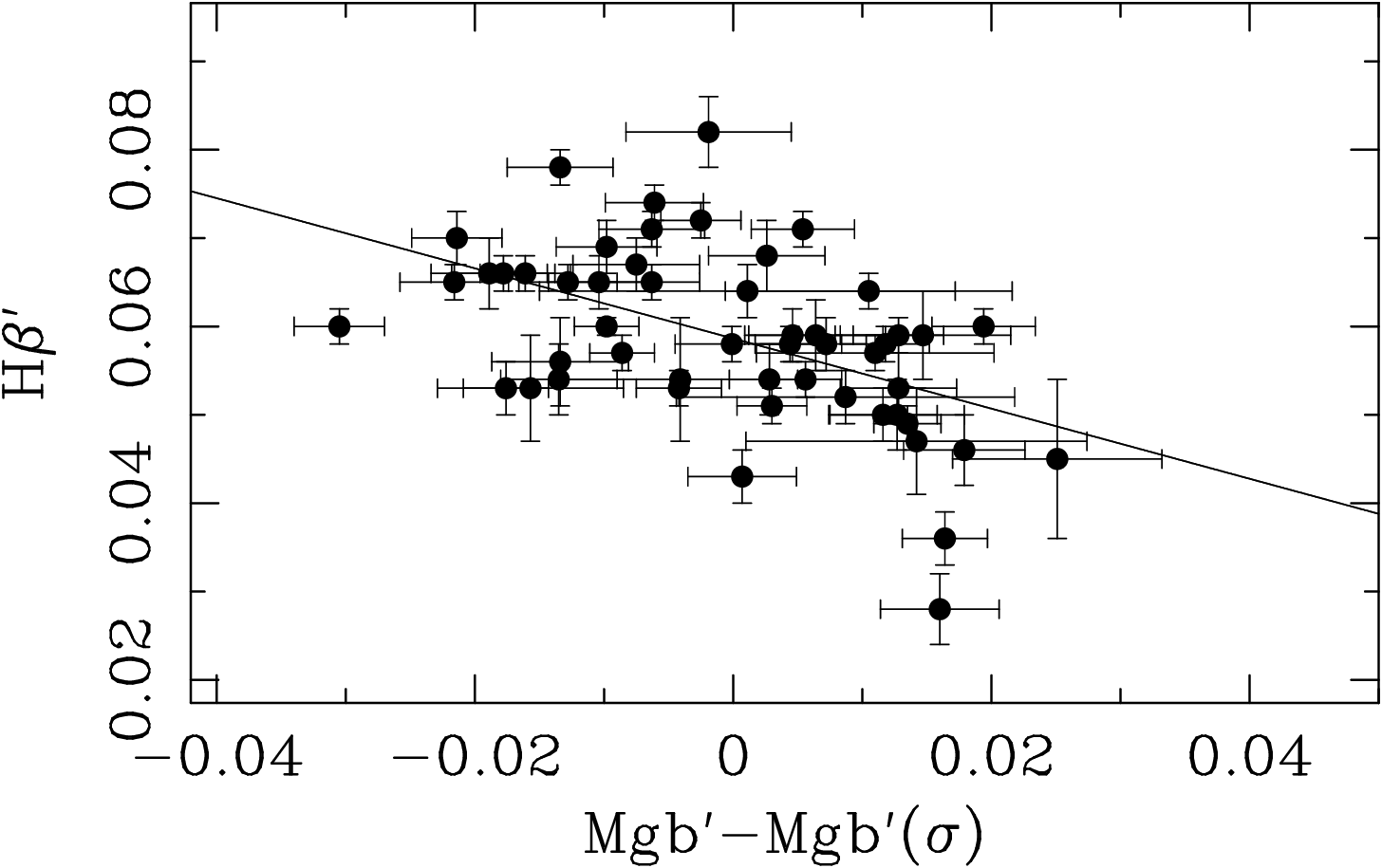}}

\vspace{3mm}

\resizebox{1.0\hsize}{!}{%
\includegraphics[angle=-90]{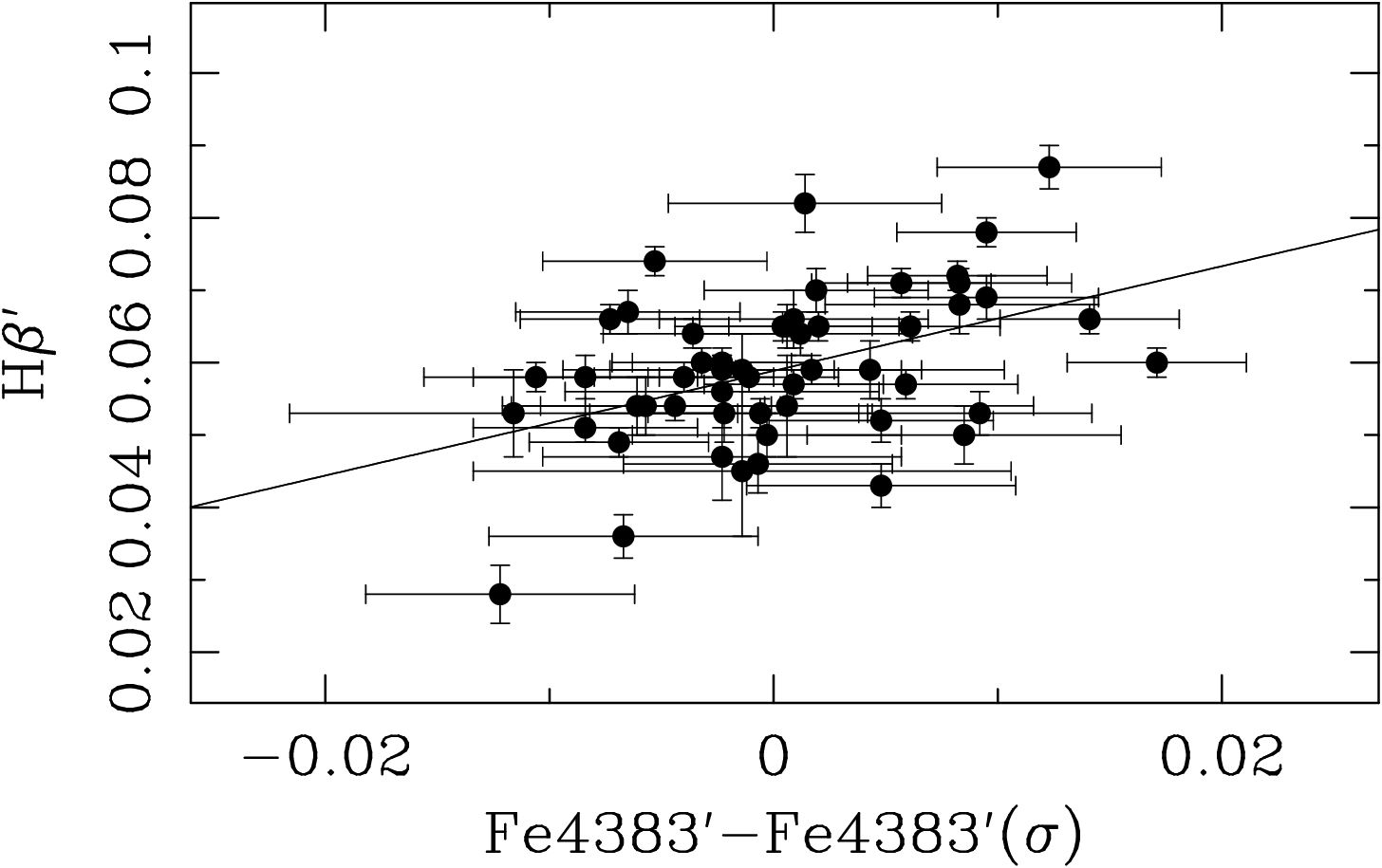}}
\caption{ H$\beta$' index as a function of the residuals of the
\mbox{Mgb'--$\log\sigma$} (Mgb$'-$Mgb$'$($\sigma$)) and
\mbox{Fe4383$'$--$\log\sigma$} (Fe4383$'$--Fe4383$'$($\sigma$)) relations for
LDEGs. The linear fits are shown.\label{fig.residuals}}
\end{figure}

In this section we investigate the scatter in the relations of the Lick indices
with the central velocity dispersion, trying to find which parameters are
varying between galaxies with the same velocity dispersion.  As we said,
several studies suggest that the scatter in the Mg--$\sigma$ relations is due
to variations in the age of the galaxies at a given $\sigma$ through the
dilution of the metallic features in a young stellar population.
To explore this possibility, we have investigated, following the GG96 analysis,
the dependence of the residuals in the index--$\sigma$ relations on the
H$\beta$ index. We performed a $t$-test to study the degree of correlation
between them.  Table \ref{tab-corr-hbeta} summarises the results. A correlation
is considered significant if the  $t$ value is   higher than 1.96. As can be
seen in the table, we confirm the result obtained by GG96 concerning the
existence of an strong anti-correlation between the residuals of the
Mgb--$\sigma$ relation and the H$\beta$ index for the LDEGs but, on the other
hand, we find a positive correlation between the residuals of the
Fe4383--$\sigma$ relation and this index. This is illustrated in
Fig. \ref{fig.residuals}. If the age were the main parameter responsible for
the scatter in the index--$\sigma$ relations, an anti-correlation of the
residuals in the Fe4383--$\sigma$ with H$\beta$ should also be observed,
contrary to our findings. Therefore, our result excludes age as the only
parameter producing the dispersion in  the index--$\sigma$ relations.  On the
other hand, if only variations of overall metallicity at a given $\sigma$ were
responsible for the dispersion, we would also expect the same behaviour on the
correlation of the residuals with H$\beta$ in both indices.  Neither of these
two parameters can be entirely responsible of the residuals from the relations.
Interestingly, in the HDEG subsample, we do not find any correlation between
the residuals and the H$\beta$ index (we only found a marginal correlation for
Fe5015 and Fe5335).
 
\begin{table} 
\caption{Statistical analysis of the correlations between the residuals of the
index--$\sigma$ relation and the H$\beta$ index. For a significance level lower
than $\alpha=0.05$, a value of $t$ higher than 1.96 confirms the existence of
correlation.\label{tab-corr-hbeta}}
\centering 
\begin{tabular}{@{}l r r@{}}
\hline\hline
               & \multicolumn{1}{c}{LDEG} & \multicolumn{1}{c@{}}{HDEG}\\ 
 Index         & \multicolumn{1}{c}{$t$}  & \multicolumn{1}{c@{}}{$t$} \\ 
\hline 
 D4000         &$ 1.00$  &$ -1.06$ \\ 
 H$\delta_A$   &$ 0.08$  &$-0.50$  \\ 
 H$\delta_F$   &$ 2.50$  &$  0.73$ \\ 
 CN$_2$        &$-0.85$  &$ 0.42$  \\ 
 Ca4227        &$ 0.34$  &$  0.36$ \\ 
 G4300         &$ 2.21$  &$ 1.04$  \\ 
 H$\gamma_A$   &$-0.29$  &$ -0.34$ \\ 
 H$\gamma_F$   &$ 2.07$  &$ -0.04$ \\
 Fe4383        &$ 3.00$  &$  0.82$ \\ 
 Fe4531        &$ 2.61$  &$ 0.90$  \\
 Ca4455        &$ 2.93$  &$  0.56$ \\ 
 C4668         &$ 1.47$  &$ 0.16$  \\
 Fe5015        &$ 1.32$  &$  2.23$ \\ 
 Mgb           &$-4.24$  &$ 0.36$  \\
 Fe5270        &$ 1.97$  &$  1.50$ \\ 
 Fe5335        &$ 2.20$  &$ 1.85$  \\
\hline  
\end{tabular} 
\end{table}

\begin{figure}
\centering
\resizebox{\hsize}{!}{\includegraphics[angle=-90]{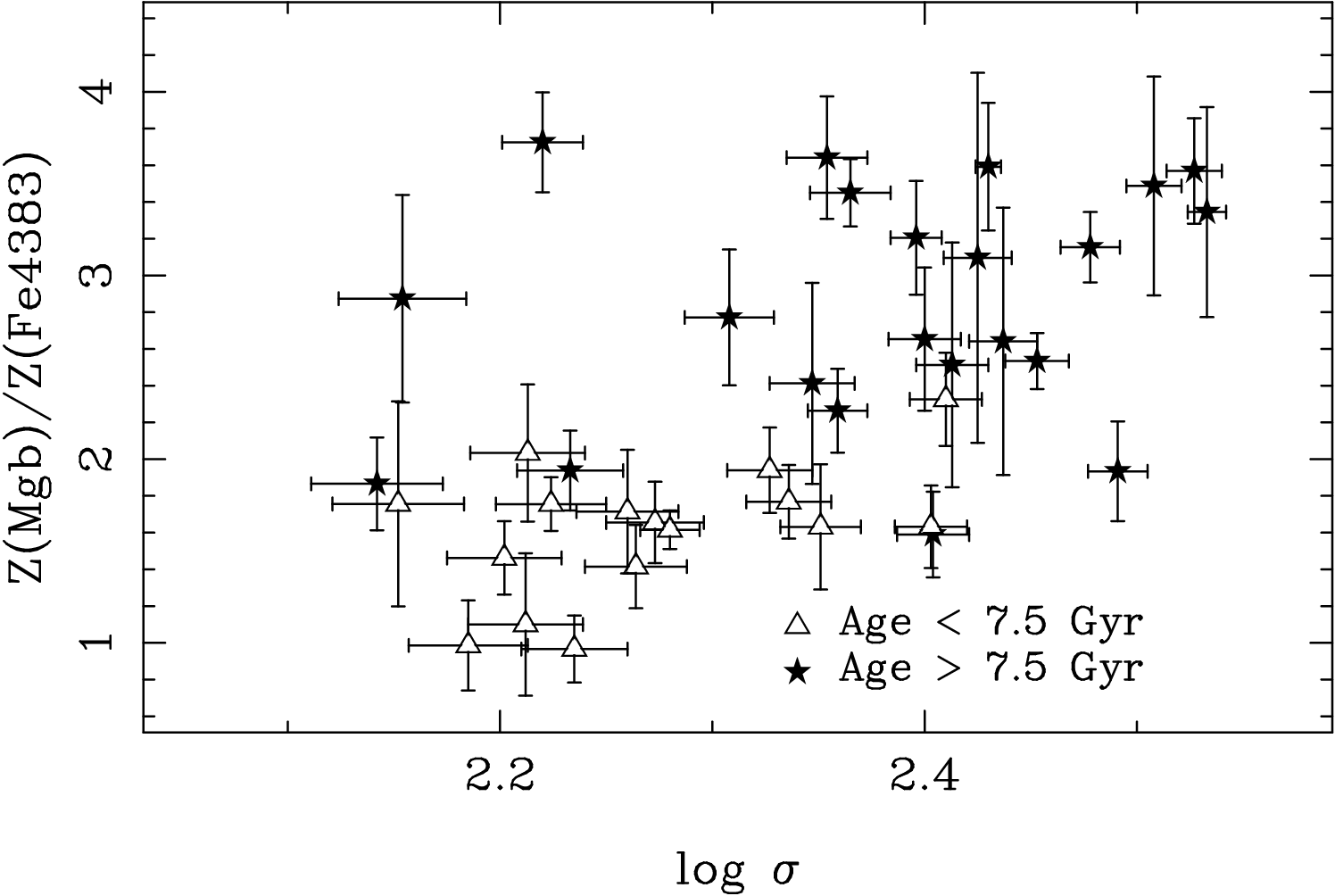}}

\vspace{3mm}

\resizebox{\hsize}{!}{\includegraphics[angle=-90]{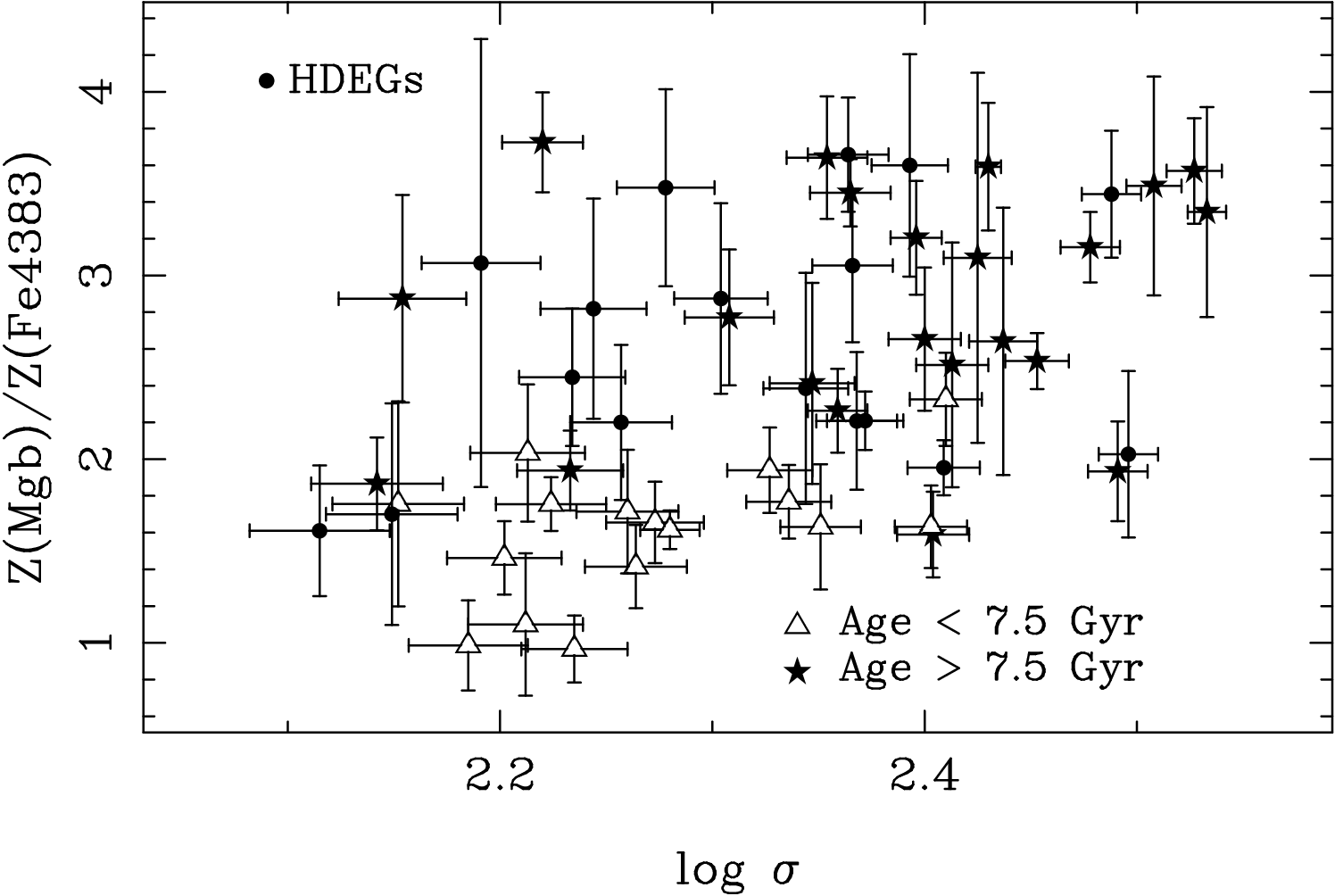}}
\caption{Ratio of the metallicities calculated in the \mbox{Mgb--H$\beta$} and
\mbox{Fe4383--H$\beta$} diagrams as a function of the velocity dispersion for
LDEGs.  Different symbols represent galaxies younger (open triangles) and older
(filled stars) than 7.5 Gyr, as derived from a [MgFe]-H$\beta$ diagram. In the
lower panel we have also included the sample of HDEGs (solid circles).
\label{figurina}}
\end{figure}

Discarding both, age and metallicity, as the only parameters responsible for
the departure of the galaxies from the index--$\sigma$ relations, we
investigate if a variation in the relative abundances between galaxies can help
to explain the observed scatter. Figure \ref{figurina} shows the quotient
between the metallicities measured in the Mgb--H$\beta$ and Fe4383--H$\beta$
diagrams (see Paper II for details) as a function of velocity dispersion,
splitting the sample into galaxies older and younger than 7.5 Gyr (according to
their position in the H$\beta$--[MgFe] plane, using V06 models).  It is clear
from this figure that the [Mg/Fe] relation with $\sigma$ depends on the
luminosity-weighted mean age of the galaxies and, therefore, on their
particular star formation history. In particular, older galaxies  have, on
average, higher [Mg/Fe] than the younger ones. In the lower panel of
Fig. \ref{figurina} we have added the galaxies from the Coma cluster. It can be
seen that they follow the trend of the older LDEGs, exhibiting, on average,
larger [Mg/Fe] ratios than younger LDEGs.

This trends indicate that the scatter in the index--$\sigma$ relations is not
an exclusive effect of a dispersion in the age, but it is a  consequence of the
variation of the element abundance ratios with this parameter. Furthermore, we
find a tendency for which older galaxies have, on average, higher [Mg/Fe]
ratios. 

Fisher, Franx \& Illingworth (1995) noted, studying a sample of 7 bright
early-type galaxies, a correlation between [Mg/Fe] and the age of the galaxies,
in the same sense as we find here. These authors, however, did not find a
correlation between the residuals of the Mgb--$\sigma$ relation and the
H$\beta$ index. This could be due  to the small sample analysed. Thomas, Maraston
\& Bender (2002)
 also find a similar relation between the
$[\alpha/$Fe$]$\footnote{$[\alpha/$Fe$]$ represents the ratio between the
$\alpha$-elements abundances and Fe.} ratio and the age, studying several
samples from the literature. Contrary to these findings, Trager et al.\ (2000b)
do not report any correlation between $[\alpha/$Fe$]$\footnote{These authors do
not use this term but instead $[{\rm E}/{\rm Fe}]$, being E the abundance of
the enhanced elements, which does not coincide completely with the $\alpha$
elements. } and the age of the galaxies. However these authors analysed the
relation between age, [Z/Fe], [E/Fe] and $\sigma$, concluding that age cannot
be the only parameter responsible for the scatter in the Mg--$\sigma$ relation,
being necessary variations of $[$E/Fe$]$ between galaxies with the same
$\sigma$, in agreement with our results. 

Since Mg is produced mainly in Type II supernovae whilst the bulk of Fe is
released by Type I supernovae  (e.g., Nomoto, Thieleman \& Wheeler 1984, Woosley \& Weaver 1995; 
Thielemann et al. 1996), the existence of a trend between [Mg/Fe] and
age favours the idea that the different element ratios are a consequence of
different star formation histories  (see, e.g. Greggio \& Renzini 1983; 
Matteucci \& Greggio 1986; Gibson 1997; Thomas et al. 1998), and that younger galaxies have suffered a
more extended star formation history, incorporating the elements produced by
low-mass stars (see, e.g. Worthey 1998).  However, there are other possibilities which we explore later
in Section \ref{overabundance.age}. Again, in this section, none of the conclusions
depends on the absolute values of the indices, therefore, they would not change if we
add a constant offset to the C4668 and Mgb indices.

\begin{figure}
 \resizebox{\hsize}{!}{\includegraphics[angle=-90]{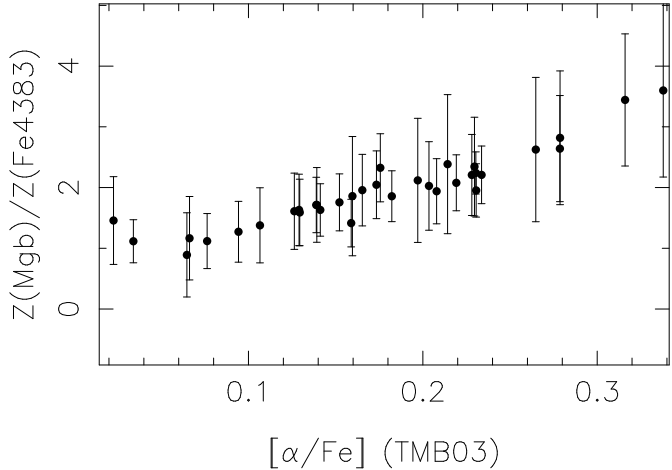}}
\caption{Comparison of Z(Fe4383)/Z(Mgb) with the values of [$\alpha/$Fe]
derived using the models by TMB03. Z(Fe4383) represent the metallicity 
measured in a Fe4383-H$\beta$ diagram, while Z(Mgb) the metallicity measured
in a Mgb-H$\beta$.\label{comparison.abundances}}
\end{figure}

We want to finish this section justifying the use of the term [Mg/Fe] in our discussion when, 
in reality, we are just comparing the metallicity measured in a Fe4383-H$\beta$ diagram
with the one obtained using the Mgb-H$\beta$ combination. Fig .\ref{comparison.abundances}
shows the values of Z(Fe4383)/Z(Mgb) compared with the parameter [$\alpha$/Fe] derived
from TMB03 models. To obtain  these values we used the indices Fe4383, H$\beta$, 
and Mgb, and followed an iterative process. We first measured age and metallicities
assuming [$\alpha$/Fe]=0 and obtained a first guess of the age that we used to determine
[$\alpha$/Fe]. With this new value, we calculated, again, age and metallicity and so on, 
until the derived parameters in two consecutive iterations were consistent within 5\% accuracy. 
As can be seen in Fig. \ref{comparison.abundances} there is an excellent correspondence
between the values of [$\alpha$/Fe] derived with TMB03 models and the 
ratio of metallicities
measured using Fe4383 and Mgb with the V06 models.  We could have used a transformation 
to derive [$\alpha$/Fe] values from Z(Fe4383)/Z(Mgb) values, but several 
aspects of the models need to be understood before 
deriving  any meaningful number. In any case, we are not making any conclusion 
based on the absolute  value of [Mg/Fe]. We prefer to use the term [Mg/Fe]
instead of the most general [$\alpha$/Fe] as different $\alpha$-elements may have different
behaviour in spheroidal systems, as it seems to be the case in the bulge stars of 
our Galaxy (see, e.g. Fulbright et al.\ 2005). 
%%%%%%%%%%%%%%%%%%%%%%%%%%%%%%%%%%%%%%%%%%%%%%%%%%%%%%%%%%%%%%%%%%%%%%%%%%%%%%%
\section{Differences in the line-strength indices between LDEG{\scriptsize s} 
and HDEG{\scriptsize s}}

Some studies have analysed the possible differences between the line-strength 
indices of galaxies in different environments.
Guzm\'an et al.\ (1992) found systematic variations in the zero-point of the
Mg$_2$--$\sigma$ relation for a sample of Coma galaxies as a function of their
distance to the cluster centre. J\o rgensen et al.\ (1996, 1997), examining a
sample of eleven clusters, detected a weak correlation between the intensity of
the Mg line and the local density within the clusters, in agreement with the
Guzm\'an et al. result.  Similar differences have been found by several other
authors (e.g.\ de Carvalho \& Djorgovski 1992) while others (e.g.\ Bernardi et
al.\ 1998) do not find evidences of an environmental influence on the
zero point of the relation of the indices with the velocity dispersion of the
galaxies. The dependence of the absorption features on the environment has been
studied, almost exclusively, with just two indices Mg$_2$ and Mgb. 

In this section we study the differences  on 17 Lick indices, at a given
$\sigma$, between LDEGs and HDEGs.
To quantify  possible systematic differences, we have performed a linear
least-squares fit to the index--$\sigma$ relations of the LDEGs subsample (see
Section \ref{sec.indexsigma}) and have measured the mean offsets (weighting
with errors) of the HDEGs from the fits. Table \ref{tab-differences} shows
these mean offsets ($d$), with their errors ($s$/$\sqrt{n_{\rm eff}}$, being
$s$ the standard deviation and $n_{\rm eff}$ the effective number of points).
We find significant differences  in the following indices: H$\delta_A$,
H$\delta_F$, Ca4455, Fe4531, Fe5015, and also in the indices CN$_2$ and C4668,
whose differences have been previously reported by S\'anchez--Bl\'azquez et
al.\ (2003). On the other hand, in agreement with other studies (e.g.\
Bernardi et al.\ 1998), we do not find any significant difference in the Mgb
index of both subsamples at a given velocity dispersion. In the next section
we discuss the possible causes of these differences.
 
\begin{table}
\caption{Mean differences in the Lick indices at a given $\sigma$ between
galaxies in different environments.  For each index, we list the mean offset
$d$ and its error $\Delta d$ and the $t$-parameter (of the statistical test
$d\ne 0$).\label{tab-differences}}
\centering
\begin{tabular}{@{}l r r r@{}}
\hline\hline
Index       &  
\multicolumn{1}{c}{$d$} &
\multicolumn{1}{c}{$\Delta d$} &
\multicolumn{1}{c@{}}{$t$}\\
\hline
D4000        &$ 0.0067$ & 0.0208     & 0.321\\  
H$\delta_A$  &$ 0.0202$ & 0.0041     & 4.904\\
H$\delta_F$  &$ 0.0158$ & 0.0032     & 4.886\\
CN$_2$       &$-0.0287$ & 0.0062     & 4.600\\
Ca4227       &$-0.0057$ & 0.0034     & 1.689\\ 
G4300        &$-0.0035$ & 0.0034     & 1.121\\
H$\gamma_A$  &$ 0.0061$ & 0.0033     & 1.843\\
H$\gamma_F$  &$ 0.0060$ & 0.0042     & 1.442\\
Fe4383       &$-0.0029$ & 0.0022     & 1.332\\ 
Ca4455       &$-0.0063$ & 0.0020     & 3.166\\
Fe4531       &$-0.0058$ & 0.0015     & 3.827\\  
C4668        &$-0.0103$ & 0.0029     & 3.535\\
H$\beta$     &$ 0.0006$ & 0.0024     & 0.259\\
Fe5015       &$-0.0053$ & 0.0014     & 3.776\\
Mgb          &$ 0.0015$ & 0.0029     & 0.500\\
Fe5270       &$-0.0022$ & 0.0010     & 2.146\\ 
Fe5335       &$ 0.0014$ & 0.0014     & 0.993\\
\hline
\end{tabular}
\end{table}

%------------------------------------------------------------------------------
\subsection{Possible interpretations of the differences}

In S\'anchez-Bl\'azquez et al.\ (2003) we briefly discussed possible mechanisms
that would  produce differences in the CN and C4668  between galaxies in
different environments. In this section, we extend this discussion including
the rest of the indices analysed.

\begin{itemize}

\item {\it IMF variations:\/} Given the high sensitivity of the CN$_2$ index to
the stellar surface gravity (giant stars have higher CN$_2$ index than  dwarfs;
Gorgas et al.\ 1993; Worthey et al.\ 1994), a lower proportion of giant stars
in HDEGs with respect to LDEGs would produce systematic differences in that and
in some other spectral characteristics between both subsamples. However,
H$\delta_A$, H$\delta_F$, C4668, Ca4455, Fe4531 and Fe5015 are not especially
sensitive to gravity variations. Therefore it does not seem probable that this
is the reason behind the reported differences. In any case, and in order to
quantify this, we calculated the variation of the IMF slope necessary to
produce the observed difference in the CN$_2$ index using the predictions of
V06 corresponding to a single stellar population  of 10 Gyr and solar
metallicity.  This variation in the slope of the IMF would also reproduce the
observed differences in C4668 and Fe5015 but, at the same time, would lead to
variations in the opposite sense in the Ca4455 and Fe4531 indices. Therefore,
although we do not discard the possibility  of differences in the IMF between
both subsamples, it cannot be the only factor responsible for the offsets of
the relations in the indices

\item {\it Differences in the mean age of galaxies in distinct environments:\/}
This would also introduce systematic variations in the line-strength indices.
Using V06 models, the observed differences in CN$_2$, C4668, Ca4455, Fe4531 and
Fe5015 could be explained by assuming an age discrepancy between both
subsamples of 9.8 Gyr, with the Coma galaxies being younger than the galaxies
in low density environments (considering a single burst with solar metallicity,
a Salpeter IMF and a variation in age from 7.94 to 17.78 Gyr). This age
difference would however produce changes in the H$\delta$ indices almost twice
as large as observed, and would also produce differences in other indices, like
H$\beta$, not only higher than observed, but in the opposite sense to the
measured ones.  Furthermore, the results reported by other authors suggest that
galaxies in denser environments are older than galaxies in the field and low
density groups (e.g.\ Kuntschner et al.\ 2002; S\'anchez-Bl\'azquez 2004;
Thomas et al.\ 2005; see also Paper II).  

\item {\it Variations in the relative abundances between LDEG and HDEG:\/}
Another explanation for the observed offsets could be the existence of
differences in the  element abundance ratios of different chemical species.
According to TB95, C4668 is extremely sensitive to variations in the carbon
abundance. CN$_2$, on the other hand, is sensitive to variations of both C and
N abundances. Furthermore, the blue band of the H$\delta$ index overlaps with
the CN bands, causing a decrease in these values as the N abundance increases
(Worthey \& Ottaviani 1997; Schiavon, Caldwell \& Rose 2004). 
Fe4531 and Fe5015 are both very sensitive to
variations of Ti. 

If we assume than the differences in CN$_2$, C4668, Fe5015 and Fe4531 between
HDEGs and LDEGs are due to variations in some chemical species with respect to
Fe, it would be required the existence of differences in C, N and Ti abundances
between both subsamples.  This is a very qualitative statement. To check if a
variation of these chemical species is compatible with the calculated offsets
for all the indices, we follow a similar approach to that in Section
\ref{sec.slopes}. In this case, instead of comparing the variation of the
indices with the velocity dispersion, we compare the offsets between galaxies
in different environments and try to reproduce these differences by varying the
abundances of different chemical species.  We separated the elements into 6
different groups: (i) $\alpha$ elements: Ne, Na, Mg, Si, S, and O; (ii) C;
(iii) N; (iv) Ti; (v) Ca; and (vi) Fe-peak elements (Fe and Cr). We have built
4 different models.  In building the models, we have not tried to fit all the 
indices by varying the ratios of different elements arbitrarily, but we have proposed
4 simple permutations which can give a rough idea of the differences in the chemical 
composition between galaxies in different environments. This rough idea though, 
can give us important clues to understand the differences in the star 
formation history of galaxies between LDEG and HDEG.
The models can be described as follows:\\
\begin{tabular}{@{}l@{}c@{$\;$}l@{}}
 \raisebox{0ex}[3ex]{\mbox{}}
 Model 1 &:& \parbox[t]{0.8\columnwidth}{the abundances of C and N are
 increased by 0.05 dex, leaving the other element abundances unchanged}\\
 \raisebox{0ex}[3ex]{\mbox{}}
 Model 2 &:& \parbox[t]{0.8\columnwidth}{the abundances of C and N are
 increased by 0.05 dex, the abundance of Ti by 0.15 dex, and the other
 elements are kept fixed}\\
 \raisebox{0ex}[3ex]{\mbox{}}
 Model 3 &:& \parbox[t]{0.8\columnwidth}{the abundances of C is varied in
 0.04 dex, N by 0.05 dex, Ti by 0.15 dex, and Ca by 0.10 dex; the remaining
 abundances are unchanged}\\
 \raisebox{0ex}[3ex]{\mbox{}}
 Model 4 &:& \parbox[t]{0.8\columnwidth}{similar to Model 3, increasing the
 abundances of all the $\alpha$ and the Fe-peak elements by 0.05 dex and the 
 N abundance by 0.15 dex}
\end{tabular}\\
\raisebox{0ex}[2.5ex]{A} summary of the 4 models is shown in 
Table \ref{modelillos}.

\begin{table*}
\caption{Differences in the abundances of the distinct chemical species used
to build the 4 models described in the text. The last column shows the
dispersion around the 1:1 relation when comparing the offsets in the indices
between galaxies in different environments and the offsets in the indices due
to the changes in the chemical species of each model. \label{modelillos}}
\centering
\begin{tabular}{@{}l rrrrrrr@{}}
 \hline\hline
 Model  &[C/H] & [N/H] & [$\alpha$/H] & [G2/H] & [Ca/H] & [Ti/H] & r.m.s.\\
\hline
 1& +0.05 & +0.05 &  +0.00       & +0.00  & +0.00  & +0.00  & 0.0066\\
 2& +0.05 & +0.05 &  +0.00       & +0.00  & +0.00  & +0.15  & 0.0060\\
 3& +0.04 & +0.05 &  +0.00       & +0.00  & +0.10  & +0.15  & 0.0035\\
 4& +0.04 & +0.15 &  +0.05       & +0.05  & +0.10  & +0.15  & 0.0034\\ 
\hline
\end{tabular}
\end{table*}

\begin{figure}
\resizebox{0.47\hsize}{!}{\includegraphics[bb=88 145 578 630,angle=-90]{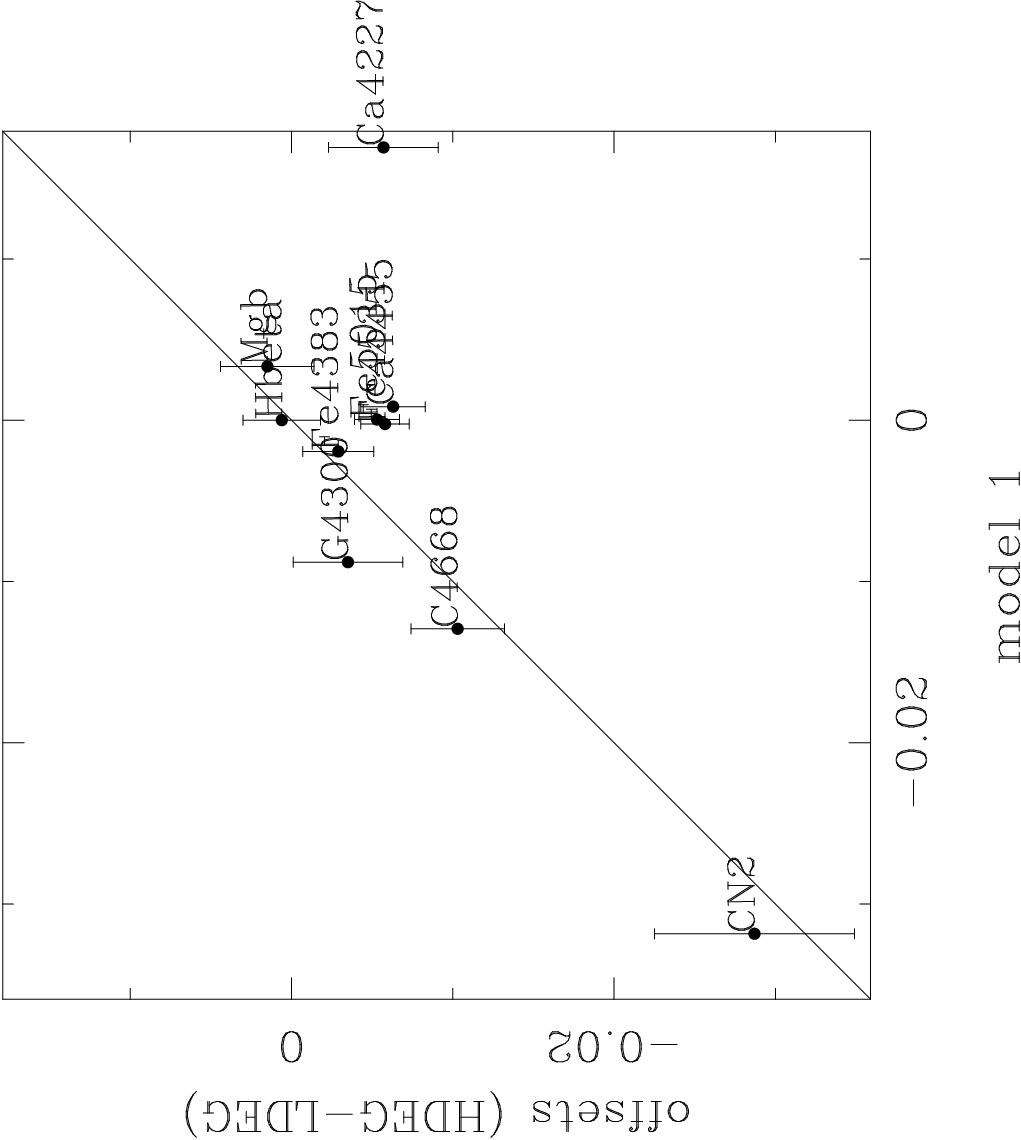}}
\hfill
\resizebox{0.47\hsize}{!}{\includegraphics[bb=88 145 578 630,angle=-90]{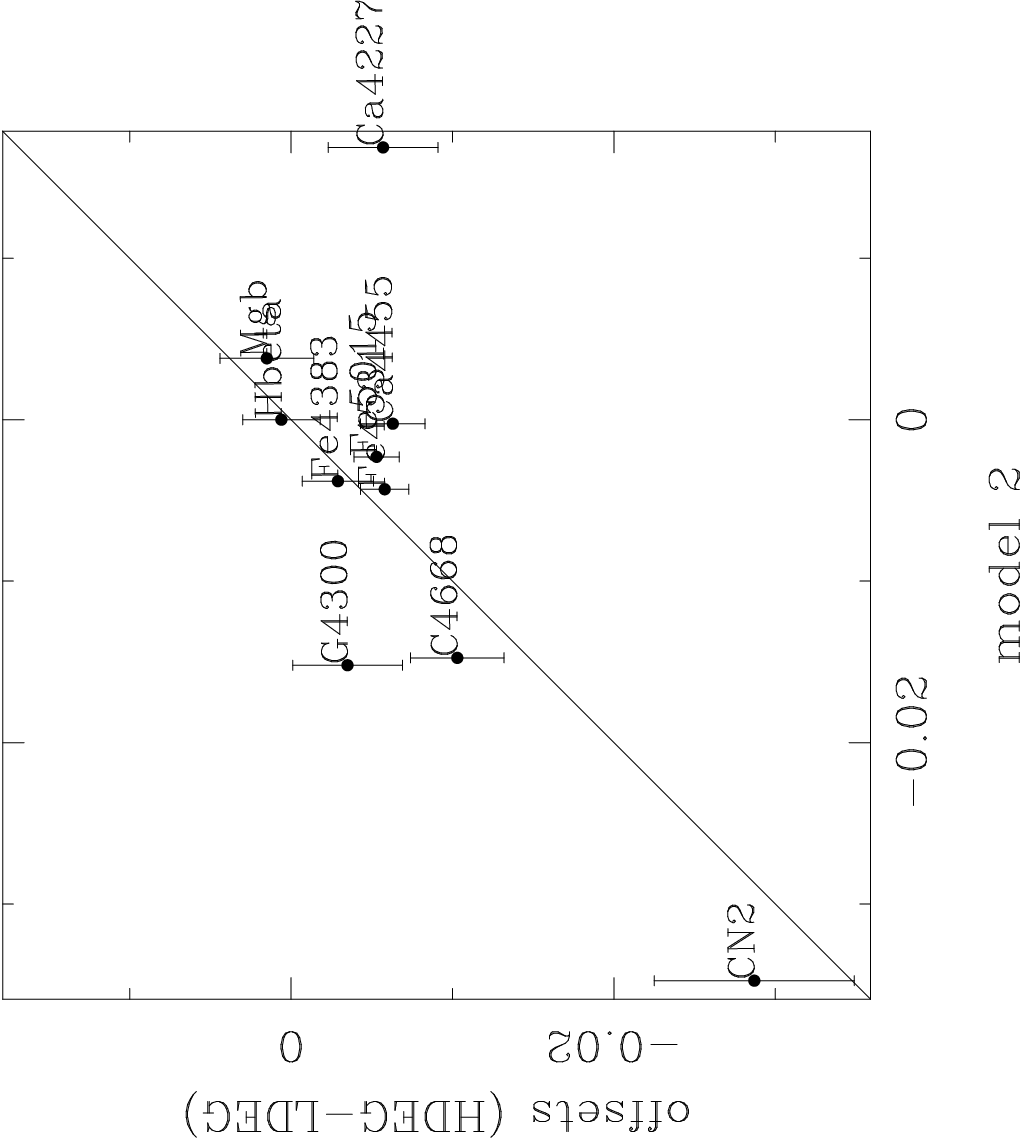}}

\resizebox{0.47\hsize}{!}{\includegraphics[bb=88 145 578 630,angle=-90]{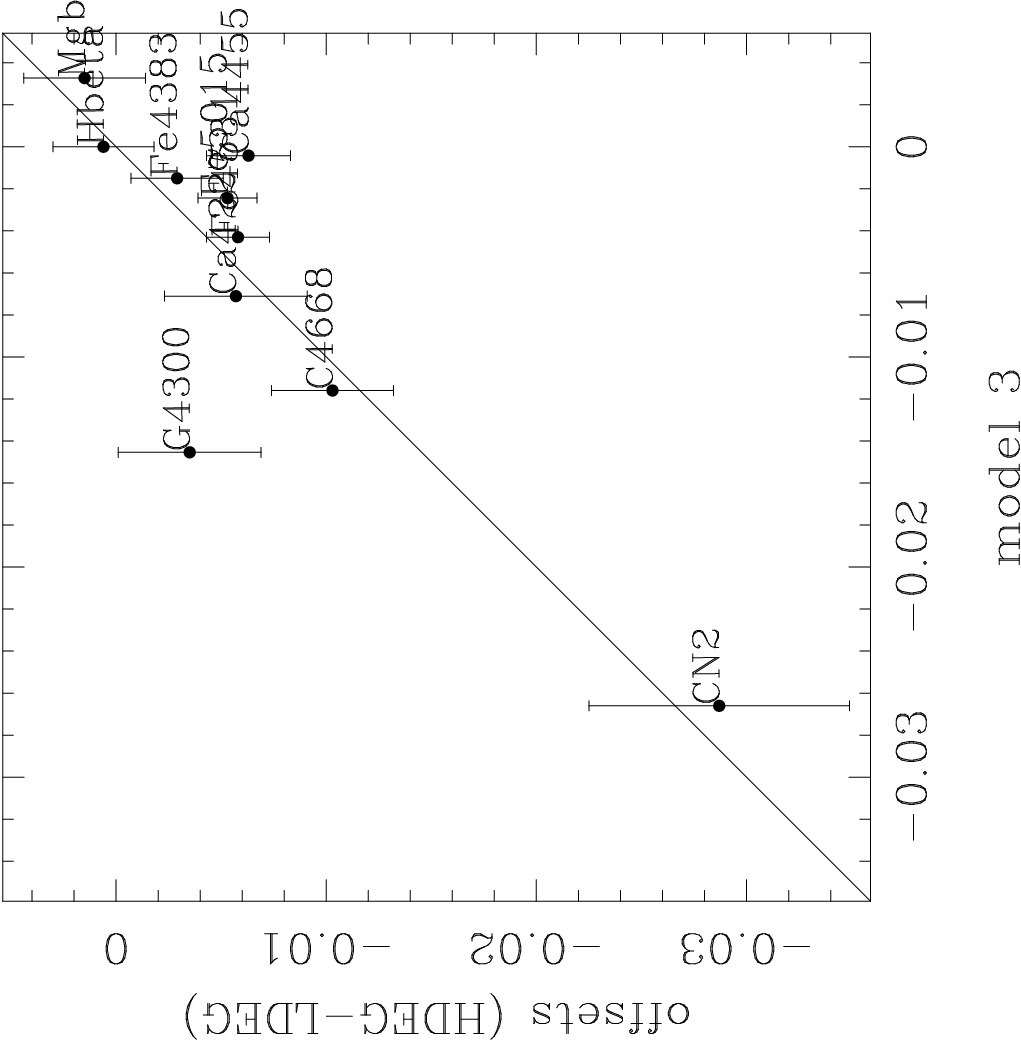}}
\hfill
\resizebox{0.47\hsize}{!}{\includegraphics[bb=88 145 578 630,angle=-90]{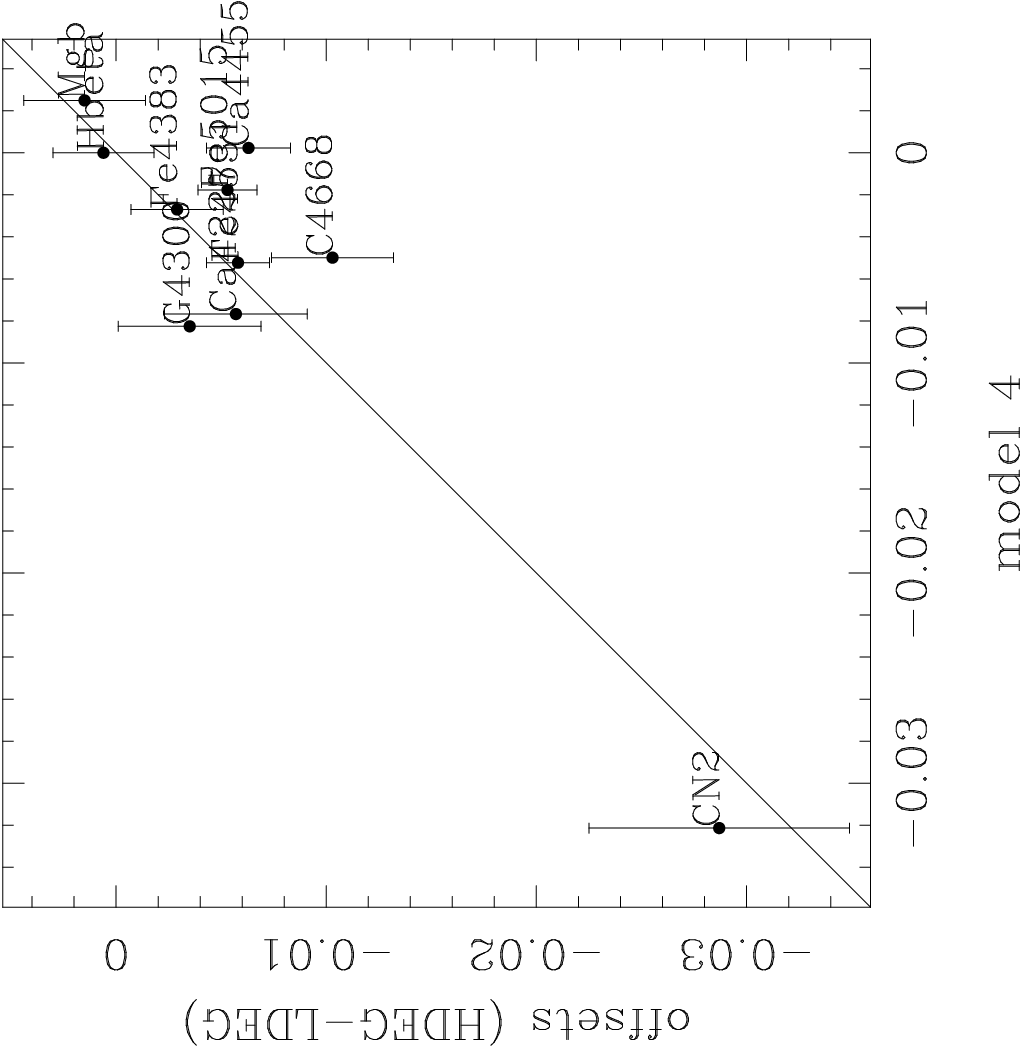}}
\caption{Comparison of the offsets between HDEGs and LDEGs with the 
differences in the indices obtained in different models in which 
we have calculated the variation  in the Lick indices due to variations
on the ratio of different  chemical elements (see text for
a more detailed description). The error bars indicate the errors in the 
offsets. \label{fig.modelillos}}
\end{figure}

Figure \ref{fig.modelillos} shows the comparison of the observed offsets
between galaxies in different environments and the ones obtained from the
different models. The lines in the figures indicate the 1:1 relation (the last
column of Table \ref{modelillos} indicates the r.m.s.\ of the dispersion around
this relation). From the analysis of the figure and the r.m.s.\ of  the
residuals we can conclude the following:
\begin{itemize}
 \item The models in which the only elements varying are C, N and Ti are not
 the best to reproduce the observed offsets in all the indices. 

 \item  Although we do not find a significant  difference in the Ca4227 index
 between galaxies in different environments, due to the dependence of this
 index on the C abundance (its value drops when the C abundance increases; see
 TB95), it is necessary a difference in the Ca abundance to reproduce the
 observed differences in this index. 

 \item The models which include a variation in all the chemical species
 reproduce best the differences in all the observed indices. 

 \item The observed differences between LDEGs and HDEGs are reproduced with a
 very small variation  in the C abundance. However, it is difficult to
 simultaneously explain both the C4668 and G4300 indices. In fact, in order to
 reproduce the differences in G4300 it would be necessary a larger difference
 in the C abundance between low- and high-density environment galaxies. Two
 possibilities, as we argued in Section \ref{sec.slopes}, are that the index
 C4668 is not so sensitive to the C abundance as reported in TB95 or that
 G4300 is not very well calibrated in the model atmospheres of TB95. 
\end{itemize}
%%%%%%%%%%%%%%%%%%%%%%%%%%%%%%%%%%%%%%
One study which may provide evidence for the existence of C differences between
galaxies in different subsamples is that of  Mobasher \& James (2000). These
authors found significant differences in the CO band at 2.3 $\mu$m between
galaxies in the centre and in the outskirts of the Coma cluster.  
These
authors interpreted the differences as evidence of the presence of a younger
population in galaxies situated in the field, due to a larger contribution of
AGB stars. Although in the present work we do not discard this possibility, the
data presented in this paper allow an interpretation in which the differences
are mainly due to changes in the C abundance. In any case, as we have shown,
the reported differences between indices can not be explained exclusively with
an age variation between samples.
%%%%%%%%%%%%%%%%%%%%%%%%%%%%%%%%%%%%%%%%%%%%%%%%%%%%%%%%%%%%

Summarising, in order to reproduce the observed differences between the indices
of LDEGs and HDEGs through variations in the abundances of different elements,
we have to assume a variation in all the chemical species, in the sense that
LDEG are, on average, more metal rich than HDEG. The relative changes in all
the chemical species, however, are not the same.  In particular, the models
that best explain the differences are those which assume a higher variation in
N, Ti, Ca, and probably C, between galaxies in different environments.

Ti and Ca are elements which are very difficult to interpret. Ca is an
$\alpha$-element, but in early-type galaxies seems to track Fe-peak elements
(Worthey 1998; TMB03; Cenarro et al.\ 2004).  Ti is an element
poorly understood. Nucleosynthesis theories predict that this element is
produced in Type II supernovae and that its abundance is similar to that of Fe.
However, in galactic bulge stars this element is found in higher proportions
than in the solar partition, even for stars of solar metallicity (McWillian \&
Rich 1994, 2004), which has not been reproduced by any nucleosynthesis model.
This element can also be overabundant with respect to Fe in massive ellipticals
(Worthey 1998). So, we do not try to discuss the origin of the differences in
these two elements between galaxies in different environments. The conclusions
of this section would not change if we were to add a constant offset to the 
indices C4668 and Mgb, as the conclusions 
are based upon differences between galaxies in 
different environments, not on the absolute values of these indices.

In the following subsection we extend the discussion already presented in
S\'anchez-Bl\'azquez et al.\ (2003) about  the possible mechanisms that can
lead to  a difference in the [C/Fe] and [N/Fe] ratios between low- and
high-density environment galaxies.

%------------------------------------------------------------------------------
\subsection{Abundances of C and N}

C is produced predominantly by the triple-alpha reaction of He, while N is
produced in the conversion of C and O during the CNO cycles. The problem is to
know the evolutionary phases in which these elements are predominantly
produced. Several studies seem to favour intermediate-mass stars (between 5 and
8 M$_{\odot}$) as the main contributors to N (Renzini \& Voli 1981; Henry,
Edmunds, \& K\"{o}ppen 2000; Chiappini, Romano \& Matteucci 2003), although it
is also predicted to be produced in massive stars (Meynet \& Maeder 2002).
There exists still more controversy over the location of the  C production.
Some authors claim that this occurs in low- and intermediate-mass stars
(Renzini \& Voli 1981; Chiappini et al.\ 2003a) based on the variation of
[C/Fe] with metallicity in the stars of the galactic disk. This ratio remains
constant from [Fe/H]$\sim -2.2$ to solar metallicities, as it would expected if
it were produced in non-massive stars. On the other hand, recent measurements
of the C/O abundances among halo and disk stars show a discontinuity around
$\log($O/H$)\sim -3.6$. Oxygen is mainly created in massive stars, therefore if
the main contributors to the C abundance were also these stars, we would not
expect to find this discontinuity. However, other authors affirm that, to
explain the abundances of C observed in stars in the galactic disk, most of the
C has to be produced in massive stars (Carigi 2000; Henry et al.\ 2000). The
conclusions of the different authors are very dependent on the adopted yields.
The works that favour massive stars as the main contributors to the C abundance
in the interstellar medium are based in the Maeder (1992) yields, which predict
a strong dependence of the C yields in massive stars on the metallicity.
However, this work has been improved in the new models of Meynet \& Maeder
(2002), which have taken into account the effects due to the stellar rotation.
Using these models, Chiappini, Matteucci \& Meynet (2003) showed that,
considering massive stars as the main producers of C, it is not possible to
explain the solar C/O ratio. In this work we consider that C is produced mainly
in low- and intermediate-mass stars ($1\leq M$/M$_{\odot}\leq 8$), while N is
mainly produced in intermediate mass stars ($5\leq M$/M$_{\odot}\leq$ 8).
During the AGB, these stars eject into the interstellar medium significant
amounts of $^{4}$He, $^{12}$C, $^{13}$C, and $^{14}$N.
Of course, the interpretation is subject to change if new results in stellar
nucleosynthesis report that these stars are not the main contributors to these
elements. 

A difference in the relative abundance of C and N between LDEGs and HDEGs may 
be due to:
\begin{itemize}
\item A difference in the yields of the stars in HDEGs and LDEGs.
\item A difference in the star formation history between both subsamples.
\end{itemize}

We start  exploring the first possibility.
The yields of low- and intermediate-mass stars change with the time between
thermal pulses during the AGB, the number of these pulses and the efficiency of
the third dredge-up. These factors depend, fundamentally, on the mass of the
stars and their metallicity. The number and the duration of the thermal pulses
in AGB stars increase with the mass of the star, as do the yields. However,
when the mass of the star is higher than 3--4 M$_{\odot}$, a process known as
hot bottom burning occurs. This process converts part of the $^{12}$C into
$^{14}$N. The efficiency of this process increases with mass and decreases with
metallicity (Marigo 2001) and has a particularly noticeable effect on 
the yields of $^{12}$C and $^{14}$N, leading to an increase in  N and a
decrease of C in the interstellar medium.  A higher metallicity in HDEG with
respect to LDEG could increase the efficiency of this phenomenon in the former,
leading to an increase in the N abundance and a decrease in the C abundance.
The differences in the indices C4668 and CN$_2$, however, require that both C
and N are in higher proportions in LDEGs with respect to the galaxies in high
density environments. A higher metallicity would also increase the amount of C
in the interstellar medium, due to more efficient mass loss  in massive stars
(Prantzos, Vangioni-Flam \& Chauveau 1994). This is caused by an increase in
the opacities of the stellar envelopes. Therefore, the more metal rich massive
stars would produce more C than the less massive ones through this mechanism
(these stars would also produce  less O and more He), which could compensate
for the decrease of C abundance in the hot bottom burning. However, the
metallicity of LDEG is, on average, higher than in HDEGs (see Paper II), which
is the opposite of what is needed to explain the observed variations between
galaxies in distinct environments through differences in the stellar yields.
Another effect which can change the atmospheric abundances in the stars is
mixing due to the rotation of the star (Norris 1981; Sweigart 1997). While it
does not have a big effect in the $^{4}$He, it may be important for  the
$^{12}$C (specially in low-mass stars) and $^{14}$N yields.  Following Meynet
\& Maeder (2002), rotation increases the size of the C and O core, due to an
increase of the mixing efficiency. The effect is particularly important at low
metallicities, where the angular velocity gradient is higher. In any case, it
is hard to imagine why 
the rotation of the stars should be different in galaxies inhabitating
environments of different galaxy density.

The second mechanism which could produce variations in the chemical abundances
of C and N between galaxies in different environments is a difference in their
star formation history. In particular, if we assumed that the main contributors
to the C and N are the intermediate- and low-mass stars, the timescale for the
release of these elements into the interstellar medium would be $\sim$
3$\times$10$^{7}$ years (this is, approximately, the lifetime of an
8 M$_{\odot}$ star). Therefore, if star formation in the HDEGs was shorter than
this, the  stars would not incorporate those elements. However, if the
timescale of the star formation in LDEGs were long enough for stars between 3
and 8 M$_{\odot}$ to complete their evolution until the AGB, and to  release
the products of their nucleosynthesis into the interstellar medium, the next
generation of stars would incorporate those elements, which would produce the
observed differences between both subsamples (LDEGs and HDEGs). These
differences in the star formation timescales were proposed in
Section \ref{sec.dispersion} to explain the trends between Z(Mgb)/Z(Fe4383) and
age. Note that the ages and metallicities that we are measuring are mean values
weighted with the luminosity of the stars. Therefore, it is not required that
all the stars of the LDEGs are formed more recently, since the differences
could be produced by a small percentage of stars formed in later bursts (see
Trager et al.\ 2000b).
   
\end{itemize} 

%%%%%%%%%%%%%%%%%%%%%%%%%%%%%%%%%%%%%%%%%%%%%%%%%%%%%%%%%%%%%%%%%%%%%%%%%%%%%%%
\section{The relation between   $[\alpha$/Fe$]$ and the  age}
\label{overabundance.age}

In Section \ref{sec.dispersion} we showed that the [Mg/Fe] ratio is related
with the age in such a way that older galaxies exhibit, on average, larger
[Mg/Fe] ratios than the younger ones.
 
Several different mechanisms have been proposed to explain the existence, in
giant elliptical galaxies, of an overabundance of [Mg/Fe] (see, for example,
Worthey, Faber \& Gonz\'alez 1992). All are based on the assumption that this
parameter quantifies the relative importance of the chemical enrichment from
Type II versus the delayed Type Ia supernovae.  We assess whether these
proposed mechanisms can explain the relation between the age and the relative
abundance of Mg with respect to  Fe.

\begin{itemize}
\item {\it Variations of the IMF:\/} a shallower slope of the IMF would lead to
a larger fraction of massive stars and, therefore, to a larger fraction of Type
II supernovae.  We could  explain the variation of [Mg/Fe] with age under this
scenario, assuming that the slope of the IMF increases over time.  Galaxies
that formed their stars earlier, therefore, would contain more massive stars
compared with galaxies which formed their stars in a more recent epoch.  The
evolution of the IMF with time was  proposed to explain the bimodal star
formation of our Galaxy (Schmidt 1963). Under this scenario, the first
generation of stars  enriched the interstellar medium very fast, out of which
other stars formed, with a mass distribution compatible with a much steeper
IMF.  The first author in proposing an IMF varying with time was Schmidt
(1963), but several authors have analysed this idea (Arnaud et al.\ 1992;
Worthey et al.\ 1992; Elbaz, Arnaud \& Vangioni-Flam 1995; Vazdekis
1999, among others).

To explore the possibility of a variation of the IMF slope with $\sigma$, we
made use of the Vazdekis et al.\ (2003) stellar population models.  In these
models, a new calibration of the CaT index in the near-infrared derived from a
new stellar library (Cenarro et al.\ 2002) was presented. These authors found
that this index has a high sensitivity to the slope of the IMF. They also
analysed a sample of galaxies with high quality observed spectra  (Cenarro et
al.\ 2003), proposing the existence of a variation of the IMF with metallicity,
in the sense that the larger the metallicity the larger the IMF slope.

This dependence of the IMF on metallicity would give rise to the observed
tendency of [Mg/Fe] with age if there were a relation between age and
metallicity, i.e. younger galaxies were also more metal rich. This relation has
been observed in several studies (Trager et al.\ 2000b; K01
among others), although it is difficult to separate the real trend from the one
caused by the correlated errors in both parameters (see, e.g.\ K01). 
A discussion of this relation is presented in Paper II. 

Figure \ref{catriplet} shows the relation between the slope of the IMF derived
from the relation given in Cenarro et al.\ (2003) versus the Z(Mgb)/Z(Fe4383)
ratio (the quotient between the metallicity measured separately with both Mgb
and Fe4383 combined with H$\beta$, which can be used as an estimation of behaviour
(not the numerical value) of the [Mg/Fe] ratio (see Sec. \ref{sec.dispersion}).  
As can be seen, the slope of the IMF
obtained in this way is higher for galaxies with larger [Mg/Fe] ratios, which
is the opposite to the expected trend.  Therefore, although we do not discard
the possibility of differences in the IMF between galaxies, these cannot be the
responsible for the relation between [Mg/Fe] abundance and  age. 

\begin{figure}
\centering
\resizebox{\hsize}{!}{\includegraphics[angle=-90]{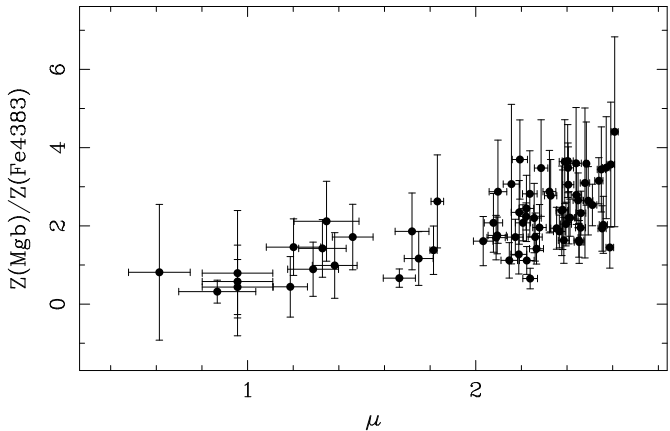}}
\caption{Slope of the IMF ($\mu$) derived from the  relation between this
parameter and $\sigma$ calculated by Cenarro et al. (2003) versus the quotient
between the metallicities determined from Mgb and Fe4383.\label{catriplet}}
\end{figure}
 
\item {\it Selective loss mechanisms:\/} the gas losses due to supernova winds
in the earlier phases of galaxy formation is another mechanism proposed to
explain the overabundances of [Mg/Fe] in elliptical galaxies (see, for example,
Worthey et al.\ 1992). In order to produce an  overabundance of [Mg/Fe], the
mechanism has to be selective, so Mg has to be retained with a higher
efficiency than Fe.  Powerful starbursts in giant ellipticals could drive out
all previously accumulated Fe-rich gas. The next generations of stars would
then be preferentially Mg-rich.  In any case, there is no reason to suggest
that gas loss should be more efficient in older galaxies than in younger ones. 

\item {\it Differences in the star formation history of the galaxies:\/} the
last possibility is that the differences in the chemical abundance ratios are
due to variations in the star formation histories. Under this scenario,
primeval, genuinely old, ellipticals preserve the $\alpha$-element
overabundances attained at their formation epoch (which should in turn be
mainly controlled by the depth of the potential well). On the other hand,
galaxies which have experienced subsequent star formation episodes (that is,
those that exhibit younger mean ages) have been able to incorporate the later
created Fe into their stars, lowering their [Mg/Fe] ratio. This scenario
explains naturally the relation between [Mg/Fe] abundances and the age of the
galaxies and therefore is our preferred one. Under this scenario, the
differences in the Mg/Fe ratios between LDEG and HDEG are due to differences in
the star formation history, where galaxies in dense clusters have suffered a
truncation of their star formation at early epochs. This was already proposed
to explain the differences in other chemical species by S\'anchez-Bl\'azquez et
al.\ (2003).  Under  this scenario we would expect to find an age--metallicity
relationship in which the younger galaxies should  also be more metal rich. We
will analyse this relation in Paper II.
\end{itemize}

%%%%%%%%%%%%%%%%%%%%%%%%%%%%%%%%%%%%%%%%%%%%%%%%%%%%%%%%%%%%%%%%%%%%%%%%%%%%%%%
\section{Conclusions}
\label{conclusions}

In this paper we have investigated the relations between line-strength indices
and velocity dispersion with the aim of understanding their origin, the causes
of the scatter, and the influence of environment. The main results can be
summarised as follows:
\begin{enumerate}
 \item The slopes in the index--$\sigma$ relations are mostly due to an
 increase in metallicity with the velocity dispersion of the galaxies. However,
 different chemical elements do not vary in lockstep along the $\sigma$
 sequence.  In particular $\alpha$ elements vary more than Fe-peak elements. On
 the other hand, the necessary variations of N and Mg to explain the slope in
 the relations are even larger than those required by the rest of $\alpha$
 elements.

 \item To explain the slope of the relation of  H$\beta$ with the velocity
 dispersion obtained for the LDEGs, a variation of the age with the velocity
 dispersion is required in the sense that low-$\sigma$ galaxies have to be also
 younger. This age variation is not required to explain the slope of the HDEGs. 

 \item Studying the residuals of the relations of the indices with the velocity
 dispersion, we conclude that they are due to a variation of the relative
 abundance of [Mg/Fe] with the age of the galaxies, in the sense that older
 galaxies show, on average, a higher content of Mg with respect to Fe. This
 relation is interpreted in terms of different histories of star formation.  If
 the younger galaxies have had a more extended star formation history, the low
 mass stars have had time to release the products of their nucleosynthesis into
 the interstellar medium out of which new stars are formed. 

 \item We have detected differences in some indices between galaxies belonging
 to distinct environments. These variations are likely a consequence of
 changes in the overall metallicity, plus differences in the abundances
 ratios of Ti, Ca, N, and probably C, between galaxies in different
 environments.
\end{enumerate}

All the differences between galaxies in distinct environments can be explained
under a common scenario in which galaxies in dense clusters have suffered a
more truncated star formation history than their counterparts in low density
environments. We will analyse this in more detail in the second paper of the
series, where we will derive ages and metallicities by comparing the
line-strength indices with synthesis stellar population models. 

%%%%%%%%%%%%%%%%%%%%%%%%%%%%%%%%%%%%%%%%%%%%%%%%%%%%%%%%%%%%%%%%%%%%%%%%%%%%%%%
%%%%%%%%%%%%%%%%%%%%%%%%%%%%%%%%%%%%%%%%%%%%%%%%%%%%%%%%%%%%%%%%%%%%%%%%%%%%%%%
\begin{acknowledgements}
We would like to thank Daisuke Kawata and Christopher Thom for the careful
reading of the manuscript and for their useful comments. We are also grateful
to Javier Cenarro, Reynier Peletier and Alexandre Vazdekis for fruitful discussions.  
We are very grateful to the referee, Jim Rose, for his very constructive report and many
useful suggestions. Finally, PSB would like to thank Brad Gibson for his continued support.
The WHT is
operated on the island of La Palma by the Royal Greenwich Observatory at the
Observatorio del Roque de los Muchachos of the Instituto de Astrof\'{\i}sica de
Canarias. The Calar Alto Observatory is operated jointly by the Max-Planck
Institute f\"{u}r Astronomie, Heidelberg, and the Spanish Instituto de
Astrof\'{\i}sica de Andaluc\'{\i}a (CSIC).  
This research has made use of the SIMBAD database operated at CDS, Strasbourg, France.
This work was supported by the
Spanish research project AYA 2003-01840, the European Community through its
NOVA fellowship program, and the Australian Research Council. We are grateful
to the CATs for generous allocation of telescope time.
\end{acknowledgements}

%%%%%%%%%%%%%%%%%%%%%%%%%%%%%%%%%%%%%%%%%%%%%%%%%%%%%%%%%%%%%%%%%%%%%%%%%%%%%%%
\appendix

\section{Comparison with the Lick library}
\label{appen.broad-lick}

\begin{figure*}
\centering
\resizebox{0.23\hsize}{!}{\includegraphics[angle=-90]{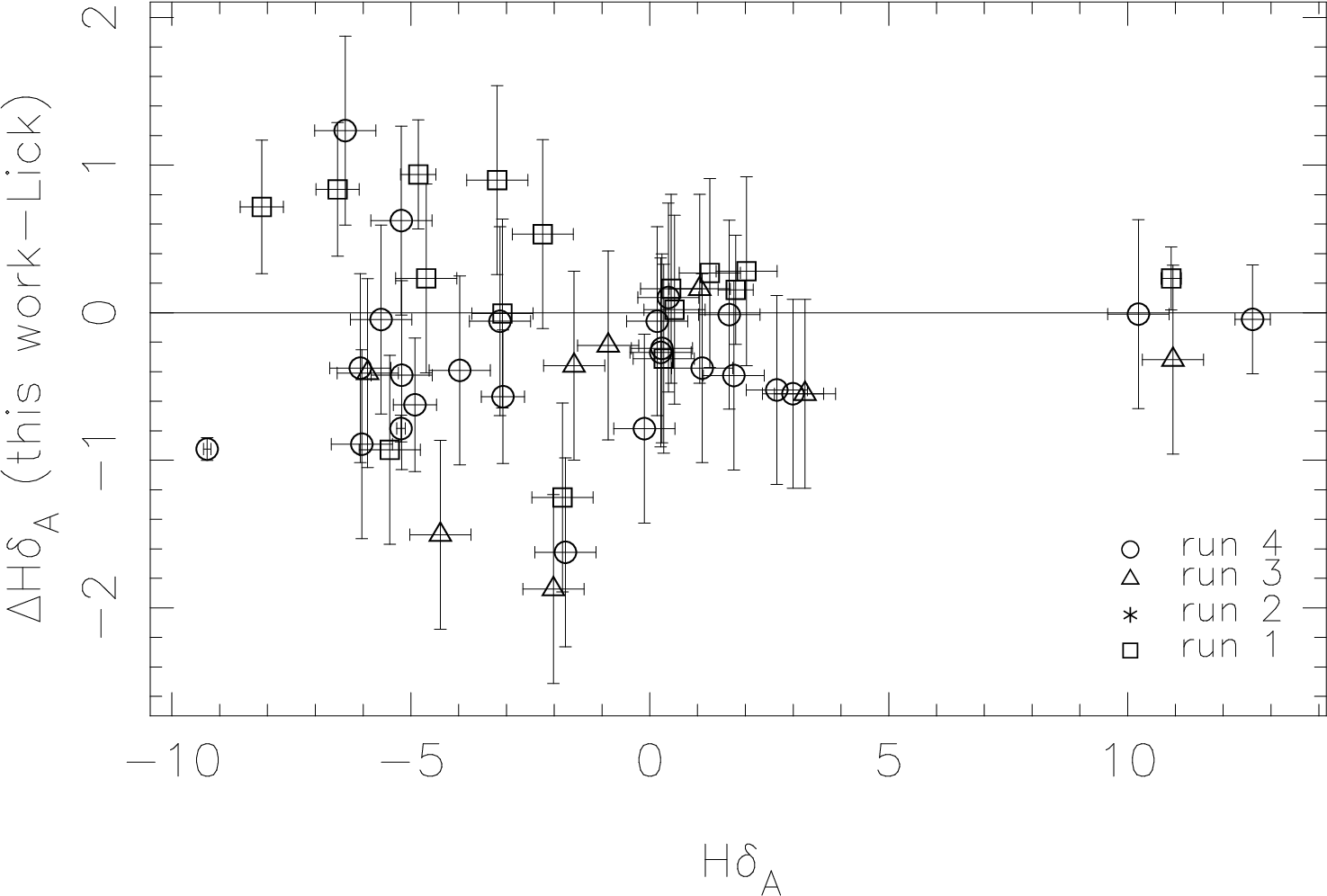}}
\hfill
\resizebox{0.23\hsize}{!}{\includegraphics[angle=-90]{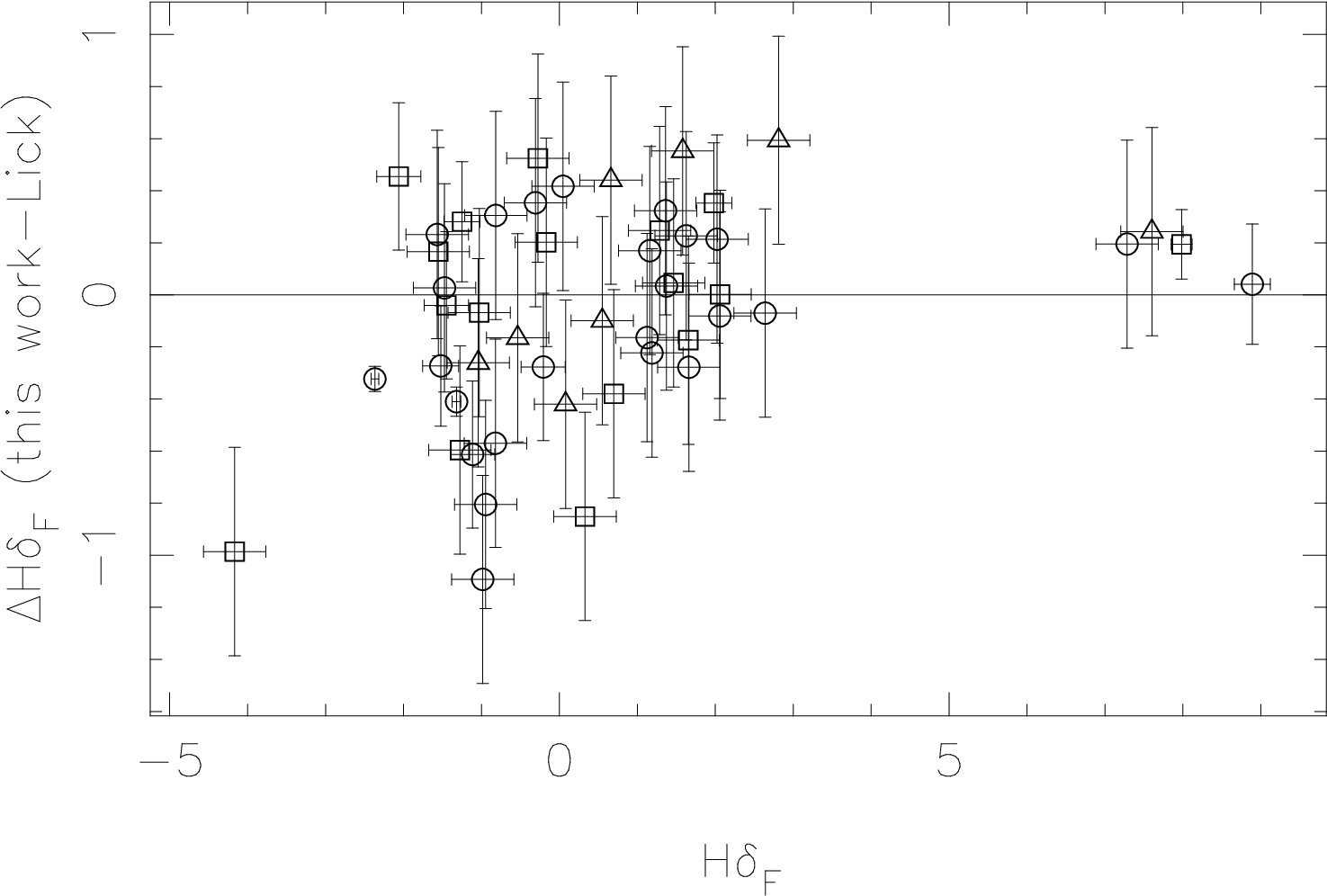}}
\hfill
\resizebox{0.23\hsize}{!}{\includegraphics[angle=-90]{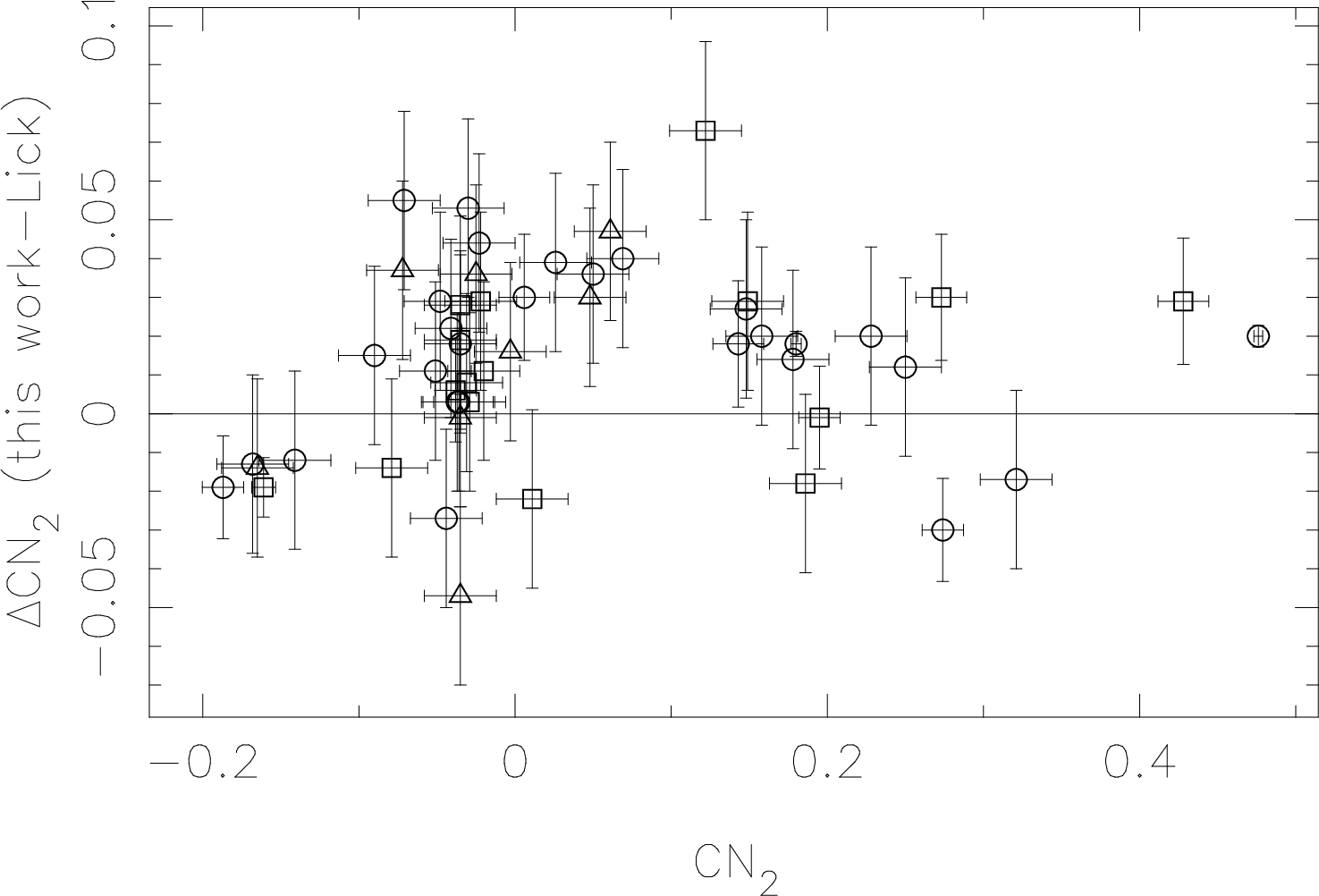}}
\hfill
\resizebox{0.23\hsize}{!}{\includegraphics[angle=-90]{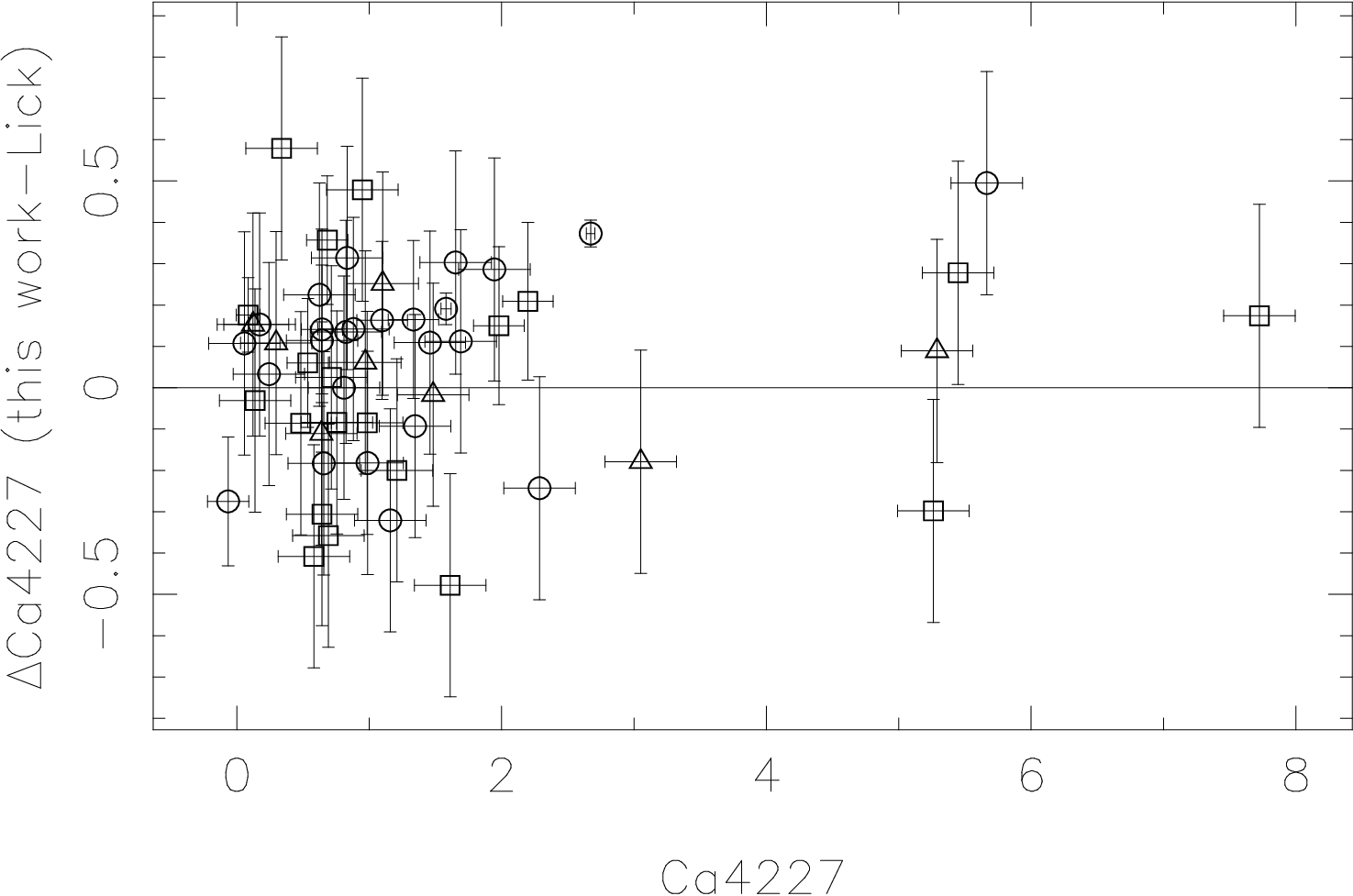}}
\vspace{3mm}

\resizebox{0.23\hsize}{!}{\includegraphics[angle=-90]{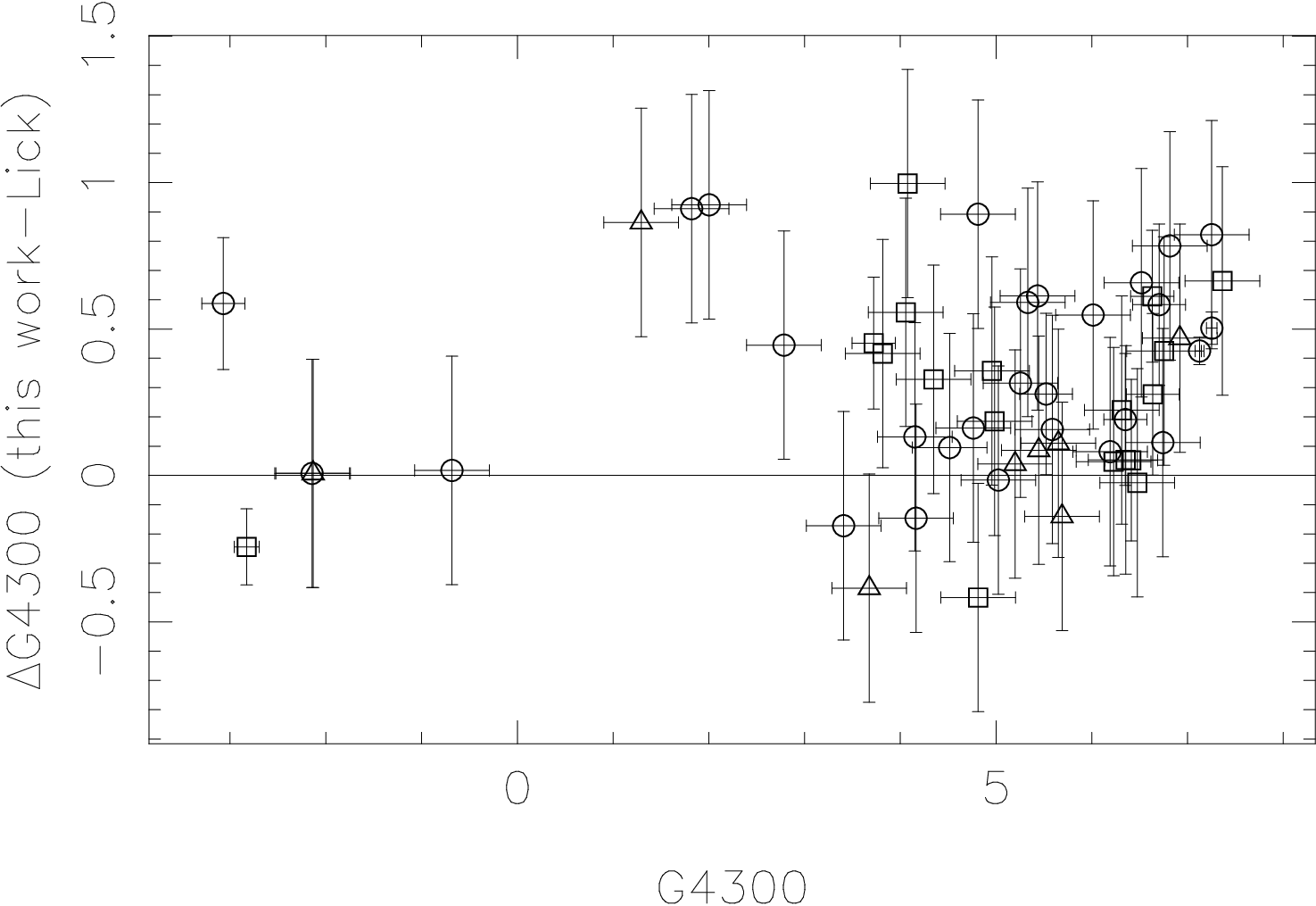}}
\hfill
\resizebox{0.23\hsize}{!}{\includegraphics[angle=-90]{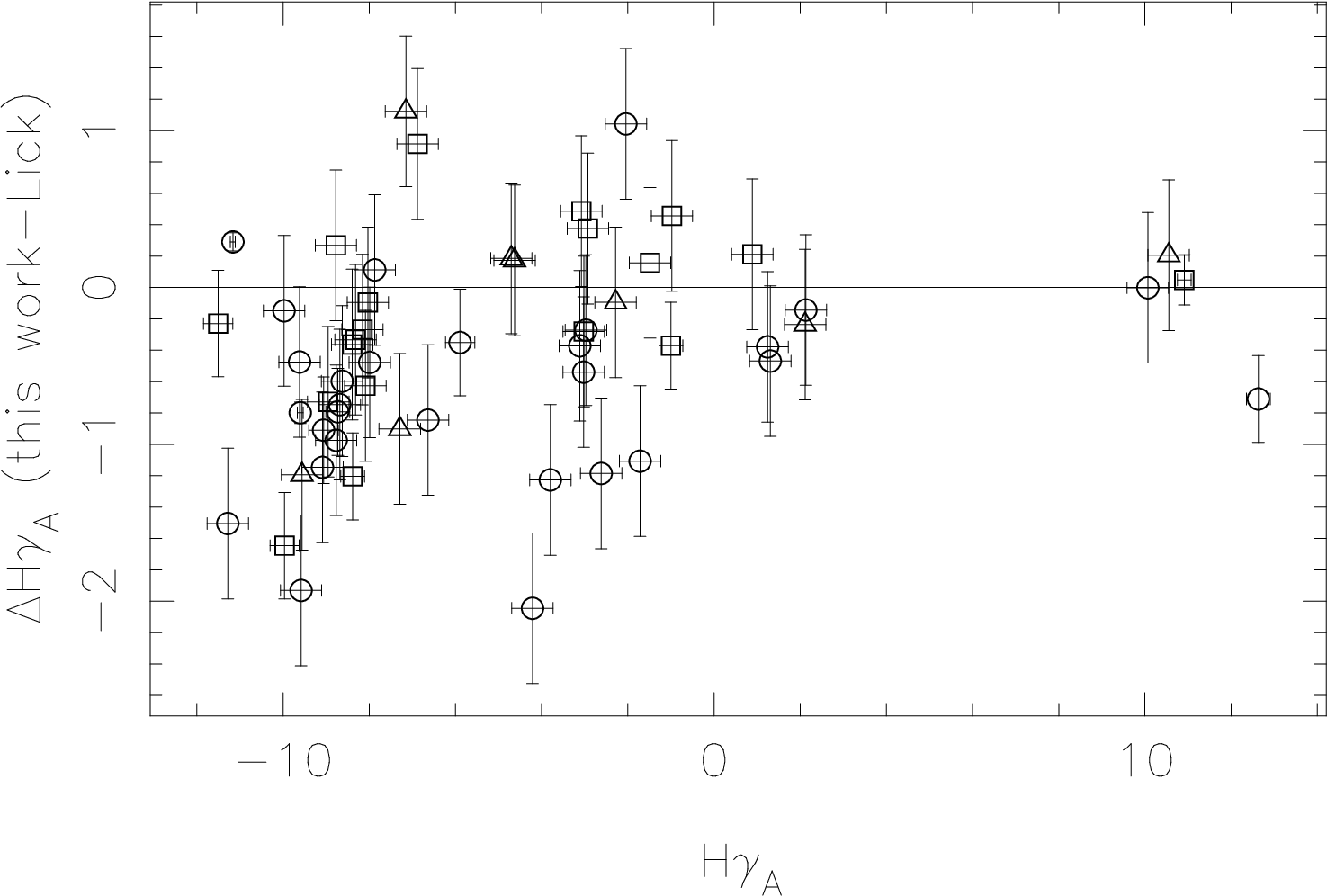}}
\hfill
\resizebox{0.23\hsize}{!}{\includegraphics[angle=-90]{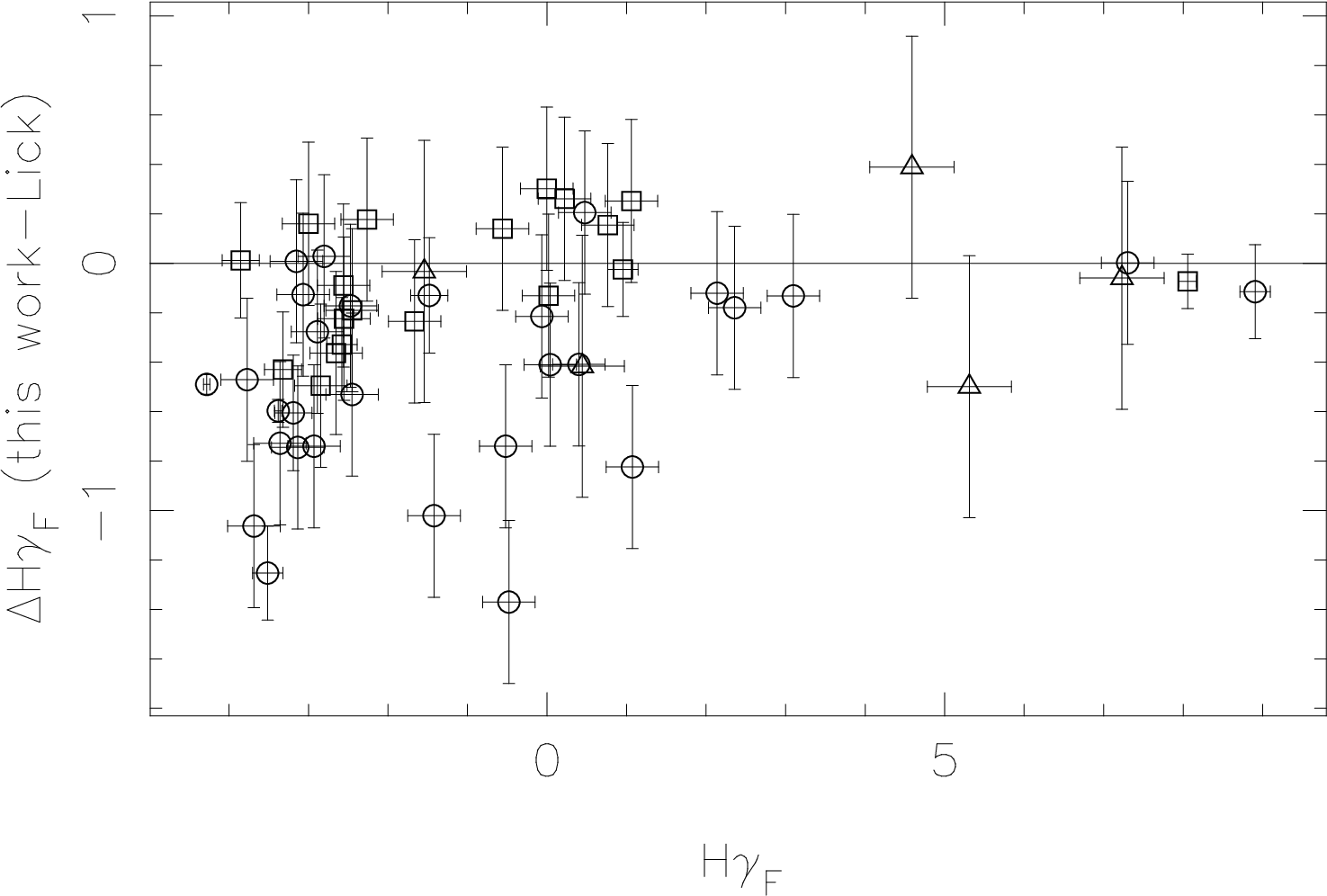}}
\hfill
\resizebox{0.23\hsize}{!}{\includegraphics[angle=-90]{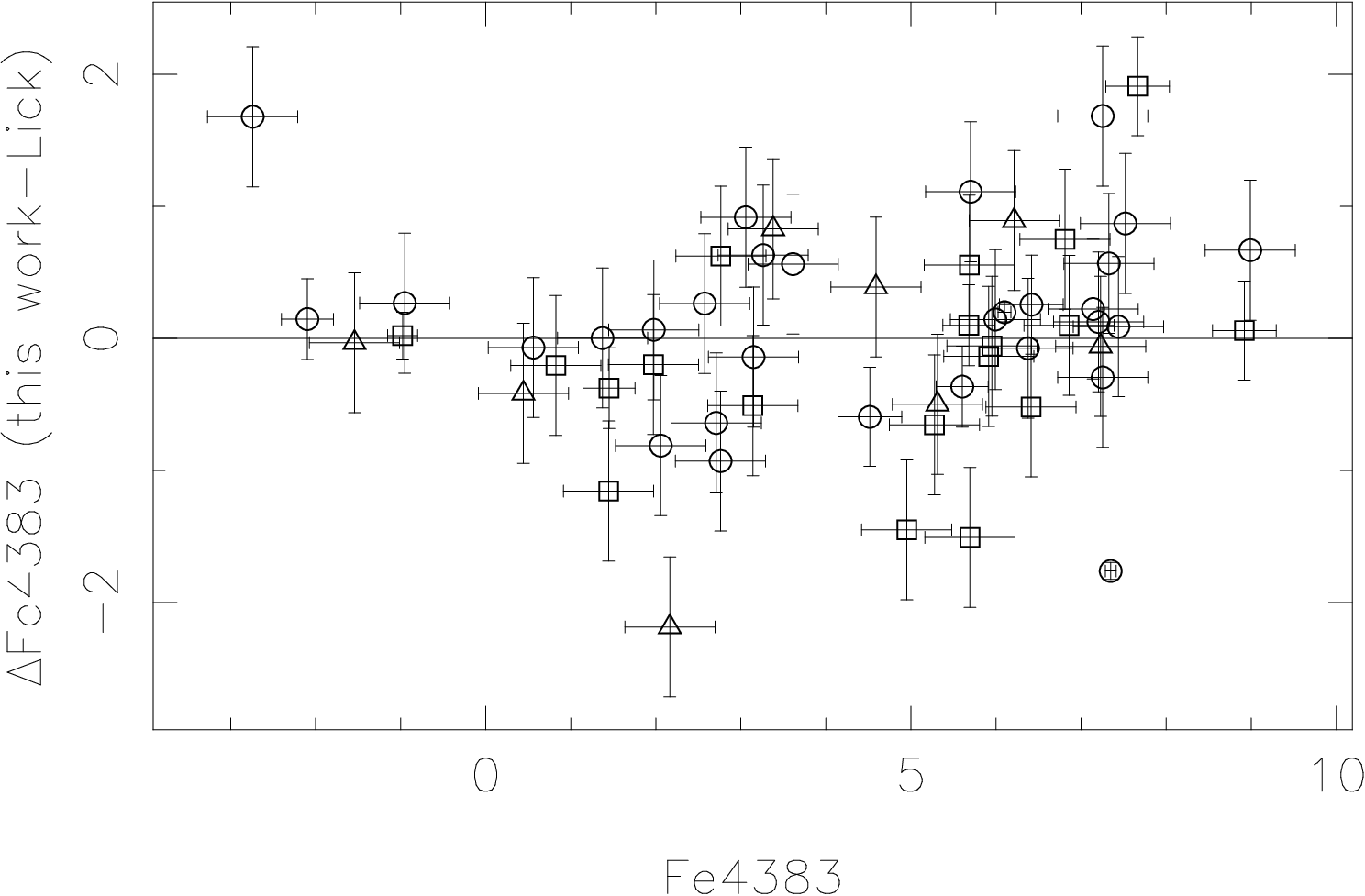}}

\vspace{3mm}

\resizebox{0.23\hsize}{!}{\includegraphics[angle=-90]{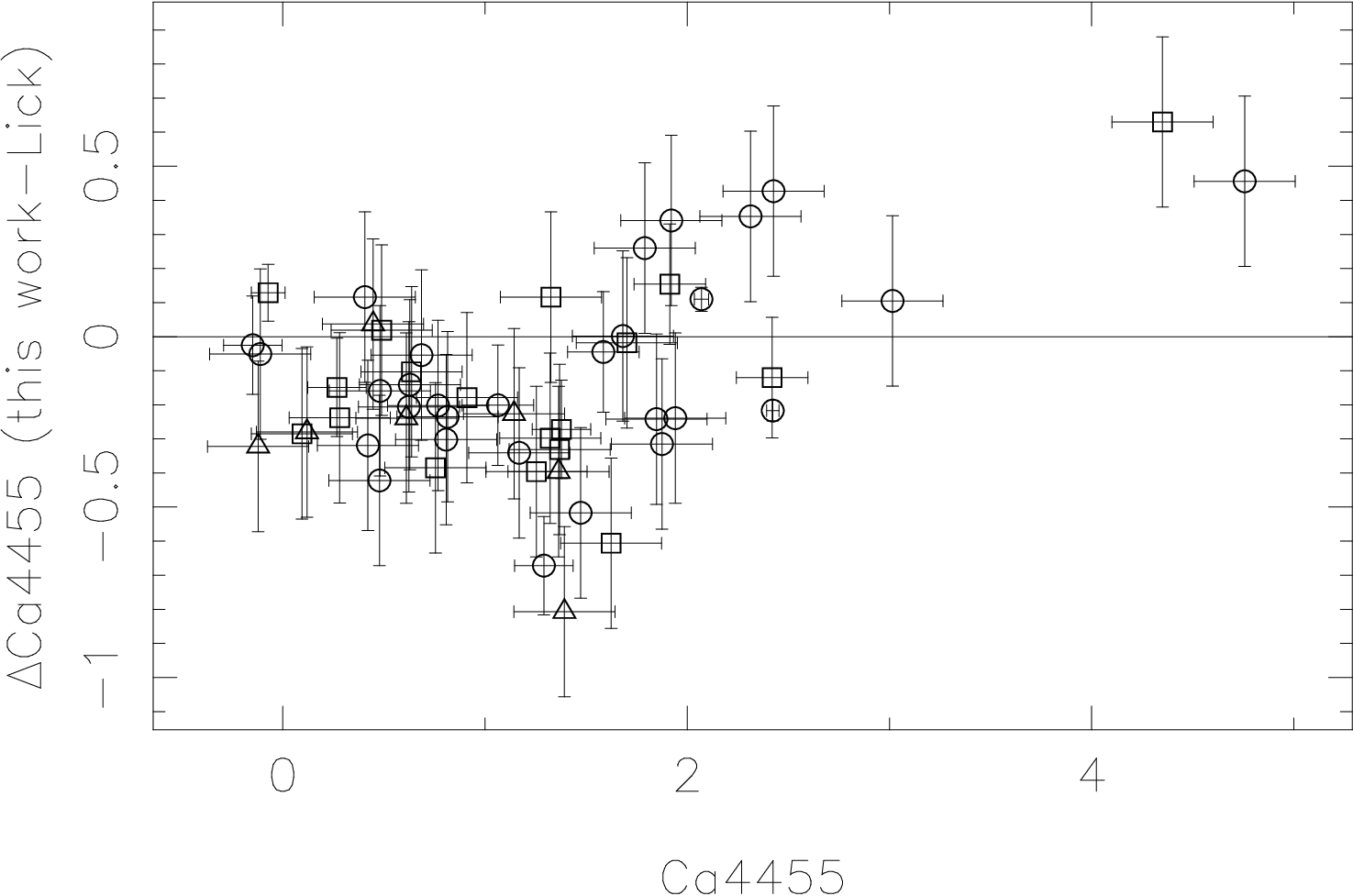}}
\hfill
\resizebox{0.23\hsize}{!}{\includegraphics[angle=-90]{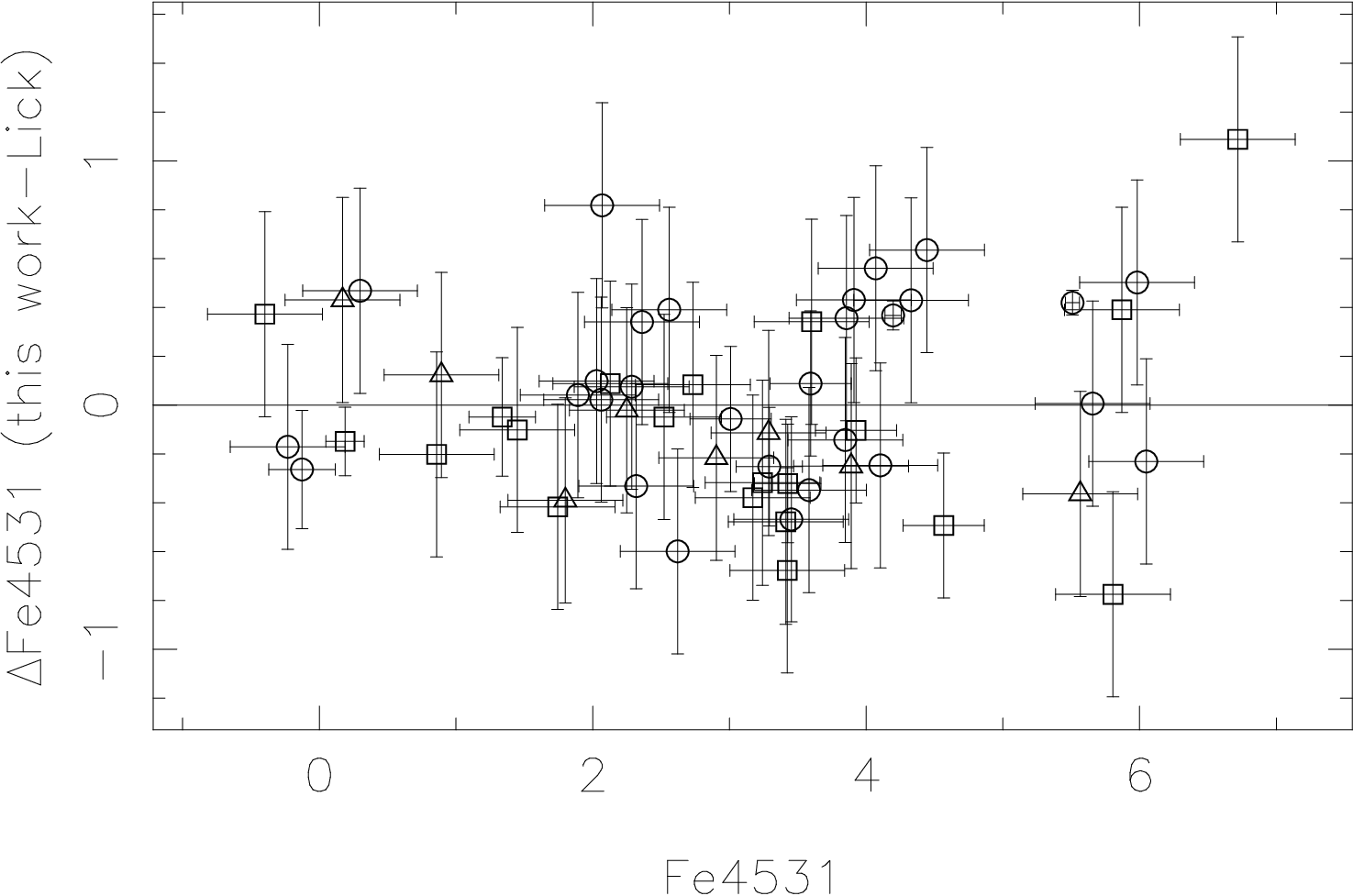}}
\hfill
\resizebox{0.23\hsize}{!}{\includegraphics[angle=-90]{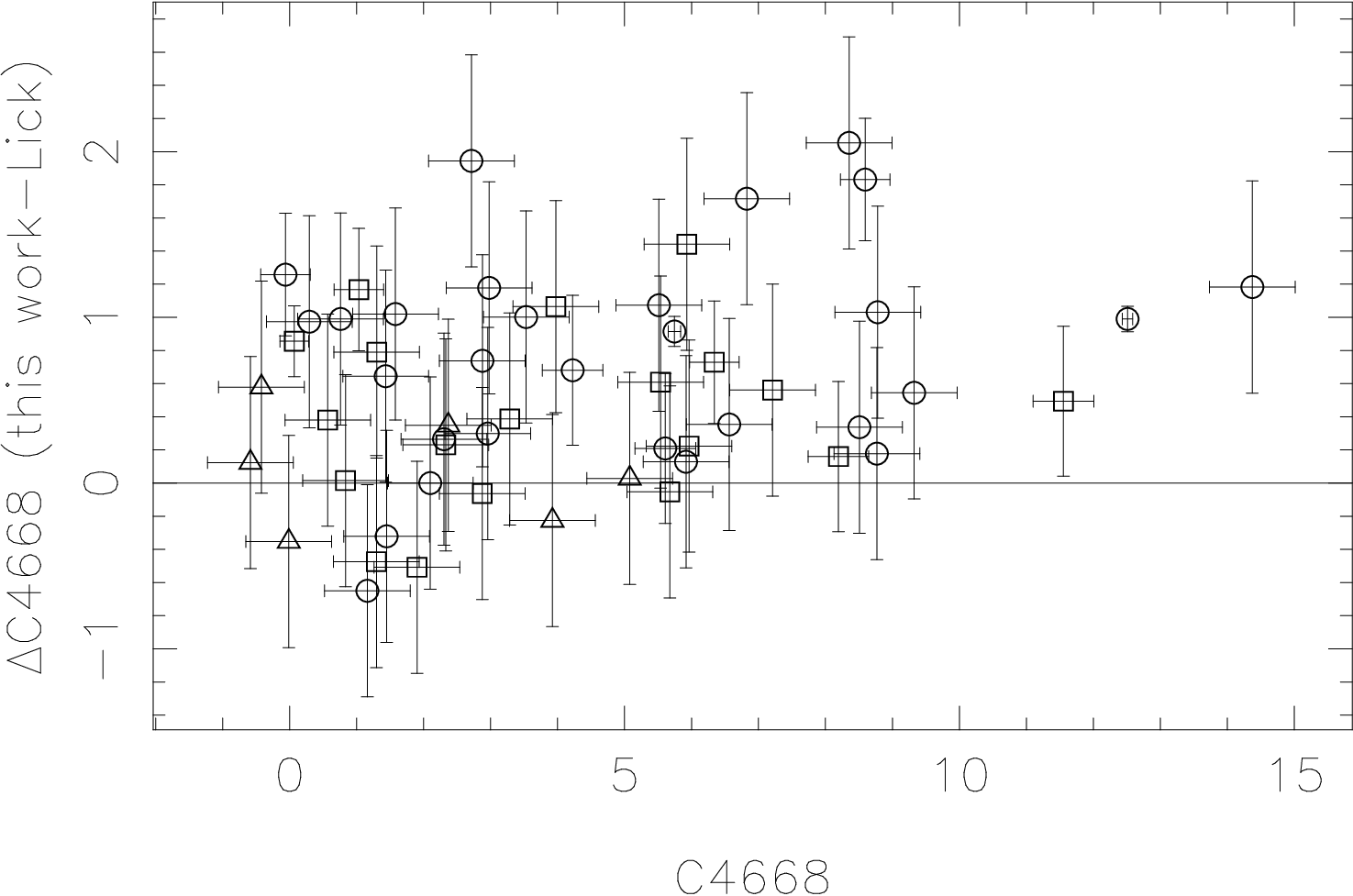}}
\hfill
\resizebox{0.23\hsize}{!}{\includegraphics[angle=-90]{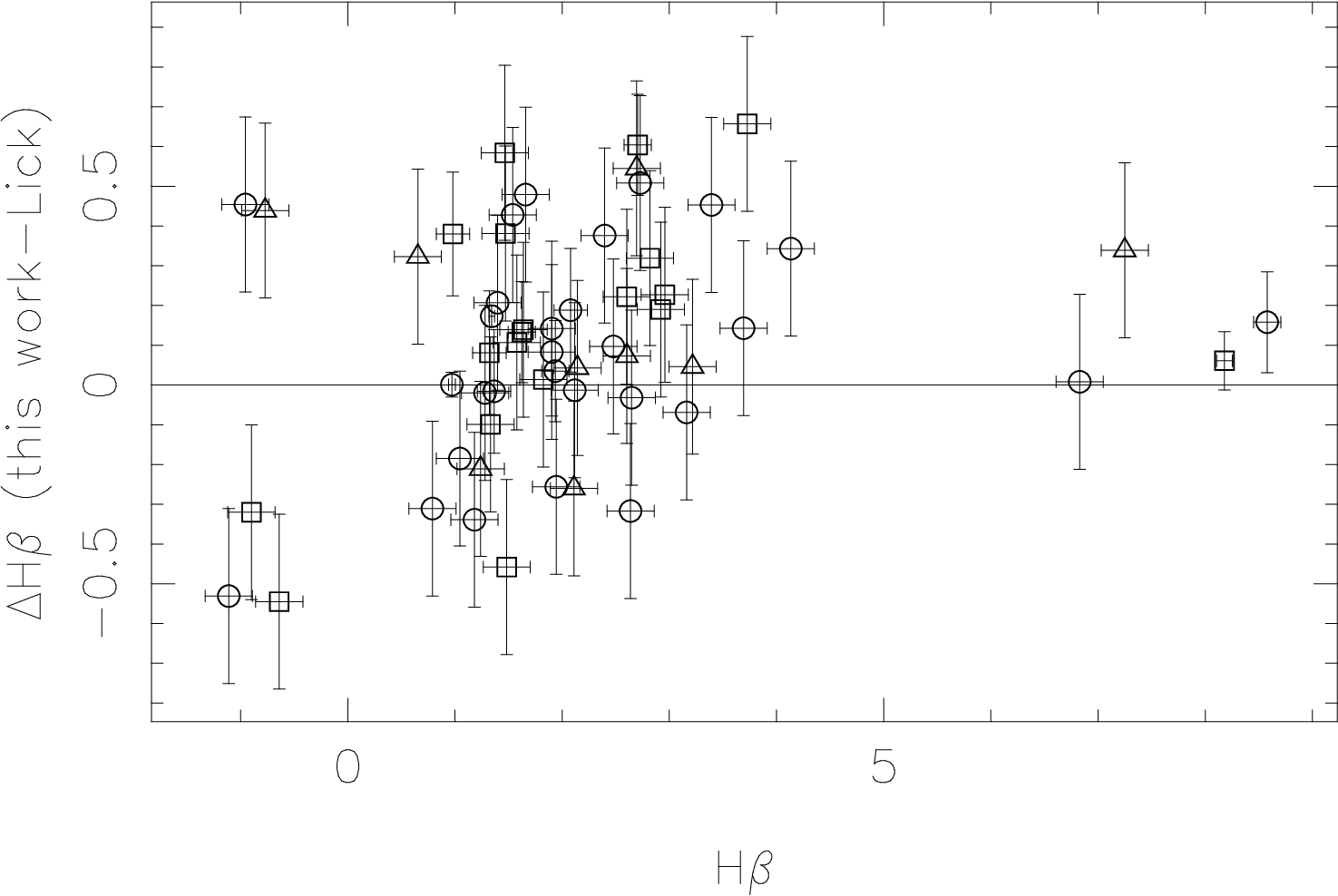}}

\vspace{3mm}

\resizebox{0.23\hsize}{!}{\includegraphics[angle=-90]{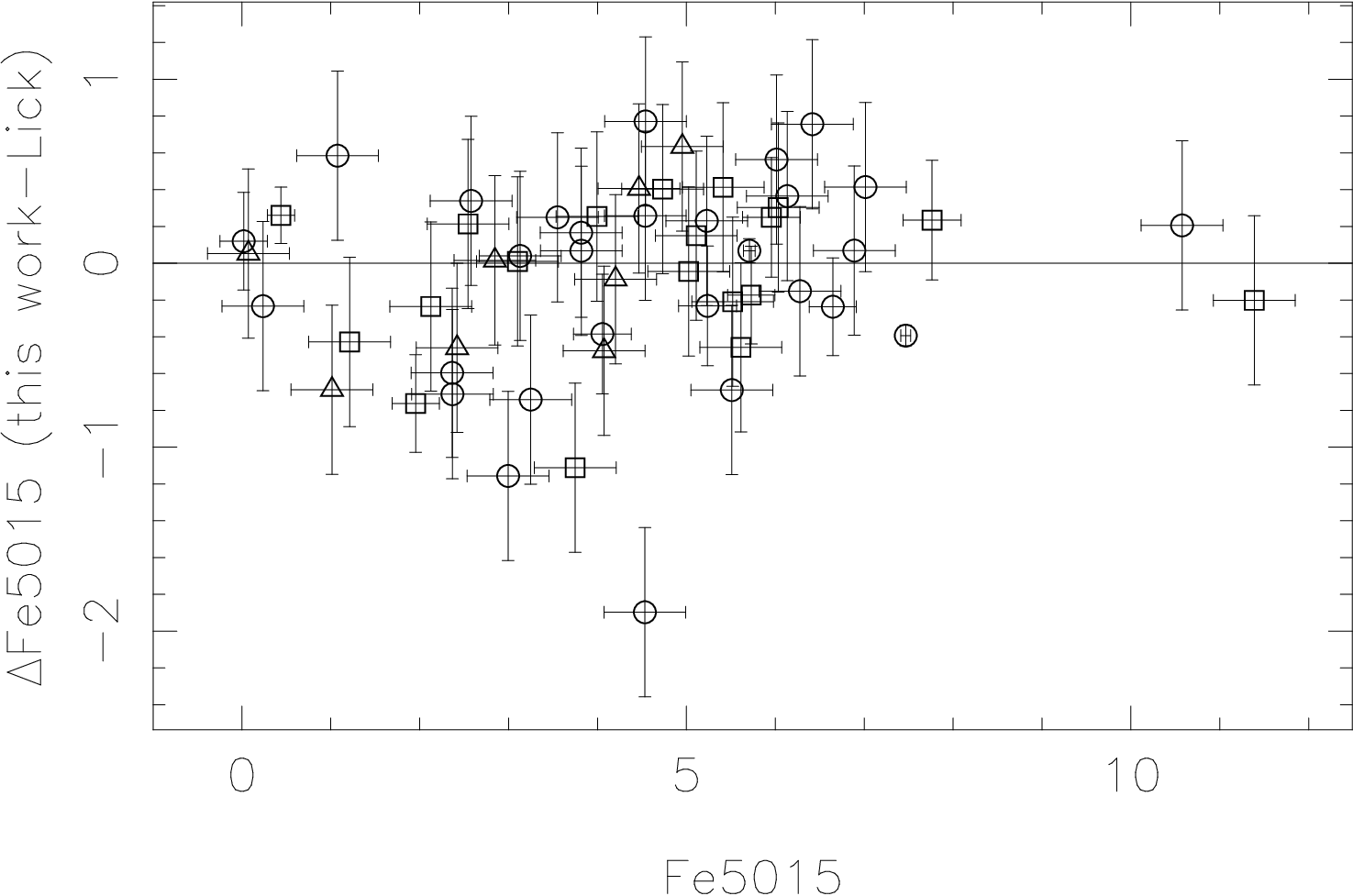}}
\hfill
\resizebox{0.23\hsize}{!}{\includegraphics[angle=-90]{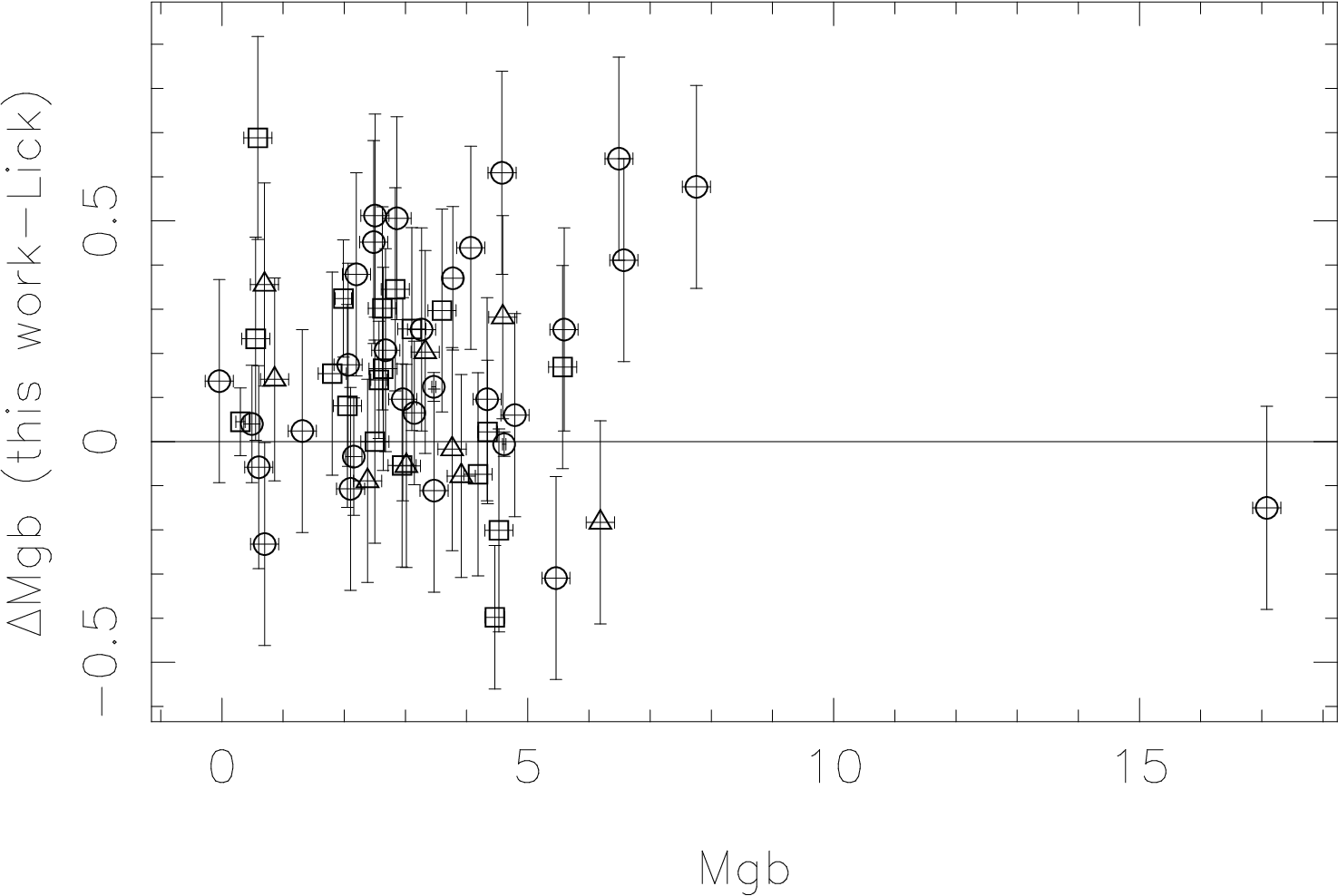}} 
\hfill
\resizebox{0.23\hsize}{!}{\includegraphics[angle=-90]{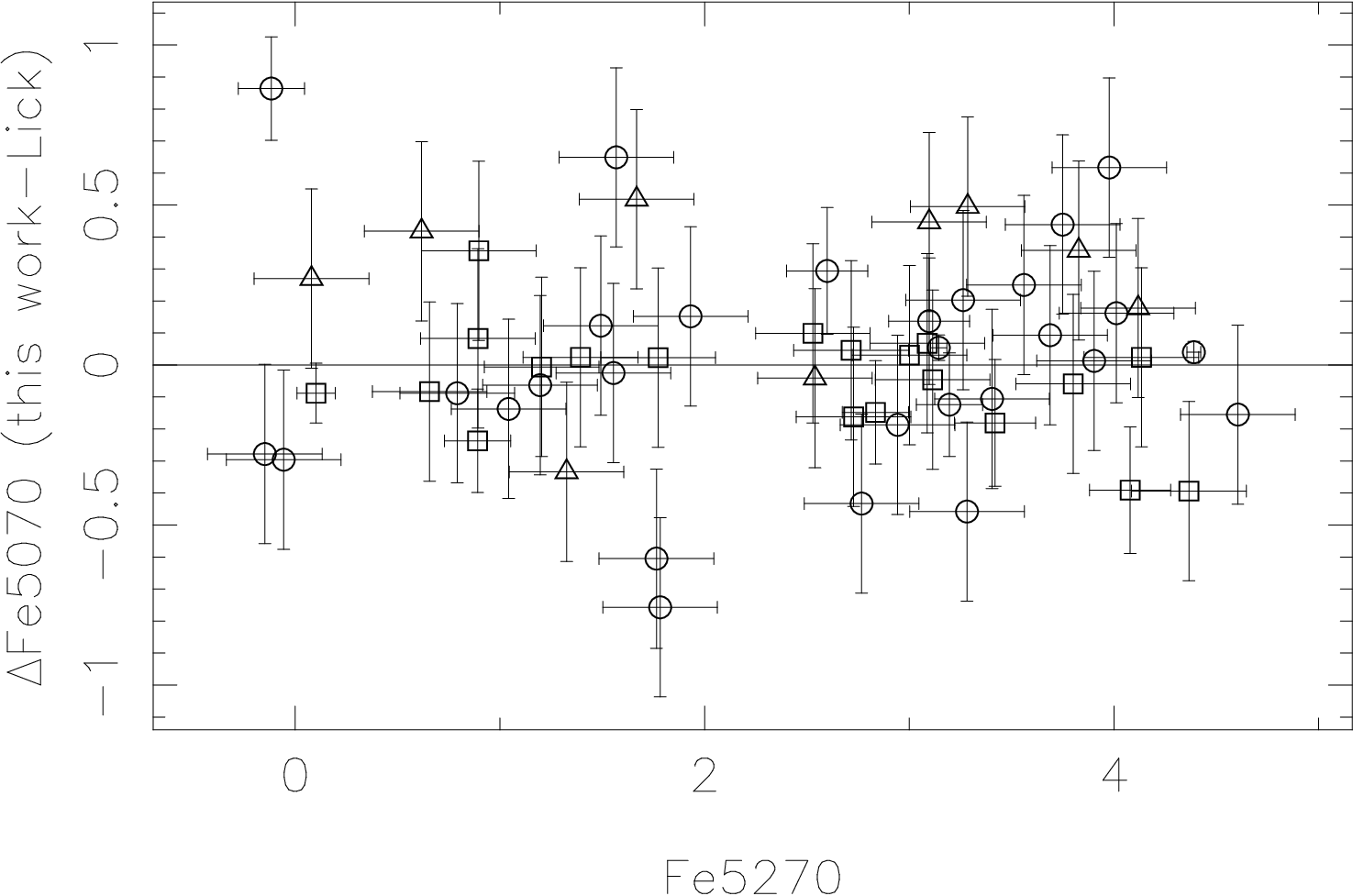}}
\hfill
\resizebox{0.23\hsize}{!}{\includegraphics[angle=-90]{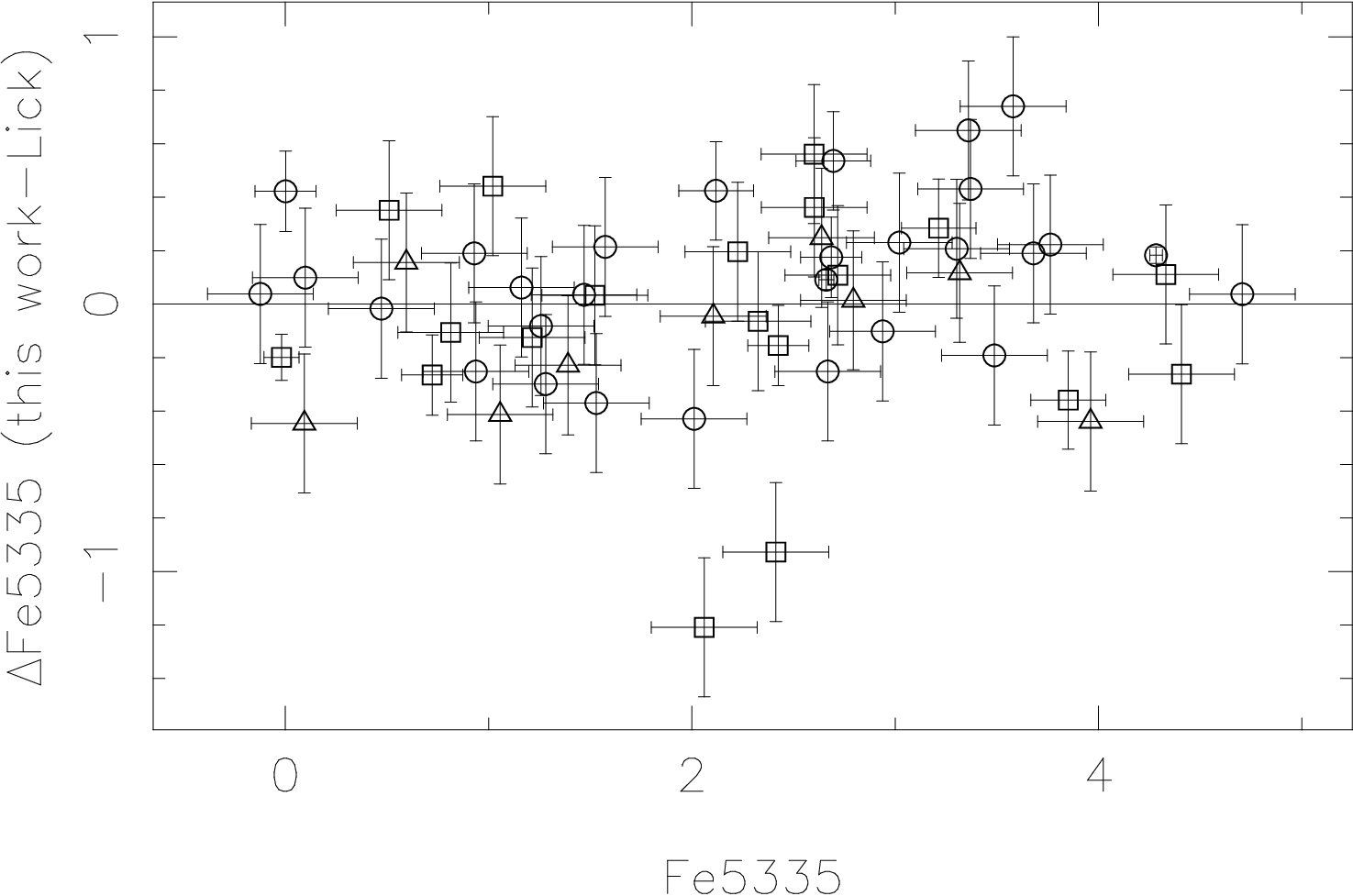}} 
\caption{Differences between the indices measured in the Lick stars and our own
measurements as a function of the later. Different symbols represent the
data obtained in the 4 observing runs; run1: squares; run2: stars; run3: triangles, 
and run4: circles.
\label{lick1}}
\end{figure*}

Figure \ref{lick1} shows, for the stars observed in the different runs, the
comparison between the original Lick/IDS index measurements and the indices
determined from our data. Table \ref{tab-comp.lick} shows the offsets obtained
in the different observing periods and the final offset adopted in this paper.

\begin{table}
\caption{\small Final Lick/IDS offsets (Lick/IDS $-$ This Work).\label{tab-comp.lick}} 
\centering
\begin{tabular}{@{}l| r@{}}
\hline\hline
Index & \multicolumn{1}{c}{Offset (Lick$-$This work)}\\
\hline
   H$_{\delta A}$  & $ 0.000\pm 0.227$ \AA\\
   H$_{\delta F}$  & $ 0.000\pm 0.130$ \AA\\
   CN$_{2}$        & $-0.014\pm 0.011$ mag\\
   Ca4227          & $-0.193\pm 0.117$ \AA\\
   G4300           & $-0.346\pm 0.160$ \AA\\
   H$_{\gamma A}$  & $ 0.568\pm 0.229$ \AA\\
   H$_{\gamma F}$  & $ 0.451\pm 0.116$ \AA \\
   Fe4383          & $ 0.000\pm 0.303$ \AA\\
   Ca4455          & $ 0.201\pm 0.132$ \AA \\
   Fe4531          & $ 0.000\pm 0.128$ \AA\\
   C4668           & $-0.682\pm 0.250$ \AA\\
   H$_{\beta}$     & $-0.104\pm 0.097$ \AA\\
   Fe5015          & $ 0.000\pm 0.162$ \AA\\
   Mgb             & $-0.157\pm 0.081$ \AA\\
   Fe5270          & $ 0.000\pm 0.075$ \AA\\
   Fe5335          & $ 0.000\pm 0.150$ \AA\\
\hline
\end{tabular}
\end{table}

\section{Comparison of stars in different runs}
\label{double.check}
\begin{figure*}
\resizebox{0.23\hsize}{!}{\includegraphics[angle=-90]{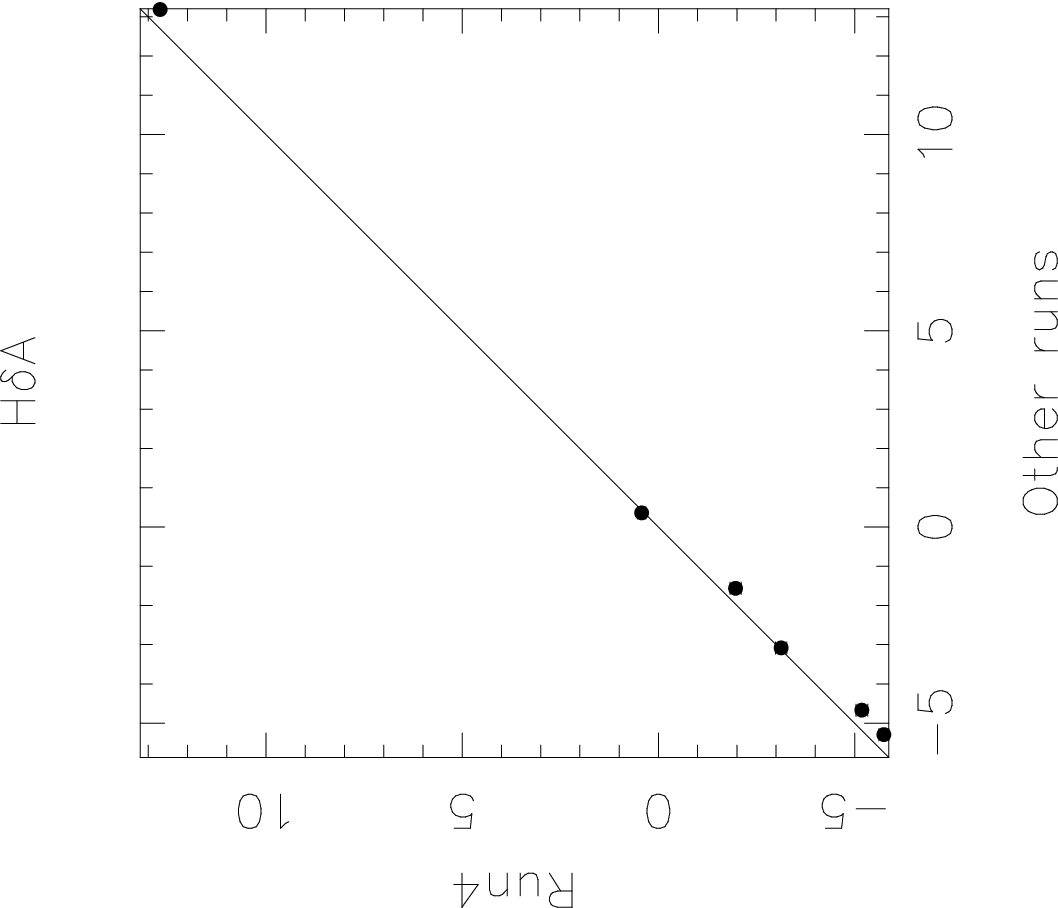}}
\hfill
\resizebox{0.23\hsize}{!}{\includegraphics[angle=-90]{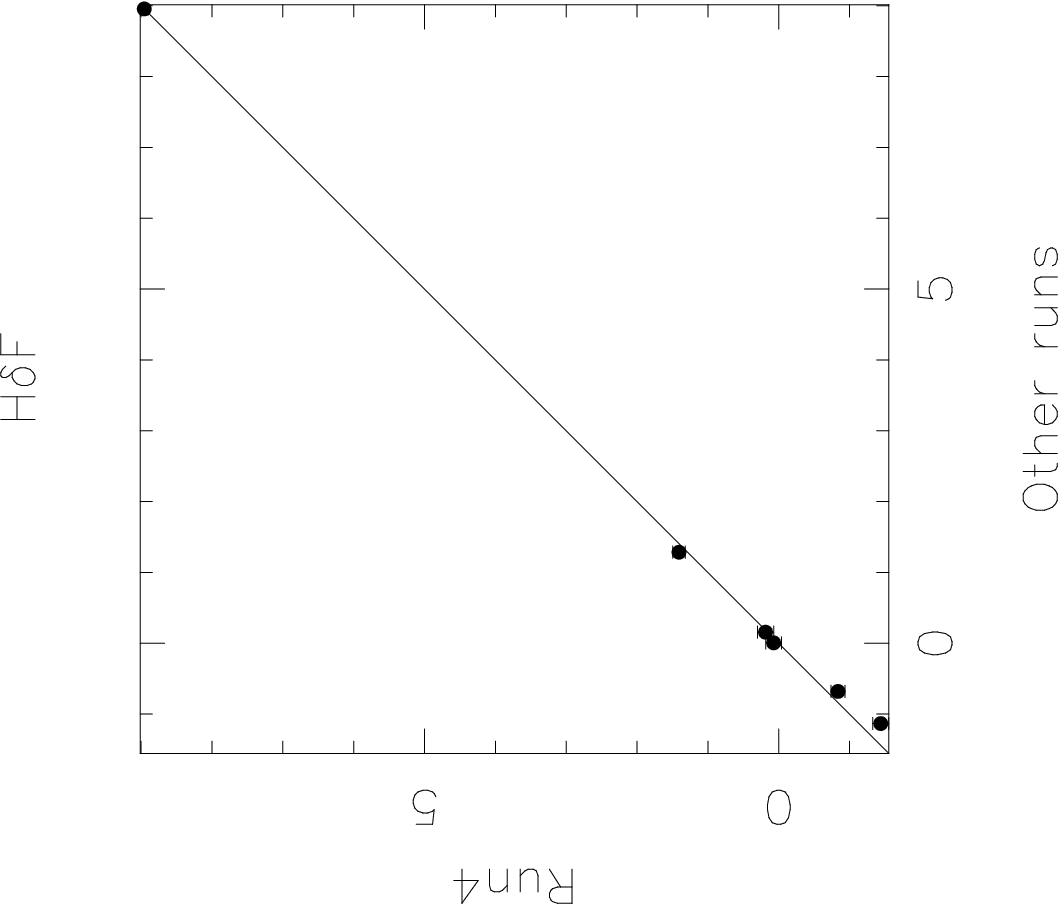}}    
\hfill
\resizebox{0.23\hsize}{!}{\includegraphics[angle=-90]{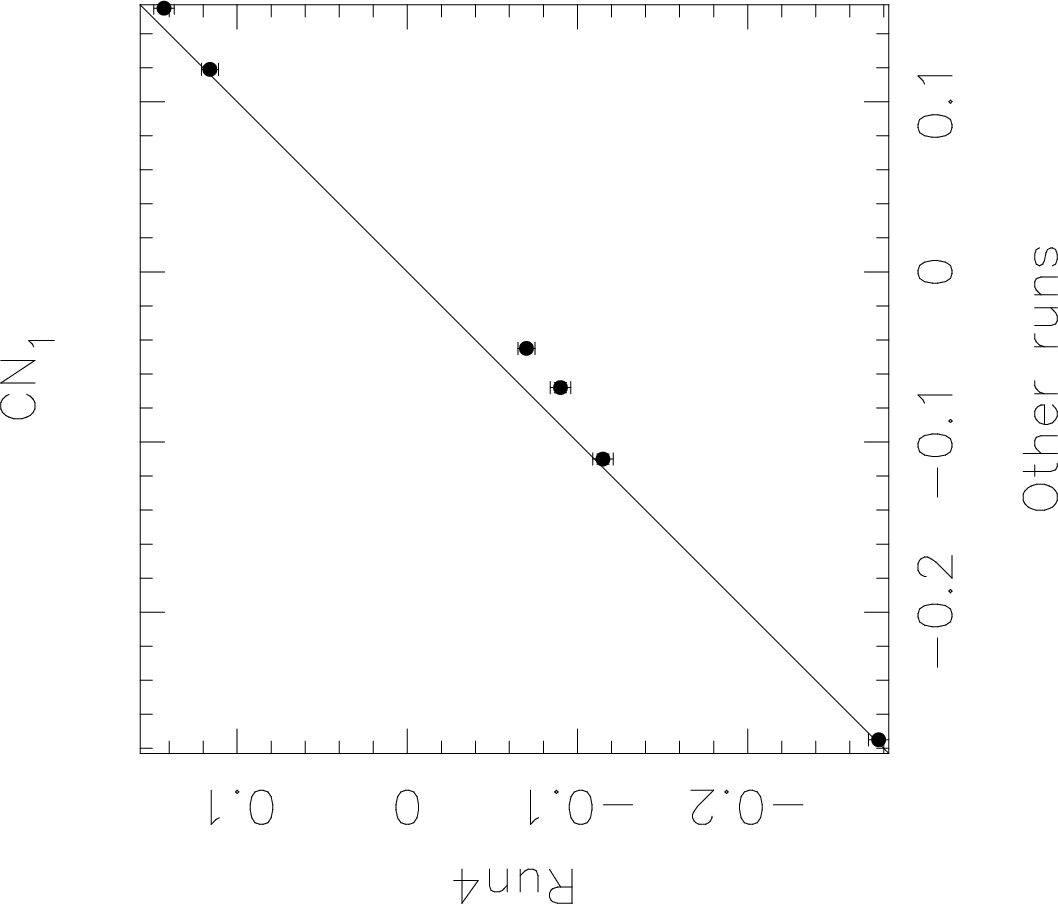}}    
\hfill
\resizebox{0.23\hsize}{!}{\includegraphics[angle=-90]{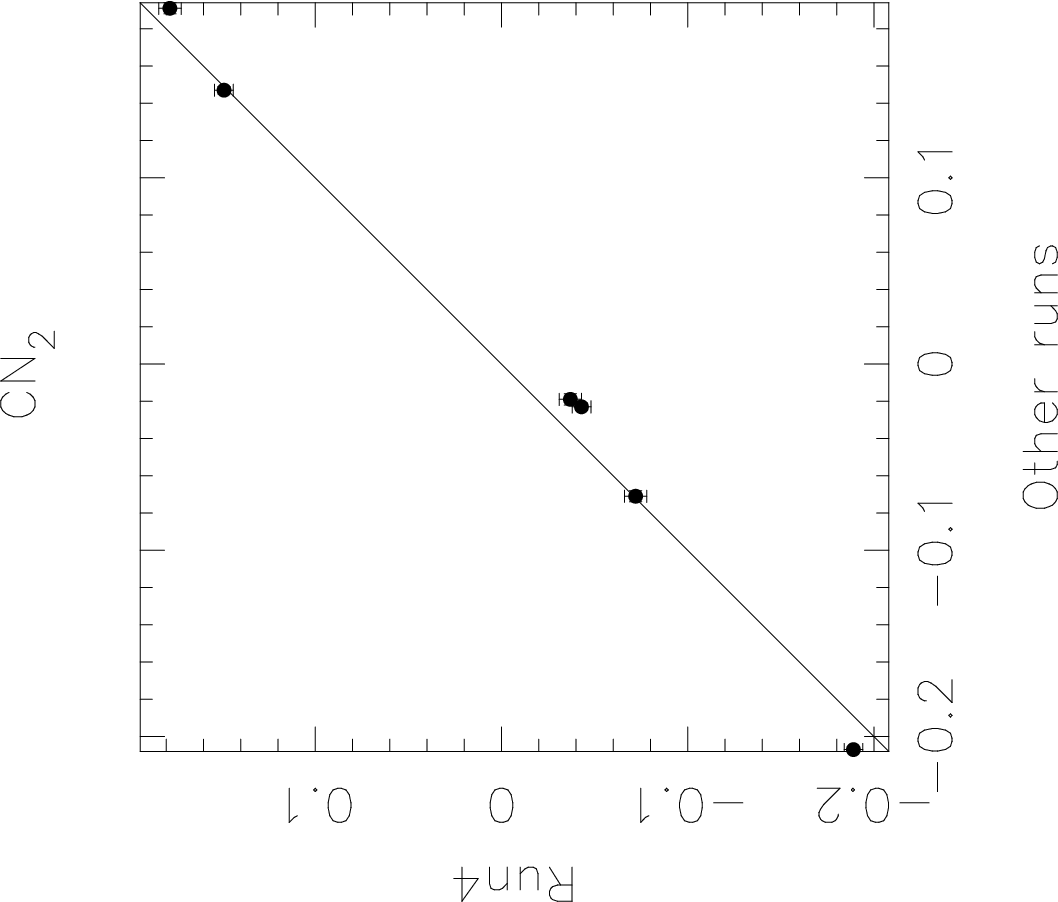}}   

\vspace{4mm}

\resizebox{0.23\hsize}{!}{\includegraphics[angle=-90]{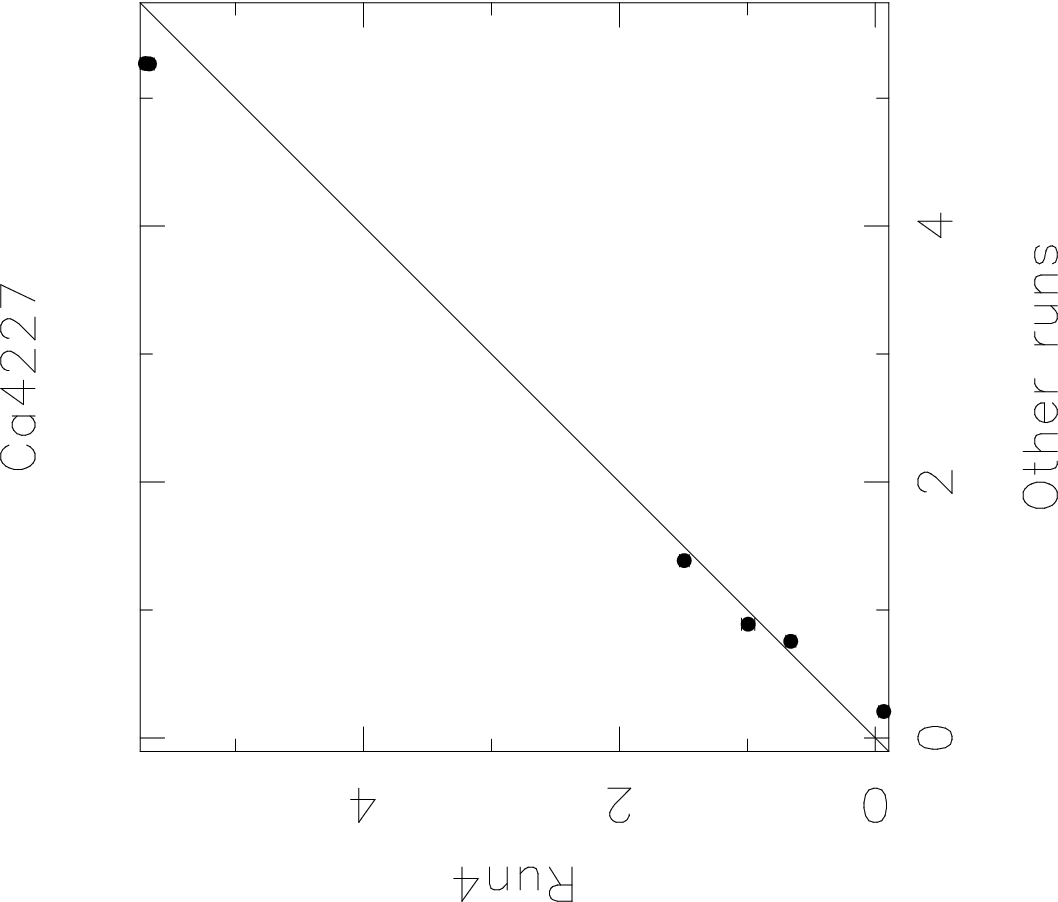}}    
\hfill
\resizebox{0.23\hsize}{!}{\includegraphics[angle=-90]{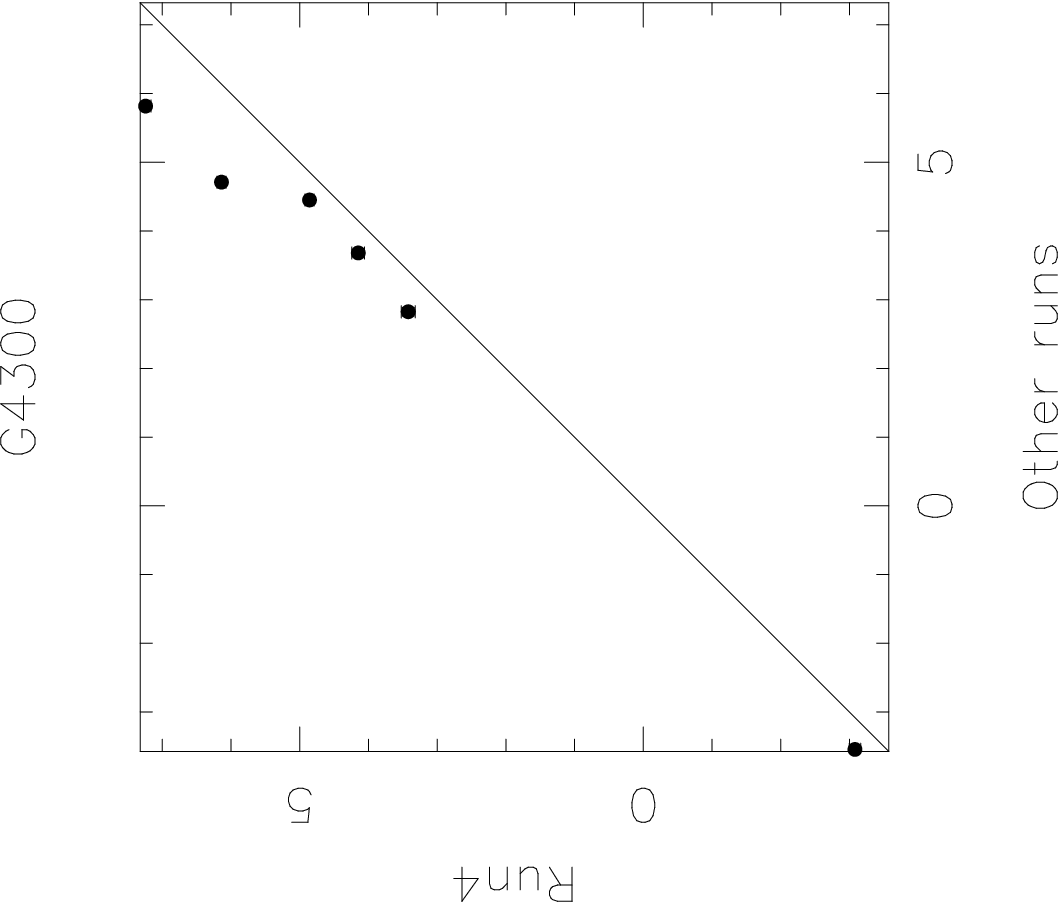}}  
\hfill
\resizebox{0.23\hsize}{!}{\includegraphics[angle=-90]{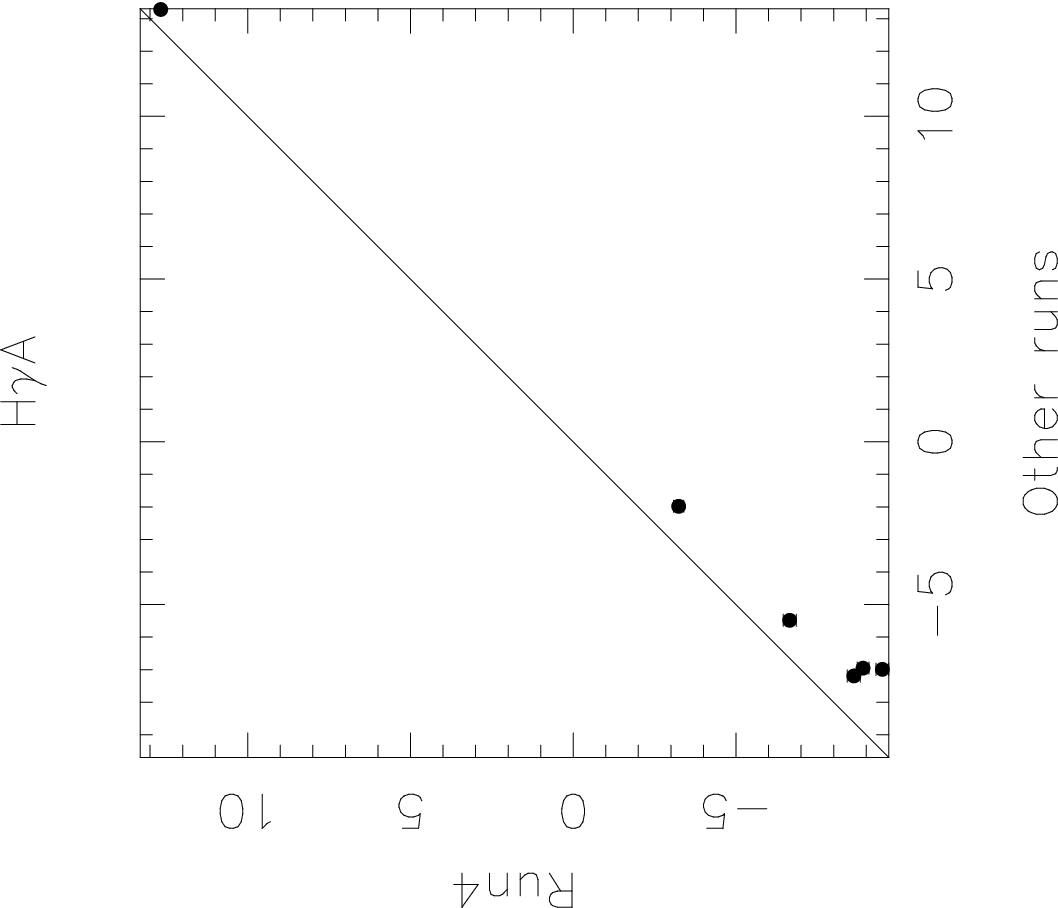}}
\hfill
\resizebox{0.23\hsize}{!}{\includegraphics[angle=-90]{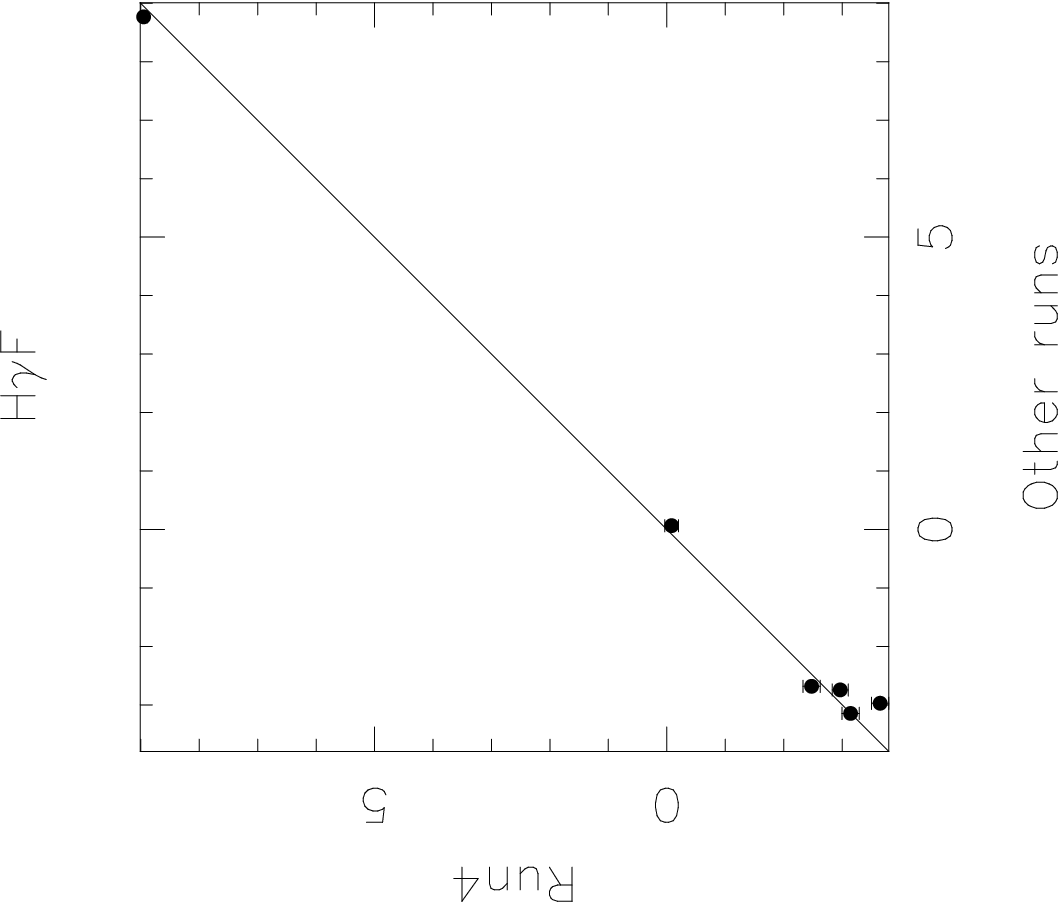}}    

\vspace{4mm}

\resizebox{0.23\hsize}{!}{\includegraphics[angle=-90]{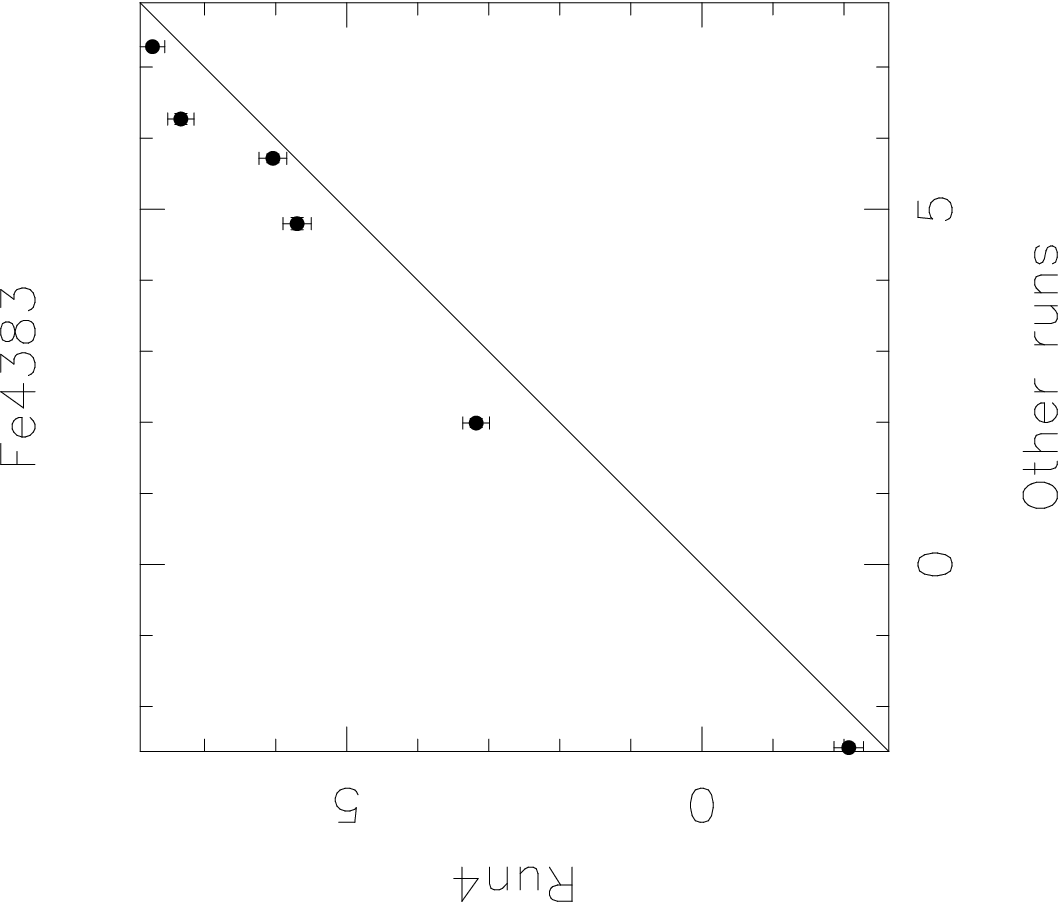}}  
\hfill
\resizebox{0.23\hsize}{!}{\includegraphics[angle=-90]{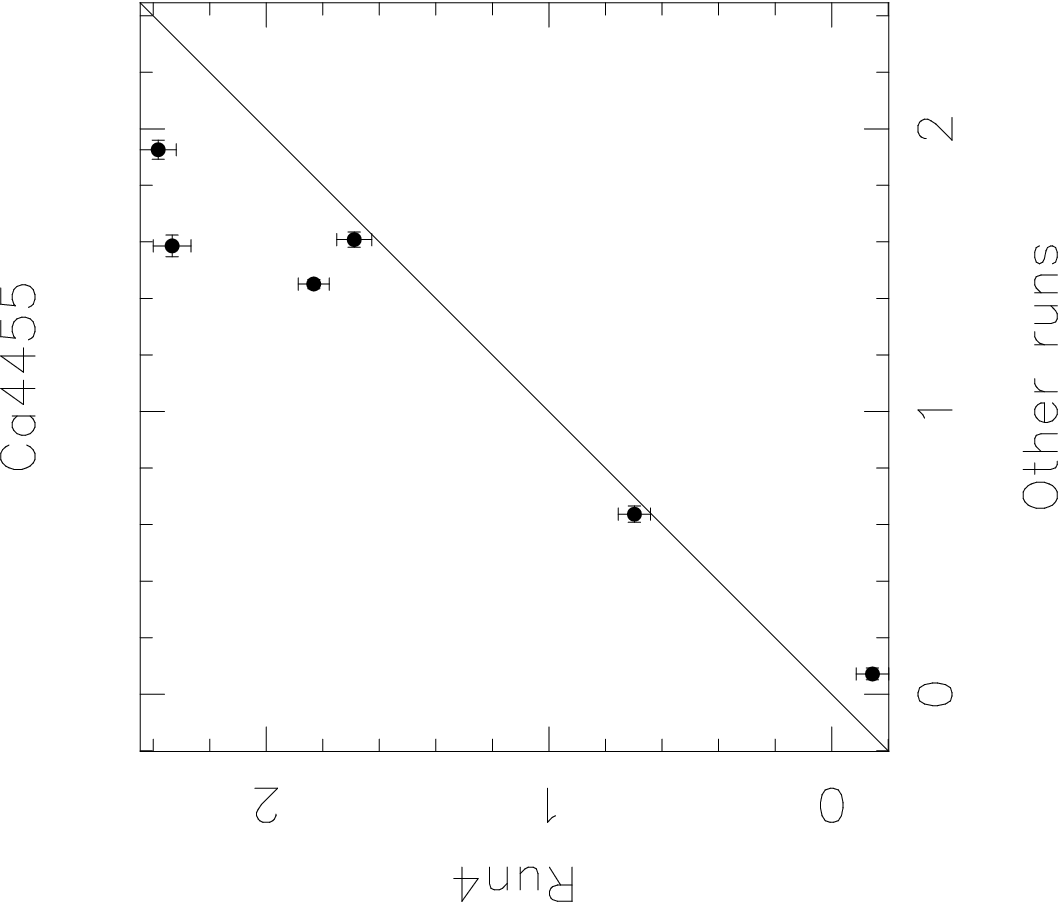}}
\hfill
\resizebox{0.23\hsize}{!}{\includegraphics[angle=-90]{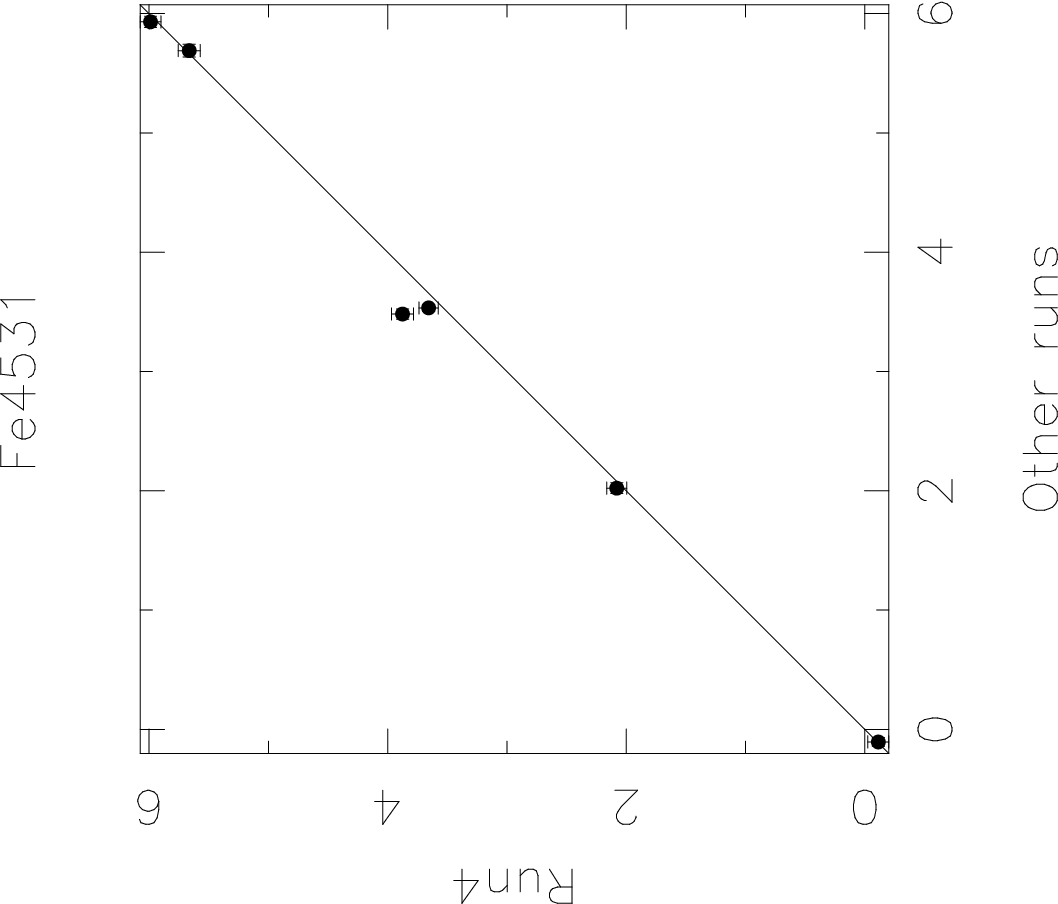}}
\hfill
\resizebox{0.23\hsize}{!}{\includegraphics[angle=-90]{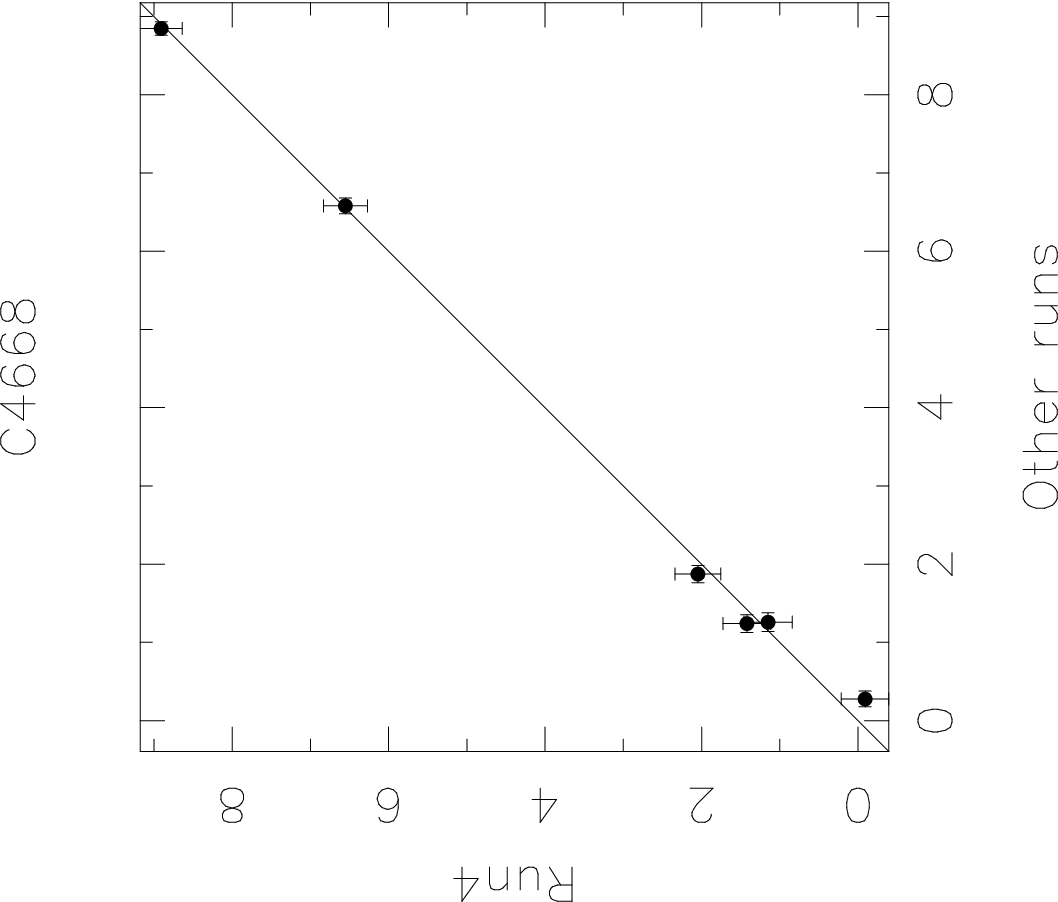}}

\vspace{4mm}

\resizebox{0.23\hsize}{!}{\includegraphics[angle=-90]{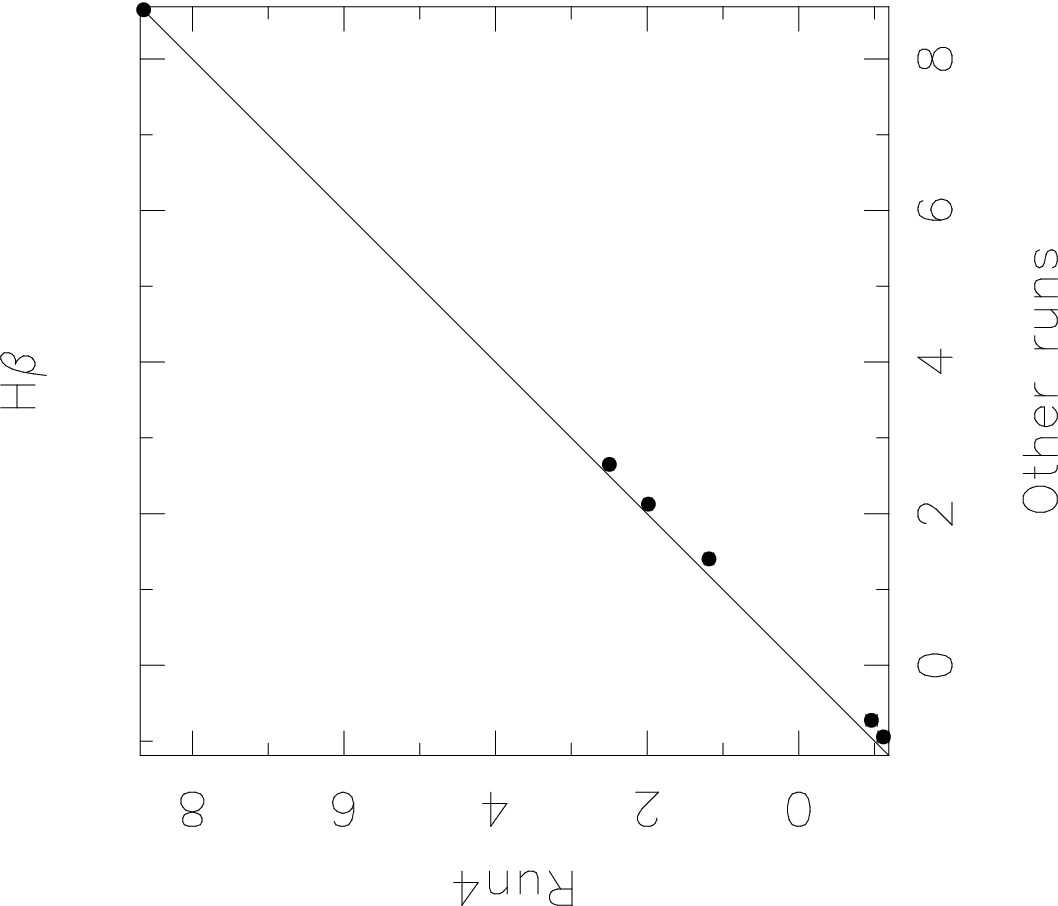}}    
\hfill
\resizebox{0.23\hsize}{!}{\includegraphics[angle=-90]{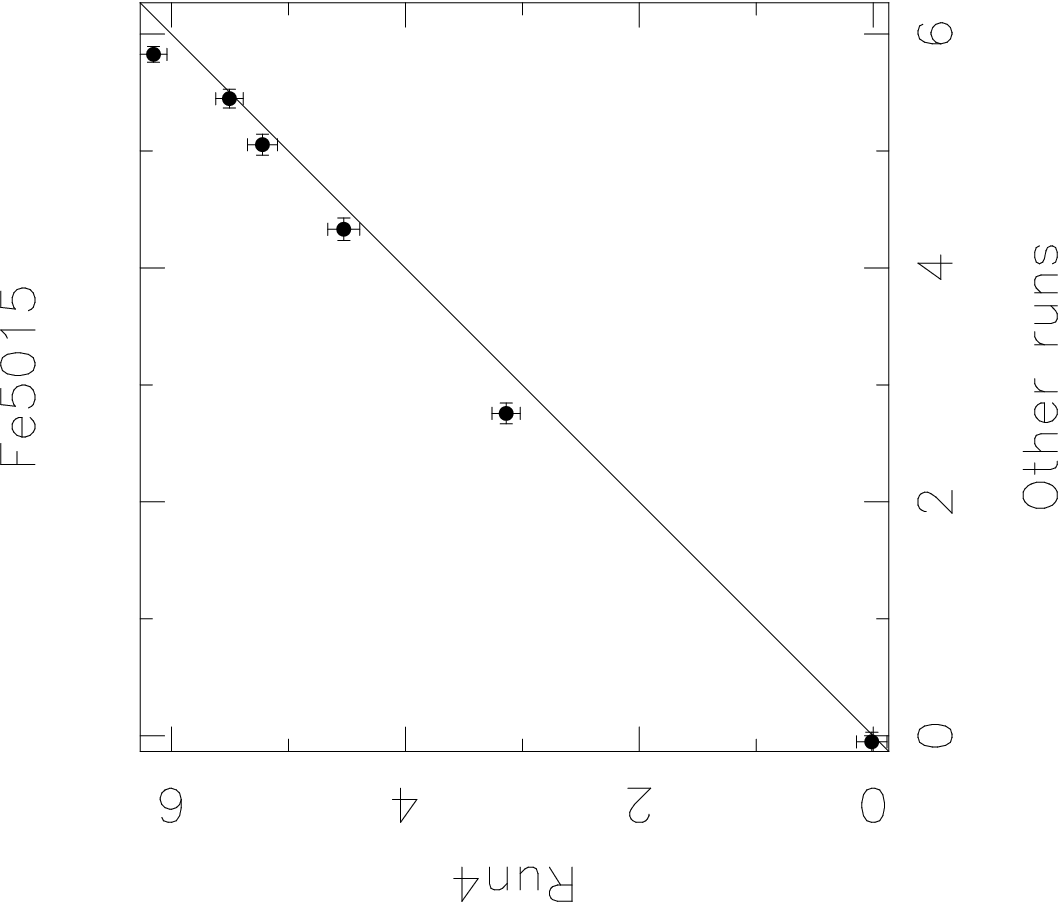}}   
\hfill
\resizebox{0.23\hsize}{!}{\includegraphics[angle=-90]{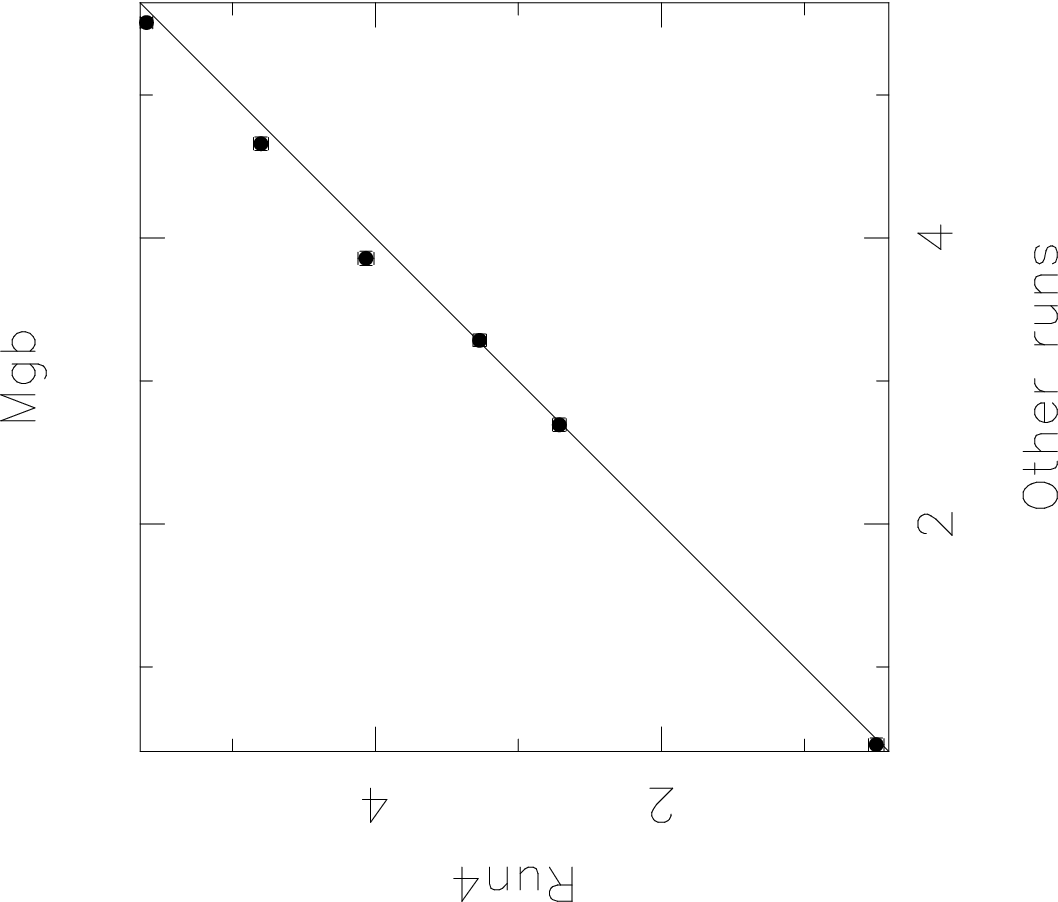}}    
\caption{Comparison of the indices measured in the common stars between 
Run 4 and the other runs.\label{fig.dc}}
\end{figure*}

Fig. \ref{fig.dc} shows the comparison of the indices
measured in stars observed in Run 4 and other runs. 
 Table \ref{tab.dc} indicates the mean offset, r.m.s dispersion,
 and r.m.s dispersion expected by errors. The last column of the 
 table shows the $z$ parameter of the comparison. A $z$-value 
 higher than 1.96 indicates that the offset is significant, with 
 a significance level lower than 0.05.
 Fe5270 and Fe5335 could not be measured in Run 2 and therefore
 are not shown in the table.
\begin{table}
\begin{tabular}{l rrrrr}
\hline\hline
Index       & offset &  $\sigma$ & $\sigma$(exp)  & $z$ \\
\hline
H$\delta_A$ &$-0.141$&  0.249    & 0.117          & 1.18\\
H$\delta_F$ &$-0.021$&  0.135    & 0.092          & 0.38\\
CN$_2$      &$-0.004$&  0.016    & 0.006          & 0.56\\
Ca4227      &$ 0.095$&  0.279    & 0.045          & 0.78\\
G4300       &$-0.017$&  0.524    & 0.087          & 0.08\\
H$\gamma_A$ &$-0.235$&  0.680    & 0.151          & 0.79\\
H$\gamma_F$ &$ 0.017$&  0.282    & 0.107          & 0.15\\
Fe4383      &$ 0.129$&  0.362    & 0.201          & 0.81\\
Ca4455      &$ 0.219$&  0.329    & 0.066          & 1.32\\
Fe4531      &$ 0.099$&  0.146    & 0.097          & 1.34\\
C4668       &$-0.010$&  0.200    & 0.310          & 0.12\\
H$\beta$    &$-0.045$&  0.091    & 0.063          & 1.06\\
Fe5015      &$ 0.004$&  0.137    & 0.149          & 0.07\\
Mgb         &$ 0.077$&  0.079    & 0.067          & 1.64\\
\hline
\end{tabular}
\caption{Mean offset, r.m.s ($\sigma$) and r.m.s expected by 
the errors ($\sigma$(exp)) in the comparison between stars observed
in Run 4 and in the other runs. Last column shows the $z$ parameter,
which indicate the significance of the mean offset. A $z$-value higher 
than 1.96 indicate that the offsets are significant with a significance 
level lower than 0.05.\label{tab.dc}}
\end{table}

%%%%%%%%%%%%%%%%%%%%%%%%%%%%%%%%%%%%%%%%%%%%%%%%%%%%%%%%%%%%%%%%%%%%%%%%%%%%%%%
\section{Velocity dispersion corrections for line-strengths}
\label{appen.index-sigma}

The observed spectrum of a galaxy is the convolution of the integrated spectrum
of its stellar population with the instrumental broadening and the distribution
of line-of-sight velocities of the stars. The broadening of the spectra causes
the indices to appear weaker than they intrinsically are.  To compare
line-strength indices between different galaxies it is necessary to correct for
the velocity dispersion and instrumental resolution broadening effects. This
was done by using the optimal template obtained in the calculation of the
velocity dispersion and radial velocity. This reference spectrum is the
composition of the template stellar spectra which best matches the spectrum of
each galaxy, corrected for differences in the intensities of the spectral
lines. We used these composite templates instead of individual spectra because
the variation of the indices with the broadening varies with their intensity
(see Kuntschner 2000).   

The templates were first broadened to the Lick resolution (which varies with
wavelength) and then convolved with a Gaussian function of widths ranging from
0 to 450 km s$^{-1}$ (in steps of 20 km s$^{-1}$) to simulate the velocity
dispersion broadening within a galaxy. Index strengths were measured for each
spectrum and a correction factor was calculated as
\begin{equation}
C(\sigma)=I(0)/I(\sigma),
\end{equation}
where $I(0)$ is the index measured in the template spectrum, previously
broadened to the instrumental resolution of Lick ($\sigma_{lick}$), 
and $I(\sigma)$ the index
measured in the same spectrum after being broadened by $\sigma$. Although a
different polynomial was calculated for each galaxy spectrum, an average
polynomial was applied to correct all the galaxies. This was done in order to
avoid possible systematic effects due to a non-perfect match between the
optimal template and the galaxy spectrum. Table \ref{correcsigma} shows the
final polynomial used to correct the indices for all the galaxies. The last
column gives an estimate of the maximum error in each index due to this
correction. This is calculated as the standard deviation of all the
coefficients $C(\sigma)$ obtained with the different templates computed at
$\sigma=400$ km s$^{-1}$. Note that all the corrections are multiplicative
except for the CN$_1$, CN$_2$, H$\delta_A$ and H$\delta_F$ indices for which the corrections are additive
$C(\sigma)=I(0)-I(\sigma)$.
\newpage 
\begin{table*}
\caption{Average polynomial coefficients for each line index correction
function, where $\sigma*=\sigma_{\rm Lick}/\sigma$. The last column shows the
maximum error due to this correction for a galaxy of $\sigma=400$ km s$^{-1}$.
\label{correcsigma}}.
\centering
\begin{tabular}{%
@{}lr@{$\;$}c@{$\;$}l@{$\times$}c@{$\;$}c@{$\;$}l@{$\times$}cr@{}}
\hline\hline
Index        & \multicolumn{7}{c}{$C(\sigma)$}& Error\\
\hline
H$\delta_A$ &$ 0.0118$&--&$0.0005      $&$\sigma  $&$-$&$1.5421 \ 10^{-6}$&$ \sigma^2$&0.175 \\
H$\delta_F$ &$-0.0013$&+ &$2.4565 \ 10^{-5}$&$\sigma  $&$+$&$1.2798 \ 10^{-6}$&$ \sigma^2$&0.068 \\
CN$_1$      &$-0.0001$&+ &$1.081 \ 10^{-5}$&$\sigma  $&$+$&$1.2259 \ 10^{-8}$&$ \sigma^2$&0.002 \\  
CN$_2$      &$-0.0008$&+ &$2.6876 \ 10^{-5}$&$\sigma  $&$+$&$6.1555 \ 10^{-8}$&$ \sigma^2$&0.006\\
Ca4227      &$ 0.7802$&--&$0.2809      $&$\sigma_*$&$+$&$0.5007$&$ \sigma_{*}^2$&0.059 \\
G4300       &$ 0.8915$&+ &$0.1000      $&$\sigma_*$&$+$&$0.0085$&$ \sigma_{*}^2$&0.014 \\
H$\gamma_A$ &$ 1.1184$&--&$0.2125      $&$\sigma_*$&$+$&$0.0940$&$ \sigma_{*}^2$&0.008 \\
H$\gamma_F$ &$ 0.8043$&+ &$0.2565      $&$\sigma_*$&$-$&$0.0608$&$ \sigma_{*}^2$&0.035 \\
Fe4383      &$ 0.8798$&--&$0.0095      $&$\sigma_*$&$+$&$0.1297$&$ \sigma_{*}^2$&0.025 \\
Ca4455      &$ 0.6385$&+ &$0.0962      $&$\sigma_*$&$+$&$0.2653$&$ \sigma_{*}^2$&0.034 \\
Fe4531      &$ 0.9133$&--&$0.0123      $&$\sigma_*$&$+$&$0.0990$&$ \sigma_{*}^2$&0.009 \\
C4668       &$ 1.0054$&--&$0.1268      $&$\sigma_*$&$+$&$0.1214$&$ \sigma_{*}^2$&0.013 \\
H$\beta$    &$ 0.9907$&+ &$0.0056      $&$\sigma_*$&$+$&$0.0037$&$ \sigma_{*}^2$&0.012 \\
Fe5015      &$ 0.7816$&+ &$0.1791      $&$\sigma_*$&$+$&$0.0393$&$ \sigma_{*}^2$&0.023 \\
Mgb         &$ 0.9645$&--&$0.0749      $&$\sigma_*$&$+$&$0.1104$&$ \sigma_{*}^2$&0.020 \\
Fe5270      &$ 0.8253$&+ &$0.1228      $&$\sigma_*$&$+$&$0.0518$&$ \sigma_{*}^2$&0.004 \\
Fe5335      &$ 0.8432$&--&$0.0814      $&$\sigma_*$&$+$&$0.2382$&$ \sigma_{*}^2$&0.051 \\
\hline
\end{tabular}
\end{table*}

\section{Comparison with other authors}
\label{appendix.offsets}

Although we have checked the possible existence of offsets between different observing
runs by comparing galaxies observed more than once, the number of available galaxies 
to make this comparison  was admittedly sparse. Therefore, in this appendix, we repeat the
comparison  with other authors subdividing the galaxies in different observing
runs. Table \ref{tab.comparison.runs} shows this comparison.

To check if the offsets with other authors were different in the 
different  observing runs, 
we have performed a $t$-test comparing the offsets of Runs 2, 3 and 4 with 
the offsets obtained in Run 1.
The results  are shown in Table \ref{t-test.oa}.
A value of $t$ larger than 1.96 indicates that the offset in the corresponding
run is significatively different than the offset obtained in Run 1 (with a
significance level lower than $\alpha=0.001$), and it would
suggest the existence of offsets between runs. The only two  cases in which that
happens are in the comparison of H$\gamma_A$ with the data of D05 (the offset in run 2 
is significatively larger than the one obtained in run1), and in the 
comparison of Fe4383 with the same reference, D05 (the offset in run 3 is
larger as well, but negative, than in the comparison with run1).
In the first case, the number of galaxies in common between  D05 and  run2 is
only 3, which makes the result very uncertain.
In the later case, we do not obtain significative differences in the 
offsets   of Fe4383 between run 3 and other runs when we compare with 
other references (different than D05).
We conclude, therefore,  that the measurements obtained in different 
observing runs do not present any systematic difference.

\newpage
\begin{table*}
\caption{Comparison of line strengths measured in this and other studies. Ref.:
reference of the comparison work (see description in the text); $N$: number of
galaxies in common; $\Delta I$: calculated offset between both studies (other
study minus this work); $\sigma$: r.m.s.\ dispersion; $\sigma_{\rm exp}$:
expected r.m.s.\ from the errors; $t$: $t$-parameter of the comparison of
means.
\label{tab.comparison.runs}}
\centering
\begin{tabular}{@{}lccrcccc@{}}
\hline\hline
Index    &  Ref. &  $N$  & \multicolumn{1}{c}{$\Delta I$ }& $\sigma$  & $\sigma_{\rm exp}$ & $t$\\
\hline
Run1     &      &       &            &           &                   &              \\
H$\delta_A$&C03 &  7    &$-0.095\pm0.129$    & 0.293     & 0.791             & 0.81  \\ 
H$\delta_A$&D05 &  13   &$ 0.188\pm 0.149$   & 0.459     & 0.531             & 1.36 \\
H$\gamma_A$&C03 &  7    &$ 0.379\pm0.135$    & 0.305     & 0.710             & 1.96   \\
H$\gamma_A$&D05 &  13   &$ 0.285\pm 0.130$   & 0.426     & 0.532             & 1.98 \\
H$\delta_F$&C03 &  7    &$-0.268\pm0.056$    & 0.126     & 0.586             & 2.25  \\
H$\delta_F$&D05 &  13   &$-0.217\pm 0.073$   & 0.185     & 0.399             & 2.68 \\
H$\gamma_F$&C03 &  7    &$ 0.415\pm0.053$    & 0.142     & 0.518             & 2.34  \\
H$\gamma_F$&D05 &  13   &$ 0.069\pm 0.128$   & 0.441     & 0.478             & 0.55 \\
CN$_2$     &T98 &  19   &$-0.006\pm0.006$    & 0.023     & 0.190             & 1.05   \\
CN$_2$    &C03 &  7     &$-0.006\pm0.023$    & 0.051     & 0.128             & 0.30  \\
CN$_2$     &D05 &  13   &$-0.014\pm 0.009$   & 0.028     & 0.093             & 1.62 \\
Ca4227     &T98 &  19   &$-0.150\pm0.064$    & 0.289     & 0.735             & 2.26    \\
Ca4227     &C03 &   7   &$-0.307\pm0.036$    & 0.082     & 0.469             & 2.38   \\ 
Ca4227     &D05 &  13   &$-0.208\pm 0.057$   & 0.141     & 0.301             & 2.90 \\
G4300      &T98 &  19   &$-0.088\pm0.120$    & 0.460     & 0.763             & 0.80   \\
G4300      &C03 &   7   &$-0.357\pm0.119$    & 0.269     & 0.607             & 2.01   \\
G4300      &D05 &  13   &$-0.168\pm 0.068$   & 0.222     & 0.474             & 2.14 \\
Fe4383     &T98 &  19   &$-0.101\pm0.144$    & 0.543     & 1.039             & 0.79   \\
Fe4383     &C03 &   7   &$ 0.066\pm0.099$   & 0.224     & 0.709             & 0.74    \\
Fe4383     &D05 &  13   &$-0.641\pm 0.120$   & 0.416     & 0.580             & 2.94 \\
Ca4455     &T98 &  19   &$-0.083\pm0.062$    & 0.236     & 0.658             & 1.44    \\
Ca4455     &D05 &  13   &$ 0.086\pm 0.060$   & 0.200     & 0.387             & 1.41 \\
Fe4531     &T98 &  19   &$-0.235\pm0.108$    & 0.417     & 0.837             & 2.13    \\
Fe4531     &D05 &  13   &$-0.212\pm 0.082$   & 0.261     & 0.482             & 2.24 \\
C4668      &T98 &  19   &$-0.716\pm0.170$    & 0.664     & 1.104             & 3.15    \\
C4668      &D05 &  13   &$-0.956\pm 0.185$   & 0.572     & 0.675             & 3.00 \\
H$\beta$   &G93 &  17   &$-0.208\pm0.028$    & 0.113     & 0.326             & 3.54     \\
H$\beta$   &T98 &  19   &$ 0.129\pm0.065$    & 0.259     & 0.586             & 1.80    \\
H$\beta$   &K01 &  8    &$-0.050\pm 0.043$   & 0.121     & 0.536             & 1.06 \\
H$\beta$   &C03 &   7   &$-0.016\pm0.049$    & 0.111     & 0.423             & 0.37    \\
H$\beta$   &D05 &  13   &$-0.179\pm 0.100$   & 0.220     & 0.362             & 2.23 \\
\hline
\end{tabular}
\end{table*}
\addtocounter{table}{-1}
\begin{table*}
\caption{\it Continued}
\centering
\begin{tabular}{@{}lccrcccc@{}}
\hline\hline
Index    &  Ref. &  $N$  & \multicolumn{1}{c}{$\Delta I$ }& $\sigma$  & $\sigma_{\rm exp}$ & $t$\\
\hline
Fe5015     &G93 &  17   &$-0.281\pm0.151$    & 0.610     & 0.626             & 1.72     \\
Fe5015     &T98 &  17   &$-0.251\pm0.149$    & 0.549     & 0.954             & 1.70    \\
Fe5015     &C03 &   7   &$ 0.328\pm 0.121$   & 0.281     & 0.674             & 1.92 \\        
Fe5015     &D05 &  13   &$-0.323\pm 0.231$   & 0.744     & 0.577             & 1.43 \\
Mgb        &G93 &  17   &$ -0.264\pm0.066$   & 0.267     & 0.375             & 2.86     \\
Mgb        &T98 &  19   &$ -0.159\pm0.063$   & 0.262     & 0.677             & 2.16    \\
Mgb        &K01 &  8    &$-0.123\pm 0.113$   & 0.316     & 0.579             & 1.02 \\
Mgb        &C03 &   7   &$  0.301\pm0.088$   & 0.207     & 0.509             & 2.07   \\
Mgb        &D05 &  13   &$-0.011\pm 0.077$   & 0.264     & 0.467             & 0.15 \\
Fe5270     &T98 &  19   &$ -0.088\pm0.060$   & 0.228     & 0.618             & 1.56    \\
Fe5270     &C03 &   7   &$ -0.188\pm0.098$   & 0.221     &  0.469            & 1.66    \\
Fe5270     &D05 &  13   &$-0.275\pm 0.057$   & 0.180     & 0.411             & 2.93 \\
Fe5335     &T98 &  19   &$-0.098\pm0.077$    & 0.287     &  0.720            & 1.41    \\
Fe5335     &D05 &  13   &$-0.243\pm 0.080$   & 0.248     & 0.446             & 2.47 \\
\end{tabular}
\end{table*}

\begin{table*}
\addtocounter{table}{-1}
\caption{Continued}
\centering
\begin{tabular}{@{}lccrcccc@{}}
\hline\hline
Index    &  Ref. &  $N$  &\multicolumn{1}{c}{$\Delta I$} & $\sigma$  & $\sigma_{\rm exp}$ & $t$\\
\hline
Run2     &       &       &            &           &                    &    &        \\
H$\delta_A$&D05&   3    &$-0.172\pm 0.251$       & 0.346             & 0.576 & 0.73\\
H$\delta_F$&D05&   3    &$-0.256\pm 0.046$       & 0.056             & 0.313 & 1.39\\
H$\gamma_A$&D05&   3    &$-0.044\pm 0.305$       & 0.477             & 0.533 & 0.16\\
H$\gamma_F$&D05&   3    &$-0.016\pm 0.390$       & 0.634             & 0.452 & 0.04\\
CN$_2$     &T98&   5    &$-0.017\pm 0.020$       & 0.035             & 0.190 & 0.99\\
CN$_2$     &D05&   3    &$ 0.008\pm 0.017$       & 0.020             & 0.104 & 0.60\\
Ca4227     &T98&   5    &$-0.265\pm 0.073$       & 0.120             & 0.709 & 1.85\\ 
Ca4227     &D05&   3    &$-0.194\pm 0.069$       & 0.088             & 0.215 & 1.33\\
G4300      &T98&   5    &$-0.103\pm 0.452$       & 0.767             & 0.769 & 0.30\\
G4300      &D05&   3    &$-0.151\pm 0.080$       & 0.134             & 0.479 & 1.14\\
Fe4383     &T98&   5    &$-0.118\pm 0.238$       & 0.408             & 1.039 & 0.62\\
Fe4383     &D05&   3    &$-0.348\pm 0.351$       & 0.549             & 0.610 & 0.87\\
Ca4455     &T98&   5    &$-0.107\pm 0.172$       & 0.279             & 0.635 & 0.99\\
Ca4455     &D05&   3    &$ 0.158\pm 0.151$       & 0.216             & 0.401 & 0.95\\
Fe4531     &T98&   5    &$-0.263\pm 0.180$       & 0.296             & 0.823 & 1.41\\
Fe4531     &D05&   3    &$-0.035\pm 0.236$       & 0.337             & 0.446 & 0.18\\
C4668      &T98&   5    &$-0.632\pm 0.521$       & 0.878             & 1.063 & 1.25\\
C4668      &D05&   3    &$-0.798\pm 0.385$       & 0.641             & 0.739 & 1.18\\
H$\beta$   &G93&   5    &$-0.314\pm 0.117$       & 0.231             & 0.329 & 1.46\\
H$\beta$   &T98&   5    &$ 0.120\pm 0.121$       & 0.204             & 0.575 & 1.10\\
H$\beta$   &D05&   3    &$-0.026\pm 0.184$       & 0.231             & 0.247 & 0.19\\
Fe5015     &G93&   4    &$-0.558\pm 0.281$       & 0.483             & 0.614 & 1.15\\
Fe5015     &T98&   3    &$-0.158\pm 0.165$       & 0.235             & 0.807 & 0.90\\
Fe5015     &D05&   3    &$-0.818\pm 0.575$       & 0.945             & 0.736 & 1.03\\
Mgb        &G93&   5    &$-0.406\pm 0.288$       & 0.572             & 0.374 & 1.10\\
Mgb        &T98&   5    &$-0.203\pm 0.342$       & 0.607             & 0.675 & 0.70\\ 
Mgb        &D05&   3    &$ 0.144\pm 0.018$       & 0.031             & 0.412 & 1.39\\
Fe5270     &T98&   5    &$-0.031\pm 0.134$       & 0.217             & 0.595 & 0.31\\
Fe5270     &D05&   3    &$-0.247\pm 0.164$       & 0.258             & 0.411 & 1.08\\
Fe5335     &T98&   5    &$ 0.094\pm 0.126$       & 0.203             & 0.680 & 1.02\\
Fe5335     &D05&   3    &$-0.044\pm 0.360$       & 0.545             & 0.464 & 0.14\\
\hline
\end{tabular}
\end{table*}
\begin{table*}
\addtocounter{table}{-1}
\caption{Continued}
\centering
\begin{tabular}{@{}lccrccc@{}}
\hline\hline
Index    &  Ref. &  $N$  &\multicolumn{1}{c}{$\Delta I$} & $\sigma$  & $\sigma_{\rm exp}$ & $t$\\
\hline
Run3     &       &       &            &           &                    &           \\
H$\delta_A$&C03&   9    &$-0.098\pm 0.132$    & 0.332      & 0.633              & 0.85\\
H$\delta_A$&D05&  12    &$ 0.266\pm 0.137$    & 0.435      & 0.577              & 1.78\\
H$\gamma_A$&C03&   9    &$ 0.453\pm 0.080$    & 0.201      & 0.572              & 2.61\\
H$\gamma_A$&D05&  12    &$ 0.567\pm 0.117$    & 0.361      & 0.569              & 2.83 \\
H$\delta_F$&C03&   9    &$-0.243\pm 0.050$    & 0.125      & 0.475              & 2.54\\
H$\delta_F$&D05&  12    &$-0.080\pm 0.065$    & 0.181      & 0.500              & 1.39 \\
H$\gamma_F$&C03&   9    &$ 0.422\pm 0.064$    & 0.476      & 0.421              & 2.66\\
H$\gamma_F$&D05&  12    &$ 0.321\pm 0.061$    & 0.182      & 0.449              & 2.91 \\
CN$_2$     &T98&  31    &$-0.017\pm 0.007$    & 0.037      & 0.203              & 2.31\\
CN$_2$     &C03&   9    &$-0.013\pm 0.007$    & 0.018      & 0.118              & 1.68\\
CN$_2$     &D05&  12    &$-0.036\pm 0.013$    & 0.033      & 0.094              & 2.49 \\
Ca4227     &T98&  31    &$-0.198\pm 0.072$    & 0.380      & 0.787              & 2.57\\
Ca4227     &C03&   9    &$-0.309\pm 0.028$    & 0.073      & 0.382              & 2.76\\
Ca4227     &D05&  12    &$-0.300\pm 0.073$    & 0.217      & 0.426              & 2.73 \\
G4300      &T98&  31    &$-0.374\pm 0.130$    & 0.695      & 0.839              & 2.63\\
G4300      &C03&   9    &$-0.331\pm 0.134$    & 0.337      & 0.490              & 2.04 \\
G4300      &D05&  12    &$ 0.009\pm 0.119$    & 0.363      & 0.513              & 0.09 \\
Fe4383     &T98&  31    &$-0.045\pm 0.115$    & 0.612      & 1.123              & 0.40\\
Fe4383     &C03&   9    &$-0.128\pm 0.063$    & 0.159      & 0.569              & 1.83\\
Fe4383     &D05&  12    &$-1.058\pm 0.121$    & 0.405      & 0.634              & 3.11\\
Ca4455     &T98&  31    &$-0.214\pm 0.059$    & 0.315      & 0.727              & 3.12\\
Ca4455     &D05&  12    &$-0.141\pm 0.048$    & 0.160      & 0.447              & 2.24\\
Fe4531     &T98&  31    &$-0.150\pm 0.077$    & 0.410      & 0.922              & 1.91\\
Fe4531     &D05&  12    &$-0.414\pm 0.066$    & 0.140      & 0.360              & 3.16 \\
C4668      &T98&  31    &$-0.606\pm 0.175$    & 0.934      & 1.150              & 3.02\\
C4668      &D05&  12    &$-0.865\pm 0.198$    & 0.571      & 0.635              & 2.80 \\
H$\beta$   &G93&  17    &$-0.057\pm 0.037$    & 0.150      & 0.333              & 1.45\\
H$\beta$   &T98&  31    &$ 0.139\pm 0.049$    & 0.265      & 0.646              & 2.57\\
H$\beta$   &K01&  17    &$-0.025\pm 0.055$    & 0.225      & 0.553              & 0.46\\
H$\beta$   &C03&   9    &$-0.083\pm 0.064$    & 0.162      & 0.346              & 1.35\\ 
H$\beta$   &D05&  12    &$-0.022\pm 0.083$    & 0.275      & 0.449              & 0.28 \\
Fe5015     &G93&  17    &$-0.451\pm 0.115$    & 0.465      & 0.612              & 2.83\\
Fe5015     &T98&  31    &$-0.322\pm 0.133$    & 0.708      & 1.067              & 2.30\\
Fe5015     &C03&   9    &$-0.268\pm 0.093$    & 0.237      & 0.530              & 2.17\\
Fe5015     &D05&  12    &$-0.543\pm 0.200$    & 0.632      & 0.601              & 2.21\\
\hline
\end{tabular}
\end{table*}
\begin{table*}
\addtocounter{table}{-1}
\caption{Continued}
\centering
\begin{tabular}{@{}lccrccc@{}}
\hline\hline
Index    &  Ref. &  $N$  &\multicolumn{1}{c}{$\Delta I$} & $\sigma$  & $\sigma_{\rm exp}$ & $t$\\
\hline
Run3     &       &       &            &           &                    &           \\
Mgb        &G93&  17    &$-0.385\pm 0.149$    & 0.601      & 0.386              & 2.20\\
Mgb        &T98&  31    &$-0.404\pm 0.100$    & 0.537      & 0.742              & 3.33\\
Mgb        &K01&  17    &$-0.168\pm 0.133$    & 0.538      & 0.606              & 1.23\\
Mgb        &C03&   9    &$ 0.186\pm 0.080$    & 0.214      & 0.440              & 0.50\\
Mgb        &D05&  12    &$-0.210\pm 0.105$    & 0.346      & 0.458              & 2.44\\
Fe5270     &T98&  30    &$ 0.049\pm 0.038$    & 0.198      & 0.693              & 1.32 \\
Fe5270     &C03&   9    &$ 0.073\pm 0.071$    & 0.176      & 0.395              & 1.14\\
Fe5270     &D05&  12    &$-0.143\pm 0.051$    & 0.162      & 0.380              & 2.25\\
Fe5335     &T98&  30    &$ 0.083\pm 0.056$    & 0.288      & 0.784              & 1.52\\
Fe5335     &D05&  12    &$-0.263\pm 0.060$    & 0.176      & 0.448              & 2.79\\
\hline
\end{tabular}
\end{table*}
\newpage
\begin{table*}
\addtocounter{table}{-1}
\caption{Continued}
\centering
\begin{tabular}{@{}lccrccc@{}}
\hline\hline
Index    &  Ref. &  $N$  &\multicolumn{1}{c}{$\Delta I$} & $\sigma$  & $\sigma_{\rm exp}$ & $t$\\
\hline
Run 4    &       &       &                    &           &                    &      \\
CN$_2$   &  T98  &  7    &$-0.010\pm 0.014$   & 0.036     & 0.216              & 0.80\\
Ca4227   &  T98  &  7    &$-0.049\pm 0.093$   & 0.237     & 0.895              & 0.54\\
G4300    &  T98  &  7    &$-0.407\pm 0.376$   & 0.954     & 0.914              & 1.03\\
Fe4383   &  T98  &  7    &$ 0.181\pm 0.478$   & 0.478     & 1.232              & 0.93\\
Ca4455   &  T98  &  7    &$-0.058\pm 0.182$   & 0.460     & 0.831              & 0.33\\
Fe4531   &  T98  &  7    &$-0.310\pm 0.174$   & 0.441     & 1.028              & 1.48\\
C4668    &  T98  &  7    &$-0.659\pm 0.327$   & 0.831     & 1.310              & 1.59\\
C4668    &  M02  &  11   &$-1.133\pm0.446$    & 1.806     & 1.131              & 2.08\\  
H$\beta$ &  T98  &  7    &$ 0.270\pm 0.099$   & 0.251     & 0.690              & 1.86\\
H$\beta$ &  M00  &  6     &$-0.428\pm 0.154$   & 0.316    & 0.418              & 1.86\\
H$\beta$ &  K01  &   9   &$-0.105\pm0.046$    & 0.138     & 0.618              & 1.77\\
Fe5015   &  T98  &  7    &$-0.068\pm 0.118$   & 0.301     & 1.190              & 0.58\\
Fe5015   &  M02  &  11   &$0.003\pm 0.248$    & 0.664     & 1.031              & 0.01\\
Mgb      &  T98  &  7    &$-0.493\pm 0.257$   & 0.657     & 0.840              & 1.54\\
Mgb      &  M00  &  6     &$-0.158\pm 0.099$   & 0.220    & 0.491              & 1.39\\
Mgb      &  K01  &   9   &$-0.112\pm0.049$    & 0.189     & 0.664              & 1.78\\  
Fe5270   &  T98  &  7    &$ 0.237\pm 0.080$   & 0.202     & 0.769              & 1.92\\
Fe5270   &  M00  &  6     &$ 0.212\pm 0.070$   & 0.153    & 0.480              & 1.87\\
Fe5270   &  M02  &  11   &$ 0.107\pm0.067$    & 0.218     & 0.490              & 1.55\\
Fe5335   &  T98  &  7    &$ 0.311\pm 0.211$   &  0.530    & 0.922              & 1.31\\
Fe5335   &  M00  &  6     &$ 0.111\pm 0.182$   & 0.398    & 0.514              & 0.65\\
Fe5335   &  M02  &  11   &$ 0.088\pm0.110$    & 0.358     & 0.607              & 0.79 \\
\hline
\end{tabular}
\end{table*}
\newpage
\begin{table*}
\caption{$t$-parameter in the comparison of the offsets between different 
observing runs and 
other authors. A $t$-value higher
than 1.96 indicates that the offsets are significatively different, which
would suggests the existence of systematic differences  between 
observing runs.}
\label{t-test.oa}
\centering
\begin{tabular}{@{}lccccc@{}}
\hline\hline
 Indice    & Ref    &  Run 1&        Run 2 &       Run 3&        Run4\\
\hline
H$\delta_A$ &(C03) &  0   &                &      0.02 &        \\
H$\delta_A$ &(D05) &  0   &         1.23   &      0.38 &        \\
H$\delta_F$ &(D05) &  0   &         0.45   &      1.40 &        \\
H$\delta_F$ &(C03) &  0   &         0.33   &           &        \\
H$\gamma_A$ &(C03) &  0   &                &      0.47 &        \\
H$\gamma_A$ &(D05) &  0   &         3.84   &      1.60 &        \\
H$\gamma_F$ &(C03) &  0   &                &      0.08 &        \\
H$\gamma_F$ &(D05) &  0   &         0.21   &      1.77 &        \\
CN$_2$      &(T98) &  0   &         0.52   &      1.19 &     0.26\\
CN$_2$      &(D05) &  0   &         1.14   &      1.39 &        \\
CN$_2$      &(C03) &  0   &                &      0.29 &        \\
Ca4227      &(T98) &  0   &         1.18   &      0.50 &        0.89\\
Ca4227      &(C03) &  0   &                &      0.04 &        \\
Ca4227      &(D05) &  0   &         0.16   &      0.99 &        \\
G4300       &(T98) &  0   &         0.03   &      1.60 &        0.81\\
G4300       &(C02) &  0   &                &      0.14 &        \\
G4300       &(D05) &  0   &         0.16   &      1.29 &        \\
Fe4383      &(T98) &  0   &         0.06   &      0.30 &        0.56\\
Fe4383      &(C03) &  0   &                &      1.65 &        \\
Fe4383      &(D05) &  0   &         0.78   &      2.44 &        \\
Ca4455      &(T98) &  0   &         0.13   &      1.53 &        0.13\\
Ca4455      &(D05) &  0   &         0.44   &      0.74 &        \\
Fe4531      &(T98) &  0   &         0.13   &      0.64 &        0.36\\
Fe4531      &(D05) &  0   &         0.71   &      1.91 &        \\
C4668       &(T98) &  0   &         0.15   &      0.45 &        0.15\\
C4668       &(D05) &  0   &         0.37   &      0.33 &        \\
H$\beta$    &(T98) &  0   &         0.06   &      0.12 &        1.19\\
H$\beta$    &(K01) &  0   &                &      0.35 &        0.87\\
H$\beta$    &(C03) &  0   &                &      0.83 &        \\
H$\beta$    &(D05) &  0   &         0.73   &      1.20 &        \\
Fe5015      &(G93) &  0   &         0.86   &      0.89 &        \\
Fe5015      &(T98) &  0   &         0.41   &      0.35 &        0.96\\
Fe5015      &(C03) &  0   &                &      0.39 &        \\
Fe5015      &(D05) &  0   &         0.80   &      0.72 &        \\
\hline
\end{tabular}
\end{table*}
\newpage

\addtocounter{table}{-1}
\begin{table*}
\caption{Continued}
\centering
\begin{tabular}{@{}lccccc@{}}
\hline\hline
 Indice    & Ref    &  Run 1&        Run 2 &       Run 3&        Run4\\
\hline
Mgb         &(G93) &  0   &         0.48   &      0.74 &        \\
Mgb         &(T98) &  0   &         0.13   &      2.07 &        1.26\\
Mgb         &(K01) &  0   &                &      0.25 &        0.09\\
Mgb         &(C03) &  0   &                &      0.96 &        \\
Mgb         &(D05) &  0   &         1.96   &      0.54 &        \\
Fe5270      &(T98) &  0   &         0.38   &      1.92 &        3.25\\
Fe5270      &(C03) &  0   &                &      2.16 &        \\
Fe5270      &(D05) &  0   &         0.16   &      1.72 &        \\
Fe5335      &(T98) &  0   &         1.30   &      1.90 &        0.94\\
Fe5335      &(D05) &  0   &         0.54   &      0.20 &        \\
\hline
\end{tabular}
\end{table*}

\end{document}